\documentclass[aps,pre,eqsecnum,showpacs,superscriptaddress,twocolumn]{revtex4}
\pdfoutput=1

\usepackage{amsmath}
\usepackage{cases}
\usepackage{bm}
\usepackage{graphicx}

\newcommand{\nc}{\newcommand}
\nc{\dmo}{\DeclareMathOperator}

\dmo{\arsinh}{arsinh}
\dmo{\Li}{Li}
\dmo{\Rp}{Re}

\nc{\bse}{\begin{subequations}}
\nc{\ese}{\end{subequations}}
\nc{\ba}{\begin{array}}
\nc{\ea}{\end{array}}
\nc{\bmp}{\begin{minipage}}
\nc{\emp}{\end{minipage}}
\nc{\bwt}{\begin{widetext}}
\nc{\ewt}{\end{widetext}}
\nc{\nn}{\nonumber}

\nc{\ds}{\displaystyle}
\nc{\ts}{\textstyle}

\nc{\scs}{\scriptsize}

\nc{\al}{\alpha}
\nc{\be}{\beta}
\nc{\ga}{\gamma}
\nc{\de}{\delta}
\nc{\ve}{\varepsilon}
\nc{\la}{\lambda}
\nc{\om}{\omega}
\nc{\vp}{\varphi}
\nc{\vt}{\vartheta}
\nc{\ze}{\zeta}
\nc{\De}{\Delta}
\nc{\Ga}{\Gamma}
\nc{\La}{\Lambda}
\nc{\Th}{\Theta}

\nc{\Ccal}{\mathcal{C}}
\nc{\Fcal}{\mathcal{F}}
\nc{\Gcal}{\mathcal{G}}
\nc{\Hcal}{\mathcal{H}}
\nc{\Ical}{\mathcal{I}}
\nc{\Kcal}{\mathcal{K}}
\nc{\Ncal}{\mathcal{N}}
\nc{\Ocal}{\mathcal{O}}

\nc{\f}[2]{\frac{#1}{#2}}
\nc{\fr}[2]{{\ts\frac{#1}{#2}}}
\nc{\ph}{\phantom}
\nc{\ptl}{\partial}

\nc{\film}{\text{film}}
\nc{\filmaniso}{\text{film,aniso}}
\nc{\bulk}{\text{bulk}}

\nc{\rz}{r_0}
\nc{\rzt}{\tilde{r}_0}
\nc{\rztp}{\tilde{r}_{0,\perp}}
\nc{\rzcf}{r_\text{0c,film}}

\nc{\Ic}[1]{\Ical_{#1}}
\nc{\Kc}[1]{\Kcal_{#1}}

\nc{\kB}{k_\text{B}}
\nc{\Tc}{T_\text{c}}
\nc{\Tcfilm}{T_\text{c,film}}
\nc{\Tcbulk}{T_\text{c,bulk}}
\nc{\tcfilm}{t_\text{c,film}}
\nc{\xie}{\xi_\text{e}}
\nc{\bt}{\tilde{b}}
\nc{\Bt}{\widetilde{B}}
\nc{\btd}{\bt_d}
\nc{\Btd}{\Bt_d}
\nc{\fb}{f_\text{b}}
\nc{\fs}{f_\text{s}}
\nc{\fns}{f_\text{ns}}
\nc{\fbs}{f_\text{b,s}}
\nc{\fsf}{f_\text{sf}}
\nc{\fbns}{f_\text{b,ns}}
\nc{\fsfns}{f_\text{sf,ns}}
\nc{\fsfs}{f_\text{sf,s}}
\nc{\fex}{f_\text{ex}}
\nc{\fexs}{f_\text{ex,s}}
\nc{\ffilm}{f_\film}
\nc{\ffilms}{f_\text{film,s}}
\nc{\uf}{U}
\nc{\ubs}{U_\text{b,s}}
\nc{\usf}{U_\text{sf}}
\nc{\usfs}{U_\text{sf,s}}

\nc{\tbc}{\text{($\tau$)}}
\nc{\ttbc}{\text{($\tau\tau$)}}
\nc{\pbc}{\text{(p)}}
\nc{\abc}{\text{(a)}}
\nc{\Nbc}{\text{(N)}}
\nc{\Dbc}{\text{(D)}}
\nc{\NNbc}{\text{(NN)}}
\nc{\DDbc}{\text{(DD)}}
\nc{\NDbc}{\text{(ND)}}
\nc{\freebc}{\text{(free)}}

\nc{\xiz}{\xi_0}
\nc{\xif}{\xi_\film}
\nc{\xifa}{\xi_\filmaniso}

\nc{\Asf}{A_\text{sf}}
\nc{\ACsf}{A_\text{$C$,sf}}

\nc{\Sr}[1]{S_{#1}}
\nc{\Sf}[1]{\hat{S}_{#1}}

\nc{\at}{\tilde{a}}
\nc{\xt}{{\tilde{x}}}
\nc{\yi}{y}
\nc{\zi}{z}
\nc{\yv}{y}
\nc{\B}{B}
\nc{\W}[1]{W_{#1}}
\nc{\Wt}[1]{\widetilde{W}_{#1}}
\nc{\FCas}{F_\text{Cas}}
\nc{\FCass}{F_\text{Cas,s}}
\nc{\Lpl}{L_\parallel}
\nc{\Jpl}{J_\parallel}
\nc{\Jpp}{J_\perp}
\nc{\Npl}{N_\parallel}
\nc{\ul}{u_L}
\nc{\Zgc}{Z}
\nc{\xb}{\bm{x}}
\nc{\yb}{\bm{y}}
\nc{\pb}{\bm{p}}
\nc{\kb}{\bm{k}}
\nc{\vpb}{\bm{\vp}}
\nc{\Ab}{\bm{A}}
\nc{\Abarb}{\bm{\bar{A}}}
\nc{\az}{a_0}

\nc{\arXiv}[3]{arXiv:#1.#2v#3}
\nc{\blank}[3]{{\bf #1}, #2 (#3)}
\nc{\ibid}[3]{{\em ibid.} {\bf #1}, #2 (#3)}
\nc{\ANYAS}[3]{Ann.\ N.Y.\ Acad.\ Sci.\ \textbf{#1}, #2 (#3)}
\nc{\APNY}[3]{Ann.\ Phys.\ (N.Y.) \textbf{#1}, #2 (#3)}
\nc{\CPC}[3]{Comp.\ Phys.\ Comm.\ \textbf{#1}, #2 (#3)}
\nc{\CRSB}[3]{C.\ R.\ Seances Acad.\ Sci., Ser.\ B \textbf{#1}, #2 (#3)}
\nc{\EPJB}[3]{Eur.\ Phys.\ J.\ B \textbf{#1}, #2 (#3)}
\nc{\EPL}[3]{Europhys.\ Lett.\ \textbf{#1}, #2 (#3)}
\nc{\IJMPB}[3]{Int.\ J.\ Mod.\ Phys.\ B \textbf{#1}, #2 (#3)}
\nc{\JChP}[3]{J.\ Chem.\ Phys.\ \textbf{#1}, #2 (#3)}
\nc{\JChSFT}[3]{J.\ Chem.\ Soc., Faraday Trans.\ II \textbf{#1}, #2 (#3)}
\nc{\JCpP}[3]{J.\ Comput.\ Phys.\ \textbf{#1}, #2 (#3)}
\nc{\JHEP}[3]{JHEP \textbf{#1}, #2 (#3)}
\nc{\JLTP}[3]{J.\ Low Temp.\ Phys.\ \textbf{#1}, #2 (#3)}
\nc{\JMP}[3]{J.\ Math.\ Phys.\ \textbf{#1}, #2 (#3)}
\nc{\JPA}[3]{J.\ Phys.\ A \textbf{#1}, #2 (#3)}
\nc{\JPAMG}[3]{J.\ Phys.\ A: Math.\ Gen.\ \textbf{#1}, #2 (#3)}
\nc{\JPB}[3]{J.\ Phys.\ B \textbf{#1}, #2 (#3)}
\nc{\JPCM}[3]{J.\ Phys.: Condens.\ Matter \textbf{#1}, #2, (#3)}
\nc{\JPF}[3]{J.~Phys.~(France) \textbf{#1}, #2 (#3)}
\nc{\JTB}[3]{J.\ Theor.\ Biol.\ \textbf{#1}, #2 (#3)}
\nc{\LP}[3]{Laser Phys.\ \textbf{#1}, #2 (#3)}
\nc{\MPLB}[3]{Mod.\ Phys.\ Lett.\ B \textbf{#1}, #2 (#3)}
\nc{\MUPB}[3]{Moscow Univ.\ Phys.\ Bull.\ \textbf{#1}, #2 (#3)}
\nc{\NCD}[3]{Nuovo Cimento D \textbf{#1}, #2 (#3)}
\nc{\NCL}[3]{Nuovo Cimento Lett.\ \textbf{#1}, #2 (#3)}
\nc{\NPB}[3]{Nucl.\ Phys.\ B \textbf{#1}, #2 (#3)}
\nc{\PA}[3]{Physica A \textbf{#1}, #2 (#3)}
\nc{\PB}[3]{Physica B \textbf{#1}, #2 (#3)}
\nc{\PJ}[3]{Physik Journal \textbf{#1}, #2 (#3)}
\nc{\PLA}[3]{Phys.\ Lett.\ A \textbf{#1}, #2 (#3)}
\nc{\PLB}[3]{Phys.\ Lett.\ B \textbf{#1}, #2 (#3)}
\nc{\PR}[3]{Phys.\ Rev.\ \textbf{#1}, #2 (#3)}
\nc{\PRL}[3]{Phys.\ Rev.\ Lett.\ \textbf{#1}, #2 (#3)}
\nc{\PRA}[3]{Phys.\ Rev.\ A \textbf{#1}, #2 (#3)}
\nc{\PRB}[3]{Phys.\ Rev.\ B \textbf{#1}, #2 (#3)}
\nc{\PRC}[3]{Phys.\ Rev.\ C \textbf{#1}, #2 (#3)}
\nc{\PRD}[3]{Phys.\ Rev.\ D \textbf{#1}, #2 (#3)}
\nc{\PRE}[3]{Phys.\ Rev.\ E \textbf{#1}, #2 (#3)}
\nc{\PREP}[3]{Phys.\ Rep.\ \textbf{#1}, #2 (#3)}
\nc{\PST}[3]{Phys.\ Scr.\ T \textbf{#1}, #2 (#3)}
\nc{\RMP}[3]{Rev.\ Mod.\ Phys.\ \textbf{#1}, #2 (#3)}
\nc{\Sc}[3]{Science \textbf{#1}, #2 (#3)}
\nc{\SPJETP}[3]{Sov.\ Phys.\ JETP \textbf{#1}, #2 (#3)}
\nc{\TMF}[3]{Teor.\ Mat.\ Fiz.\ \textbf{#1}, #2 (#3)}
\nc{\ZN}[3]{Z.\ Naturforsch.\ \textbf{#1}, #2 (#3)}
\nc{\ZPBCM}[3]{Z.\ Phys.\ B: Condens.\ Matter \textbf{#1}, #2 (#3)}

\begin{document}

\title{Finite-size effects in film geometry
with nonperiodic boundary conditions:\\
Gaussian model and renormalization-group theory at fixed dimension}

\author{Boris Kastening}
\email{bkastening@matgeo.tu-darmstadt.de}
\affiliation{Institute for Theoretical Physics,
RWTH Aachen University, 52056 Aachen, Germany}
\affiliation{Institute for Materials Science,
TU Darmstadt, 64287 Darmstadt, Germany }
\author{Volker Dohm}
\email{vdohm@physik.rwth-aachen.de}
\affiliation{Institute for Theoretical Physics,
RWTH Aachen University, 52056 Aachen, Germany}

\date{11 May 2010}

\begin{abstract}
Finite-size effects are investigated in the Gaussian model with isotropic
and anisotropic short-range interactions in film geometry with nonperiodic
boundary conditions (b.c.)\ above, at, and below the bulk critical
temperature $\Tc$.
We have obtained exact results for the free energy and the Casimir force
for antiperiodic, Neumann, Dirichlet, and Neumann-Dirichlet mixed b.c.\ in
$1<d<4$ dimensions.
For the Casimir force, finite-size scaling is found to be valid for all
b.c..
For the free energy, finite-size scaling is valid in $1<d<3$ and $3<d<4$
dimensions for antiperiodic, Neumann, and Dirichlet b.c., but logarithmic
deviations from finite-size scaling exist in $d=3$ dimensions for Neumann
and Dirichlet b.c..
This is explained in terms of the borderline dimension $d^*=3$, where the
critical exponent $1-\al-\nu=(d-3)/2$ of the Gaussian surface energy
density vanishes.
For Neumann-Dirichlet b.c., finite-size scaling is strongly violated above
$\Tc$ for $1<d<4$ because of a cancelation of the leading scaling terms.
For antiperiodic, Dirichlet, and Neumann-Dirichlet b.c., a finite film
critical temperature $\Tcfilm(L)<\Tc$ exists at finite film thickness
$L$.
Our results include an exact description of the dimensional crossover
between the $d$-dimensional finite-size critical behavior near bulk $\Tc$
and the $(d{-}1)$-dimensional critical behavior near $\Tcfilm(L)$.
This dimensional crossover is illustrated for the critical behavior of the
specific heat.
Particular attention is paid to an appropriate representation of the free
energy in the region $\Tcfilm(L)\leq T\leq\Tc$.
For $2<d<4$, the Gaussian results are renormalized and reformulated as
one-loop contributions of the $\vp^4$ field theory at fixed dimension~$d$
and are then compared with the $\ve=4-d$ expansion results at $\ve=1$ as
well as with $d=3$ Monte Carlo data.
For $d=2$, the Gaussian results for the Casimir force scaling function are
compared with those for the Ising model with periodic, antiperiodic, and
free b.c.; unexpected exact relations are found between the Gaussian and
Ising scaling functions.
For both the $d$-dimensional Gaussian model and the two-dimensional Ising
model it is shown that anisotropic couplings imply nonuniversal
scaling functions of the Casimir force that depend explicitly on
microscopic couplings.
Our Gaussian results provide the basis for the investigation of
finite-size effects of the mean spherical model in film geometry with
nonperiodic b.c.\ above, at, and below the bulk critical temperature.

\end{abstract}



\maketitle

\section{Introduction and Summary}
\label{G.introduction}

Critical phenomena in confined systems have remained an important topic
of research over the past decades.
Much interest has been devoted to systems confined to film geometry which
are well accessible to accurate experiments, e.g., measurements of the
critical specific heat and of the critical Casimir force in superfluid
films near the $\la$ transition of $^4$He and $^3$He-$^4$He mixtures
\cite{gasparini,GaCh99} and in binary wetting films near the demixing
critical point \cite{fukuto}.
To some extent, these phenomena have been reproduced by Monte Carlo (MC)
simulations of lattice models in finite-slab geometries
\cite{schultka_hasenbusch09,Hu07_VaGaMaDi07,VaGaMaDi08}.
While progress has been made in the theoretical understanding of these
phenomena {\it above and at} the bulk critical temperature $\Tc$ of
three-dimensional systems
\cite{huhn_schmolke,KrDi92a,KrDi92b,Kr94,dohm93_sutter_mohr,sutter,%
toepler,DiGrSh06,GrDi07,Borjan}, there exists a substantial lack of
knowledge in the analytic description of three-dimensional systems in
film geometry {\it below} bulk $\Tc$, except for the case of periodic
boundary conditions (b.c.) \cite{Do08b}, except for the study of
qualitative features of the critical Casimir force
\cite{zandi04_zandi07_mac}, and except for the study of dynamic surface
properties \cite{frank}.
Also for two-dimensional systems in strip geometry, only a few analytical
results have been known for the critical Casimir force
\cite{Indekeu,Pr90,priv,Kr94,Evans} in the past.
Analytic expressions for the Casimir force scaling functions of the
two-dimensional Ising model are known for free and fixed b.c.~\cite{Evans}
and only since very recently for periodic and antiperiodic
b.c.~\cite{RuZaShAb10}.
On the other hand, to the best of our knowledge, no complete analytic
results for the free energy finite-size scaling functions are available
for the elementary Gaussian model in strip and film geometries,
respectively, in two and three dimensions for
nonperiodic boundary conditions.

There are several reasons for this lack of knowledge.
One of the reasons is that realistic b.c., such as Dirichlet
or Neumann b.c.\ for the order parameter, imply considerable
technical difficulties in the analytic description of finite-size effects
below bulk $\Tc$ even at the level of one-loop approximations.
A second reason is the dimensional crossover between finite-size effects
near the {\it three}-dimensional bulk transition at $\Tc$ and the
{\it two}-dimensional film transition at the separate critical temperature
$\Tcfilm(L)<\Tc$ of the film of finite thickness $L$.
An appropriate description of this dimensional crossover constitutes an as
yet unsolved problem even for the simplest case of film systems in the
Ising universality class with unrealistic {\it periodic} b.c..
A third reason is the inapplicability of ordinary renormalized
perturbation theory to the $\varphi^4$ model in {\it two} dimensions
(i.e., either at fixed dimension $d=2$ or within an $\varepsilon$
expansion in $4-\varepsilon$ dimensions extrapolated to $\varepsilon=2$)
because of the large value of the fixed point of the renormalized
four-point coupling at $d=2$.
No special reason exists, on the other hand, as to why no attention has
been paid in the literature to the Gaussian model in $d=3$ film or $d=2$
strip geometries with several different boundary conditions, although this
model is exactly solvable and does provide valuable and interesting
information on various aspects of the free energy and the Casimir force,
as we shall demonstrate in this paper.
A short summary of our main results is given below.

(i){\it Gaussian model as the basis for the mean spherical model :}
The exactly solvable mean spherical model (MSM) \cite{BaFi77} has played
an important role in the analysis of finite-size effects near critical
points where, however, the free energy and the critical Casimir force
have been studied, for a long time, only for periodic
b.c.\ \cite{brankov}.
A calculation of the critical Casimir force in the MSM for
{\it nonperiodic} b.c.\ was performed recently \cite{KaDo08}, with a few
results in film geometry in $2<d\leq3$ dimensions.
Clearly these results need to be extended to a more complete investigation.
A serious shortcoming of the MSM is the pathological behavior of the
surface and finite-size properties in $d \geq 3$ dimensions
\cite{BaFi77,brezin} with logarithmic deviations from scaling in $d=3$
dimensions.
Such logarithms were also found in the Casimir force and the free energy
\cite{KaDo08,Chamati}.
A profound understanding of these pathologies is important for the
appropriate interpretation of the deviations from finite-size scaling in
the MSM.
It was suggested earlier \cite{cardy} that the pathologies in the MSM
should be attributed to the {\it effective long-range interaction} induced
by the constraint.
The earlier analyses for nonperiodic b.c.\ (see \cite{brankov}), however,
were restricted to integer dimensions $d=3,4,\ldots$.
A more recent study \cite{ChDo03} of the full {\it continuous} range of
$2<d<4$ dimensions revealed the absence of pathologies for $d<3$ and
identified the origin of the nonscaling features for $d\geq 3$ as a
consequence of the properties of the ordinary Gaussian model with
short-range interactions.
The crucial point is that the MSM can be considered as a Gaussian model
with a constraint and that there exists a borderline dimension $d^*=3$
in the Gaussian model above which the Gaussian surface energy density has
a nonuniversal finite cusp at bulk $\Tc$.
This cusp causes all nonscaling effects for $d>3$, while for $d^*=3$ the
logarithmic divergence of the Gaussian surface energy density explains the
logarithmic deviations from scaling in the three-dimensional MSM
\cite{ChDo03}.
Both pathologies enter the MSM through the Gaussian surface terms of the
constraint equation.
The long-range interaction induced by the constraint does not yet
introduce a nonuniversal parameter but it is rather the combination with
the borderline dimension $d^*=3$ of the Gaussian model with
{\it short-range} interactions that is the origin of the nonuniversal
nonscaling features for $d\geq 3$.
The analysis of \cite{ChDo03} was restricted to the regime $t\geq0$ with
$t\equiv(T-\Tc)/\Tc$ for Dirichlet b.c., without considering the critical
Casimir force.
Our goal is to fully explore the finite-size critical behavior of the free
energy and the critical Casimir force of the MSM both above and below
$\Tc$ for five different b.c.\ and to properly explain the expected
deviations from finite-size scaling in three dimensions as well as to
study the scaling functions for all b.c.\ in $2<d<3$ dimensions.
It is our conviction that this goal must be based on a profound analysis
of the Gaussian model as a first step, before turning to the MSM.
The appropriateness of this strategy was demonstrated earlier in
\cite{ChDo03}.
In the present paper we perform this first step.
Our main results that will be relevant to our forthcoming analysis of the
MSM are as follows.
(a) Our results provide an exact description of the dimensional crossover
from the $d$-dimensional finite-size critical behavior near bulk $\Tc$ to
the $(d{-}1)$-dimensional critical behavior near $\Tcfilm$, which is
illustrated in Sec.~\ref{G.specific.heat} for the critical behavior of the
specific heat.
This dimensional crossover will constitute the basis for describing the
corresponding crossover from bulk $\Tc$ to $T \to 0$ for $d\leq 3$ in the
MSM.
(b) Our exact calculation includes nonnegligible logarithmic non-scaling
lattice effects in $d=3$ dimensions for the case of Neumann b.c.\ and
Dirichlet b.c.\ that have not been captured by the method of dimensional
regularization used in Ref.~\cite{KrDi92a}.
Such effects will be important for the interpretation of the logarithmic
nonscaling behavior in the $d=3$ MSM model.
(c) For the case of mixed Neumann-Dirichlet (ND) b.c., a strong power-law
violation of scaling is found in general dimensions $1<d<4$ that has an
important impact on the scaling structure of the free energy density in a
large part of the $L^{-1/\nu}$-- $t$ planes of both the Gaussian model
and the MSM and that is expected to imply unusually large corrections to
scaling in the $\varphi^4 $ theory.

(ii){\it Gaussian model scaling functions as one-loop
renormalization-group (RG) scaling functions:}
There is another important reason for studying finite-size effects of
the Gaussian model.
After appropriate renormalization, the Gaussian results for the free
energy, Casimir force, and specific heat can be reformulated as one-loop
contributions of the $\vp^4$ field theory.
From previous work \cite{krause_larin} it is known that, within the
minimal subtraction scheme in $d=3$ dimensions \cite{dohm}, the
one-loop bulk amplitude function of the specific heat provides a
reasonable approximation above $\Tc$ and that the one-loop finite-size
contributions for Dirichlet b.c.\ \cite{dohm93_sutter_mohr,sutter} yield
good agreement with specific-heat data \cite{nissen,gasparini} of confined
$^4$He in film geometry above and at the superfluid transition.
This suggests to determine the one-loop results for the free energy and
the critical Casimir force within the minimal subtraction scheme at
fixed dimension $d$ and to compare these results with $\ve=4-d$ expansion
results at $\ve=1$ \cite{KrDi92a,KrDi92b,DiGrSh06,GrDi07}, with recent
MC data \cite{DaKr04,Hu07_VaGaMaDi07,VaGaMaDi08}, and with the recent
result of an improved $d=3$ perturbation theory \cite{Do08b} in an
${\Lpl^2\times L}$ slab geometry with a finite aspect ratio
$\rho=L/\Lpl=1/4$.
As suggested by the earlier successes \cite{sutter,Do08a,Do08b}, the
minimally renormalized $\varphi^4$ theory at fixed $d$ is expected to
constitute an important alternative in the determination of the Casimir
force scaling function in comparison to the earlier $\ve$
expansion approach \cite{KrDi92a,KrDi92b,GrDi07}.
It is one of the central achievements of this paper that our $d=3$
one-loop RG results shown in Fig.~\ref{G.GcalX} below indeed support
this expectation.

(iii){\it Casimir force scaling functions in two dimensions:}
Most of our Gaussian results are valid in $1<d<4$ dimensions.
This permits us to study the interesting case $d=2$ and to compare it
with the exact results of the two-dimensional Ising model
\cite{Evans,RuZaShAb10,Indekeu}.
As a totally unexpected result we find (in Sec.~\ref{G.2d.Ising})
surprising relations between the Casimir scaling functions of the
Gaussian model with periodic (antiperiodic) b.c.\ and those of the
Ising model with antiperiodic (periodic) b.c..
Our comparison between these models also identifies the magnitude of
non-Gaussian fluctuation effects in the two-dimensional $\varphi^4$
model for several b.c..

(iv){\it Nonuniversal anisotropy effects:}
It has often been stated in the earlier and recent literature
\cite{Kr94,brankov,dan-8,Hu07_VaGaMaDi07,DiGrSh06,GrDi07,RuZaShAb10} that
the critical Casimir force scaling functions are universal, i.e.,
``independent of microscopic details''.
In view of these claims we briefly study the case of a simple
example of anisotropic couplings, i.e., two different nearest-neighbor
couplings $\Jpl$ and $\Jpp$ in the horizontal and vertical directions,
respectively.
Our exact results for the Gaussian model show that these anisotropic
couplings imply nonuniversal scaling functions of the Casimir force
that depend explicitly on $\Jpl$ and $\Jpp$ for all b.c., as predicted
by Chen and Dohm \cite{ChDo04,Do08a,Do06,Do09} and recently confirmed by
Dantchev and Gr\"uneberg \cite{dan09} for the case of antiperiodic
b.c.\ in the large-$n$ limit for $2<d<4$.
In particular, we verify for all b.c.\ the exact relation
\cite{ChDo04,dan09}
$\De_\text{aniso}=(\Jpp/\Jpl)^{(d-1)/2}\De_\text{iso}$
between the Casimir amplitudes of the isotropic and anisotropic
film system within the $d$-dimensional Gaussian model.
We also extend this kind of relation to the two-dimensional Ising model
for periodic and antiperiodic b.c.\ in the form $\Delta_\text{aniso}
=(\xi_{0,\perp}/\xi_{0,\parallel})\;\;\Delta_\text{iso}$, where
$\xi_{0,\perp}$ and $\xi_{0,\parallel}$ are the correlation-length
amplitudes perpendicular and parallel to the boundaries of the Ising
strip.
For the case of free b.c.\ at $\Tc$, such a relation was found earlier
by Indekeu et al.\ \cite{Indekeu}.
It would be interesting to test such nonuniversal anisotropy effects by
MC simulations for the critical Casimir force, in addition to those for
the critical Binder cumulant \cite{selke}.

As a general remark we note that the Gaussian model does not have upper
or lower critical dimensions; for this reason many of our results are
valid for arbitrary $d>0$ except for certain integer $d$ where logarithms
appear (at even integer $d$ for bulk properties and odd integer $d$ for
surface properties).

The outline of our paper is as follows.
In Sec.~\ref{G.section} we define our model, review the relevant bulk
critical properties in $d>0$ dimensions, and give a short account
of what effects arise if the model is anisotropic.
In Sec.~\ref{G.film.crit}, we consider the film critical behavior in
$2\leq d<4$ dimensions.
In Sec.~\ref{G.free.energy}, we derive and discuss the singular
contributions to the free energy density in $1<d<4$ dimensions.
In Secs.~\ref{G.Casimir.force}--\ref{G.2d.Ising} the Casimir force is
considered, in Sec.~\ref{oneloopphi4} our results are reformulated as
one-loop RG results of the $\varphi^4$ field theory and are compared to
other RG and MC results, while in Sec.~\ref{G.specific.heat} we focus on
the specific heat and its crossover from $d$ to $d-1$ dimensions.
The Appendix is reserved for details of our calculations.

\section{Gaussian model in film geometry}
\label{G.section}

\subsection{Lattice Hamiltonian and basic definitions}
\label{G.Hamiltonian}

We start from the Gaussian lattice Hamiltonian (divided by $\kB T$)
\begin{align}
\label{G.H}
\Hcal=\at^d\left[\f{\rz}{2}\sum_{\xb}\Sr{\xb}^2
+\f{1}{2\at^2}\sum_{\xb,\xb'}J_{\xb,\xb'}(\Sr{\xb}-\Sr{\xb'})^2\right],
\end{align}
with $\Sr{\xb}^2=\sum_{\al=1}^n(\Sr{\xb}^{(\al)})^2$ and with
couplings $J_{\xb,\xb'}$ between the continuous $n$-component vector
variables \mbox{$\Sr{\xb}=(\Sr{\xb}^{(1)},\ldots,\Sr{\xb}^{(n)})$} on the
lattice points $\xb$ of a $d$-dimen\-sional simple-cubic lattice with
lattice spacing $\at$.
The components $\Sr{\xb}^{(\al)}$ vary in the range
\mbox{$-\infty<\Sr{\xb}^{(\al)}<+\infty$}.
Unless stated otherwise, we shall assume an isotropic nearest-neighbor
ferromagnetic coupling $J_{\xb,\xb'}=J>0$, $J_{\xb,\xb'}=0$ for
$|\xb-\xb'|>\at$.
In the discussion of our results we shall also comment on the case of
anisotropic short-range interactions $J_{\xb,\xb'}$ with a positive
definite anisotropy matrix $\Ab$ \cite{Do08a} as defined in
Eqs.~(\ref{G.A}) and (\ref{G.A3}) below.
The only temperature dependence enters via
$\rz=a_0 t\equiv a_0(T-\Tc)/\Tc$, $a_0>0$, where $\Tc$ is the {\it bulk}
critical temperature.
We assume $\Ncal\equiv\Npl^{d-1}\times N$ lattice points in a finite
rectangular box of volume $V=\Lpl^{d-1}\times L=\Ncal\at^d$, where
$\Lpl\equiv\Npl\at$ and $L\equiv N\at$ are the lattice' extension in the
$d-1$ ``horizontal'' directions and in the one ``vertical'' direction,
respectively.
Thus we have $N$ layers each of which has $\Npl^{d-1}$ fluctuating
variables.
The lattice points are labeled by $\xb=(\yb,z)$ with
$\yb=(y_1,\ldots,y_{d-1})$.
We assume periodic b.c.\ in the horizontal $(\yb)$ directions.
As we shall take the film limit $\Npl\to\infty$, the relevant b.c.\ are
those in the vertical $(z)$ direction.
The top and bottom surfaces have the coordinates $z_1=\at$ and $z_N=L$,
respectively.
It is convenient to formulate the vertical b.c.\ by adding two fictitious
layers with vertical coordinates $z_0=0$ and $z_{N+1}=L+\at$ below the
bottom surface and above the top surface, respectively, for each value
of the $d-1$ horizontal coordinates.
Then we may define periodic \pbc, antiperiodic \abc,
Neumann-Neumann \NNbc, Dirichlet-Dirichlet \DDbc, and
Neumann-Dirichlet \NDbc~b.c.\ by
\bse
\label{G.bcs}
\begin{alignat}{3}
\text{p}:~~&&&~~~
\Sr{z_{N+1}}=&\Sr{z_1},~~&\\
\text{a}:~~&&&~~~
\Sr{z_{N+1}}=&-\Sr{z_1},~~&\\
\text{NN}:~~&&
 \Sr{z_0}&= \Sr{z_1}, & \Sr{z_{N+1}}&=\Sr{z_N}, \\
\text{DD}:~~&&
 \Sr{z_0}&=0, & \Sr{z_{N+1}}&=0, \\
 \text{ND}:~~&&
 \Sr{z_0}&=\Sr{z_1}, & \Sr{z_{N+1}}&=0,
\end{alignat}
\ese
where we have omitted the {\bf y} coordinates.
We use the representation
\begin{align}
\label{G.Sxb}
\Sr{\yb,z}=\sum_{\pb,q}\Sf{\pb,q}
\ul^\tbc(z,q)\prod_{i=1}^{d-1}u_{\Lpl}^\pbc(y_i,p_i),
\end{align}
\bwt
\bse
\label{G.ul}
\begin{alignat}{3}
\label{G.ul.p}
\ul^\pbc(z,q_m)
&=
\f{1}{\sqrt{N}}
\begin{cases}
\ph{\sqrt{2}}
\cos q_m z=1 &\quad m=0\\
\sqrt{2}\cos q_m z &\quad 1\leq m<N/2\\[1ex]
\ph{\sqrt{2}}
\cos q_m z=\cos\f{\pi z}{\at}
&\quad m=N/2\\[1ex]
\sqrt{2}\sin q_m z &\quad N/2<m\leq N-1\\[1ex]
\end{cases}
&~~ q_m
&=
\f{2\pi m}{L},
\\[1ex]
\label{G.ul.a}
\ul^\abc(z,q_m)
&=
\f{1}{\sqrt{N}}
\begin{cases}
\sqrt{2}\cos q_m z &\quad 0\leq m<(N-1)/2\\[1ex]
\ph{\sqrt{2}}
\cos q_m z=\cos\f{\pi z}{\at}
&\quad m=(N-1)/2\\[1ex]
\sqrt{2}\sin q_m z &\quad (N-1)/2<m\leq N-1\\[1ex]
\end{cases}
&~~ q_m
&=
\f{2\pi(m+\f{1}{2})}{L},
\\[1ex]
\label{G.ul.NN}
\ul^\NNbc(z,q_m)
&=
\f{1}{\sqrt{N}}
\begin{cases}
\ph{\sqrt{2}}
\cos q_m(z-\f{\at}{2})=1 &\quad m=0
\\[1ex]
\sqrt{2}\cos q_m(z-\f{\at}{2}) &\quad m=1,\ldots,N-1
\end{cases}
&~~ q_m
&=
\f{\pi m}{L},
\\[1ex]
\label{G.ul.DD}
\ul^\DDbc(z,q_m)
&=
\sqrt{\f{2}{N+1}}\;\sin q_m z
\qquad\qquad\qquad m=0,\ldots,N-1
&~~ q_m
&=
\f{\pi(m+1)}{L+\at},
\\[1ex]
\label{G.ul.ND}
\ul^\NDbc(z,q_m)
&=
\sqrt{\f{2}{N+\f{1}{2}}}\;\cos q_m(z-\fr{\at}{2})
\qquad\quad m=0,\ldots,N-1
&~~ q_m
&=
\f{\pi(m+\f{1}{2})}{L+\f{1}{2}\at},
\end{alignat}
\ese
\ewt
with the Fourier amplitudes $\Sf{\pb,q}$ and the complete set $\ul^\tbc$
of real orthonormal functions, where, for the $d-1$ horizontal directions,
the $\ul^\pbc(z,q_m)$ are used with the replacements $L\to\Lpl$,
$z\to y_i$, and $q_m\to p_{i,m_i}=2\pi m_i/\Lpl$.
The $m=N/2$ mode for periodic b.c.\ (the $m=(N{-}1)/2$ mode for
antiperiodic b.c.) is only present if $N$ is even (if $N$ is odd).
The above mode functions are equivalent to those in
\cite{BaFi77,ChDo03,DaBr9703}, where complex mode functions for periodic
and antiperiodic b.c.\ have been used instead of our real mode functions.

The functions (\ref{G.ul}) satisfy the orthonormality conditions
\bse
\begin{align}
\label{G.orthnorm1}
\sum_{z_j}\ul(z_j,q_m)\ul(z_j,q_{m'})
&=
\de_{m,m'},
\\
\label{G.orthnorm2}
\sum_{q_m}
\ul(z_j,q_m)\ul(z_{j'},q_m)
&=
\de_{j,j'},
\end{align}
\ese
with $z_j\equiv j\at$, $j=1,\ldots,N$.
For the case of isotropic nearest-neighbor couplings $J>0$, this yields
the diagonalized Hamiltonian
\begin{align}
\label{G.Hk}
\Hcal=\f{1}{2}\at^d\sum_{\pb,q}
\left(\rz+J_{\pb,d-1}+J_{q}\right)\Sf{\pb,q}^2,
\end{align}
\bse
\label{G.Jk}
\begin{align}
\label{G.Jp}
J_{\pb,d-1}
&\equiv
\f{4J}{\at^2}\sum_{i=1}^{d-1}\left(1-\cos p_i\at\right),
\\
\label{G.Jq}
J_q
&\equiv
\f{4J}{\at^2}\left(1-\cos q\at\right).
\end{align}
\ese
Equations (\ref{G.Jk}) reflect the cubic anisotropy of the lattice.
The lowest modes have $\pb=\bm{0}$ and are homogeneous ($q_0=0$) for
periodic and NN b.c., whereas they are $z$-dependent with
$q_0=\pi/(L+\at)$ for DD b.c.\ and $q_0=\pi/(2L+\at)$ for ND b.c..
For antiperiodic b.c., there is a twofold degeneracy of the lowest modes
with $q_0=\pi/L$ and $q_{N-1}=-\pi/L+2\pi/\at$, since
$J_{q_0}=J_{q_{N-1}}$.
This has important consequences for the behavior of the free energy
and specific heat near the film critical temperature, see
Secs.~\ref{G.film.crit}, \ref{G.F.pa.2-4}, and
\ref{G.specific.heat} below.
A corresponding twofold degeneracy of the ground state is known for the
mean spherical model with antiperiodic b.c.\ \cite{dan09}.

We note that the boundary conditions assumed in (\ref{G.bcs}) do not
depend on any nonuniversal parameter.
They are conceptually simple and represent only a small subset of a
large class of more complicated boundary conditions.
The latter may exist in the presence of an anisotropic lattice structure
whose symmetry axes are not orthogonal to the boundaries but have
{\it skew} directions relative to the boundaries.
Such more complicated systems (which, however, belong to the same bulk
universality class as standard spin models---such as Ising models with
nearest-neighbor couplings on simple-cubic lattices)
indeed exist, e.g., among real magnetic materials with a non-orthorhombic
lattice structure.
Models of such systems may also arise after a shear transformation has
been performed to an isotropic system \cite{Do08a} if the
original lattice model has non-cubic anisotropies.
In this case the transformed boundary conditions depend on the original
anisotropy parameters and therefore give rise to nonuniversal finite-size
effects.
We shall come back to such skew nonuniversal boundary conditions in
the context of the discussion of two-scale factor universality in
Sec.~\ref{G.Gauss.cont}.

The dimensionless partition function is
\begin{align}
\label{G.Zftbc}
\Zgc(t,\Lpl,L)
&=
\left[\prod_{\yb,z}\int_{-\infty}^{+\infty}
\f{d^n\Sr{\yb,z}}{\at^{(2-d)n/2}}\right]\exp(-\Hcal)
\nn\\
&=
\left[\prod_{\pb,q}\int_{-\infty}^{+\infty}
\f{d^n\Sf{\pb,q}}{\at^{(2-d)n/2}}\right]\exp(-\Hcal)
\nn\\
&=
\prod_{\pb,q}
\left(\f{2\pi}{\at^2\left(\rz+J_{\pb,d-1}+J_q\right)}\right)^{n/2},
\end{align}
where we have used that, due to the orthonormality of the $\ul$,
the linear transformation $\Sr{\yb,z}\to\Sf{\pb,q}$ has a Jacobian
$|(\ptl\Sr{\yb,z}/\ptl\Sf{\pb,q})|=1$.

The film limit is defined for $d>1$ by letting $\Lpl\to\infty$ while
keeping $L$ finite.
In this limit the Gaussian free energy per component and per unit volume
divided by $\kB T$ is given for $\rz\geq\rzcf(L)$ by
\begin{align}
\label{G.ftbc}
&f(t,L)=
-\f{1}{n}\lim_{\Lpl\to\infty}\f{1}{\Lpl^{d-1}L}\ln \Zgc(t,\Lpl,L)
\nn\\
&=
-\f{1}{2\at^d}\ln(2\pi)
+\f{1}{2L}\sum_q\int_{\pb}^{(d-1)}\!\!
\ln\left[\at^2\left(\rz{+}J_{\pb,d-1}{+}J_q\right)\right],
\end{align}
where
$\int_{\pb}^{(d-1)}
\equiv\prod_{i=1}^{d-1}\int_{-\pi/\at}^{+\pi/\at}dp_i/(2\pi)$.
A {\it film} critical point exists at $\rz=\rzcf(L)$, where the argument
of the logarithm on the right hand side of (\ref{G.ftbc}) vanishes
for $\pb=0$ and $q=q_0$.

As a shortcoming of the Gaussian model, the bulk critical value
$r_\text{0c}=0$ and the film critical value $\rzcf(L)$ are independent
of $d$ and $n$, and no low-temperature phase exists.
Furthermore, the Gaussian $r_\text{0c}$ is not affected by lattice
anisotropies, in contrast to $\rzcf(L)$, which depends explicitly on the
anisotropic couplings $J_{\xb,\xb'}$ (see Sec.~\ref{G.film.crit}).
For antiperiodic, DD, and ND b.c., $\rzcf(L)$ is negative, thus the free
energy (\ref{G.ftbc}) exists for negative values of $\rz$ in these cases.
The region $\rzcf(L)<\rz\leq0$ will be of particular interest for the
study of the mean spherical model {\it below} the bulk transition
temperature \cite{KaDo10}.
The film critical behavior of the Gaussian model will be discussed in
more detail in Sec.~\ref{G.film.crit}.

The bulk limit is obtained by letting $L\to\infty$,
$L^{-1}\sum_q\to\int_q\equiv\int_{-\pi/\at}^{+\pi/\at}dq/(2\pi)$.
The bulk free energy density per component divided by $\kB T$ is,
for $t\geq0$,
\begin{align}
\label{G.fb}
\hspace{-6pt}
\fb(t)
&\equiv
f(t,\infty)
\nn\\
&=
-\f{1}{2\at^d}\ln(2\pi)
+\f{1}{2}\int_{\kb}^{(d)}
\ln\left[\at^2\left(\rz+J_{\kb,d}\right)\right].
\end{align}
In the long-wavelength limit, the cubic anisotropy does not matter and
$J_{\kb,d}=2Jk^2+O(k^4)$ becomes isotropic which justifies to define a
single second-moment bulk correlation length $\xi$ above $\Tc$,
\begin{align}
\label{G.xi}
\xi^2= \lim_{L\to\infty}\frac{1}{2d}
\frac{\sum_{\xb,\xb'}(\xb-\xb')^2\langle S_{\xb}S_{\xb'}\rangle}%
{\sum_{\xb,\xb'}\langle S_{\xb}S_{\xb'}\rangle}.
\end{align}
The latter is given by
\begin{align}
\label{G.xi0.nu}
\xi&=(2J/\rz)^{1/2}=\xiz t^{-\nu}, &
\xiz&=\left(2J/\az\right)^\nu, &
\nu&=1/2.
\end{align}

In the presence of NN or DD b.c., there are surface free energy
densities per component $2\fsf^\Nbc(t)$ and $2\fsf^\Dbc(t)$ for $t>0$
as defined by
\begin{align}
\label{G.f.sf}
\fsf(t)=\f{1}{2}\lim_{L\to\infty}\left\{L[f(t,L)-\fb(t)]\right\}.
\end{align}
In the presence of ND b.c., the total surface free energy density per
component is
\begin{align}
\label{G.fsf.ND}
2\fsf^\NDbc(t)=\fsf^\Nbc(t)+\fsf^\Dbc(t).
\end{align}
For periodic and antiperiodic b.c.\ there exist no surface contributions.

For small $t>0$, the bulk and surface free energy densities
will be decomposed into
singular and nonsingular parts as
\bse
\label{G.fbsf}
\begin{align}
\label{G.fbsns}
\fb(t)=\fbs(t)+\fbns(t),
\\
\label{G.fsfsfsfns}
\fsf(t)=\fsfs(t)+\fsfns(t),
\end{align}
\ese
where $\fbns(t)$ and $\fsfns(t)$ have an expansion in positive integer
powers of $t$.
For small $t$ and large $L$, it is expected \cite{Pr88,Pr90,priv} that,
for the Gaussian model (\ref{G.H}) in film geometry, the free energy
density can be decomposed as
\begin{align}
\label{G.f.s.f.ns}
f(t,L)=\fs(t,L)+\fns(t,L),
\end{align}
\begin{align}
\label{G.fns}
\fns(t,L)=\fbns(t)+L^{-1}[\fsfns^\text{top}(t)+\fsfns^\text{bot}(t)],
\end{align}
where ``top'' and ``bot''
refer to the top and bottom surfaces of the film.
In the absence of logarithmic bulk singularities \cite{Pr90}, i.e.,
for $d\neq2$ and $d\neq4$, and in the absence of logarithmic surface
singularities \cite{ChDo03}, i.e., for $d\neq3$
(or periodic or antiperiodic b.c.), the singular part is
expected to have the finite-size scaling form \cite{PrFi84}
\begin{align}
\label{G.fscal}
\fs(t,L)=L^{-d}\Fcal(C_1tL^{1/\nu}),
\end{align}
with a nonuniversal parameter $C_1$.
For given b.c., the scaling function $\Fcal(\xt)$ is expected to be
universal only within the subclass of isotropic systems but nonuniversal
for the subclass of anisotropic systems of noncubic symmetry within
the same universality class \cite{ChDo04,Do08a}, see
(\ref{G.fscal-aniso})--(\ref{G.fscal-anisospec-neu2}) below.
A convenient choice of the scaling variable $\xt$ is
\begin{align}
\label{G.scalvar}
\xt = t(L/\xiz)^{1/\nu},
\end{align}
i.e., $C_1=\xiz^{-1/\nu}$.
The bulk singular part
\begin{align}
\label{G.fbs}
\fbs(t)&=Y_d\xi^{-d}, & d&>0, & d&\neq2,4,6,\ldots,
\end{align}
see Sec.~\ref{G.bulk.film.crit} below, with a universal bulk amplitude
$Y_d$ is included in Eq.~(\ref{G.fscal}) through
$Y_d=\lim_{\xt\to\infty}\xt^{-d\nu}\Fcal(\xt)$ for $1<d<4$, $d\neq2$.
For the surface free energy density, (\ref{G.fscal}) implies
\begin{align}
\label{G.Asurface}
\fsfs(t)=\Asf\xi^{1-d},
\end{align}
with a universal surface amplitude
$\Asf=\lim_{\xt\to\infty}\xt^{-(d-1)\nu}[\Fcal(\xt)-Y_d\xt^{d\nu}]$.

For fixed $t>0$ and large $L$ it is expected \cite{Pr90,priv,ChDo99,Do08a}
that the free energy density can be represented as
\begin{align}
\label{G.fexdef}
f(t,L)
&=
\fb(t)+L^{-1}[\fsf^\text{top}(t)+\fsf^\text{bot}(t)]+L^{-d}\Gcal(\xt)
\nn\\
&\ph{=}
+O(e^{-L/\xie}).
\end{align}
In (\ref{G.fexdef}), $\xie$ is the exponential {\it bulk}
correlation length in the direction of one of the cubic axes
\cite{ChDo99,Do08a}
\begin{align}
\label{G.xie}
\xie\equiv\left(\f{2}{\at}\arsinh\f{\at}{2\xi}\right)^{-1}.
\end{align}
Its deviation from $\xi$ for finite $\at$ causes scaling to be
violated \cite{ChDo99,Do08a} for fixed $t>0$ and large
$L\agt24\xi^3/\at^2$, i.e., $\xt\agt576(\xi/\at)^4$.
We recall that this scaling violation is a general consequence of the
exponential structure of the excess free energy for large $L$ at
fixed $\xi$ and is a lattice (or cutoff) effect that is predicted to
occur not only in the Gaussian model but quite generally in the
$\varphi^4$ lattice (or field) theory for systems with short-range
interactions \cite{ChDo99,Do08a}.
This effect is different in structure from additional nonscaling
effects that occur in the presence of subleading long-range (van der
Waals type) interactions \cite{DR2001,Do08a}.

In the absence of long-range interactions, no contributions
$\sim L^{-m}$ with $m>1$, $m\neq d$ should exist in (\ref{G.fexdef})
for film geometry.
The representation (\ref{G.fexdef}) separates the finite-size part
$\sim L^{-d}$ from the surface parts $\sim L^{-1}$.
The latter do not contribute to the Casimir force scaling function
$X(\xt)$ to be discussed in Sec.~\ref{G.Casimir.force}.

If (\ref{G.fscal}) and (\ref{G.fbs})--(\ref{G.fexdef}) are valid,
the connection between $\Fcal$, $\Asf$, and $\Gcal$ is, for $\xt>0$,
\begin{align}
\label{G.Fcal}
\!\Fcal(\xt)
=
Y_d\xt^{d\nu}
+(\Asf^\text{top}+\Asf^\text{bot})\xt^{(d-1)\nu}
+\Gcal(\xt).
\end{align}
In Sec.~\ref{G.free.energy} we shall examine the range of validity of the
structure of (\ref{G.fscal}), (\ref{G.fexdef}), and (\ref{G.Fcal}) for
the Gaussian model for various b.c.\ and calculate the scaling functions.

In Sec.~\ref{G.specific.heat} we shall also discuss the specific heat
(heat capacity per unit volume) divided by $\kB$
\begin{align}
\label{G.spec-heat.u}
C(t,L) = \partial\uf(t,L)/\partial T,
\end{align}
where $\uf(t,L)=-T^2\partial f(t,L)/\partial T$ is the energy density
(internal energy per unit volume) divided by $\kB$, with the singular
bulk part
\begin{align}
\label{G.Us}
\ubs(t)=-\Tc\xiz^{-1/\nu}d\nu Y_d \xi^{-(1-\al)/\nu}.
\end{align}
The surface part of the energy density is
$\usf(t)=-T^2\partial\fsf(t)/\partial T$, with the singular part
\begin{align}
\label{G.Using}
\usfs(t)
&=
-\Tc\xiz^{-1/\nu}(d-1)\nu\Asf\xi^{-(1-\al-\nu)/\nu}.
\end{align}
In (\ref{G.Us}) and (\ref{G.Using}) we have used the hyperscaling
relation $d\nu=2-\al$, with the Gaussian exponent
\begin{align}
\label{G.alpha}
\al&=(4-d)/2, & d&<4.
\end{align}
In the presence of NN, ND, and DD b.c., logarithmic deviations
from the scaling structure of (\ref{G.fscal}), (\ref{G.Asurface}),
(\ref{G.Fcal}), and (\ref{G.Using}) are expected for the Gaussian model
in the borderline dimension $d^*=3$ \cite{ChDo03} because of the vanishing
of the critical exponent
\begin{align}
\label{G.surface.exponent}
1-\al-\nu=(d-3)/2
\end{align}
of the singular part of the surface energy density (\ref{G.Using}).
(This is similar to the logarithmic deviations for systems with
periodic b.c.\ \cite{PrRu86} at $d=4$, where the specific-heat exponent
$\al$ vanishes.)
In this case, $\Fcal(\xt)$ and $\Asf$ do not exist, but $\Gcal(\xt)$
and $X(\xt)$ remain well defined.
The positivity of the exponent (\ref{G.surface.exponent}) for $d>3$
implies a nonuniversal cusp that is responsible for the nonscaling
features in the MSM for $d>3$ \cite{ChDo03}.

Moreover, logarithmic deviations from the structure of (\ref{G.fbs})
and (\ref{G.Us}) are expected for the Gaussian model in the borderline
dimension $d=2$ because of the vanishing of the critical exponent
\begin{align}
\label{G.bulk.exponent}
1-\al=(d-2)/2
\end{align}
of the singular part of the bulk energy density (\ref{G.Us}).

\subsection{\boldmath Bulk critical properties}
\label{G.bulk.film.crit}

In contrast to real systems with short-range interactions, the Gaussian
model has a bulk phase transition at $\rz=0$ for any dimension $d>0$
including $d=1$.
In the following we present both the singular and nonsingular parts of
the bulk critical behavior of the free energy since they will be needed
in the context of the mean spherical model in a subsequent part of the
present work \cite{KaDo10}.
The exact result for the bulk free energy density for $\rz\geq0$ in
$d>0$ dimensions is
\begin{align}
\label{G.fbulk}
\fb(t)=\f{1}{2\at^d}\left[\ln\f{J}{\pi}+\Wt{d}(\rzt)\right],
\end{align}
where $\rzt\equiv\rz\at^2/(2J)$ and
\begin{align}
\label{G.Wtdz}
\Wt{d}(z)\equiv
\int_0^\infty\f{d\yi}{\yi}\left[e^{-\yi/2}-e^{-z\yi/2}\B(\yi)^d\right],
\end{align}
with
\begin{align}
\label{G.Bfunc}
\B(\yi)\equiv e^{-\yi}I_0(\yi),
\end{align}
and where $I_0$ is a Bessel function of order zero,
$I_0(z)=\pi^{-1}\int_0^\pi d\vp\,\exp(z\cos\vp)$.
From the large-$\yi$ behavior (\ref{G.Byg}) of $B(\yi)$, the
universal amplitude of (\ref{G.fbs})
in \mbox{$d>0$}, $d\neq2,4,6,\ldots$ dimensions is derived as
\begin{align}
\label{G.Yd}
Y_d&=-\f{\Ga(-d/2)}{2(4\pi)^{d/2}},
\end{align}
with $Y_3=-(12\pi)^{-1}$.
The nonsingular bulk part $\fbns(t)$ has an expansion in integer
powers of $\rzt$,
\begin{align}
\label{G.fbns}
\!\!\fbns(t)
&=
\fbns(0)+\f{1}{2\at^d}\left[f_1\rzt+O(\rzt^2)\right],
\end{align}
where
\begin{align}
\label{G.fbns0}
\fbns(0)&=\fb(0)=\f{1}{2\at^d}\left[\ln\f{J}{\pi}+\Wt{d}(0)\right],
\end{align}
\begin{align}
f_1=
\begin{cases}\ds
\f{1}{2}\int_0^\infty d\yi\left[B(\yi)^d-(2\pi\yi)^{-d/2}\right],
& 0<d<2,
\\[2ex]
\W{d}(0), & d>2,
\end{cases}
\end{align}
with the generalized Watson function \cite{BaFi77}
\begin{align}
\label{G.Wdz}
\W{d}(z)
&\equiv\Wt{d}'(z)=\f{1}{2}\int_0^\infty d\yi\,e^{-z\yi/2}\B(\yi)^d.
\end{align}

In order to appropriately interpret the critical behavior of the
{\it three-dimensional} system in film geometry in
Secs.~\ref{G.film.crit}--\ref{G.specific.heat} below it is important
to first consider the {\it bulk} critical behavior in {\it two}
dimensions.
While $\Wt{2}(0)=4G/\pi$ with Catalan's constant $G\approx0.915966$ is
finite, both $Y_d=-1/[4\pi(d-2)]+O((d-2)^0)$ and
$W_d(0)=1/[2\pi(d-2)]+O((d-2)^0)$ diverge as $d\to2_+$.
However, the sum of the respective contributions to the singular and the
nonsingular part of the free energy remains finite and we obtain the bulk
free energy per unit area
\begin{align}
\label{G.fbs.2}
\fb(t)
&=
\fb(0)+\frac{\ln(\xi/\at)}{4\pi\xi^2}
+\f{\ln2-1}{16\pi J}\rz+O(\rz^2),
& d&=2,
\end{align}
with the singular part
\begin{align}
\label{G.fbs.ln}
\fbs(t)&=\frac{\ln(\xi/\at)}{4\pi\xi^2},
& d&=2.
\end{align}
The logarithmic structure is related to the vanishing of $1-\alpha$ for
$d=2$, see (\ref{G.bulk.exponent}).
In Sec.~\ref{G.2d.Ising} we shall compare the Casimir force scaling
function of the Gaussian model with that of the two-dimensional Ising
model.
This comparison will be restricted to the regime $T\geq\Tc$.
Correspondingly, we comment here on the Ising bulk free energy only
for this case.
For $T>\Tc$ the bulk correlation length of the $d=2$ Ising model
is, asymptotically, $\xi=\xi_{0+}t^{-\nu}$ with $\nu=1$.
In terms of this length, the singular part of the bulk free energy density
of the d=2 Ising model (on a square lattice with lattice spacing $\at$)
has the same form as given by (\ref{G.fbs.ln}) but with a negative
amplitude $-1/(4\pi)$ instead of $1/(4\pi)$ for the $d=2$ Gaussian model.

In contrast to the universal power-law structure (\ref{G.fbs}) for
$d>0$, $d\neq2,4,6,\ldots$, the logarithmic structure (\ref{G.fbs.ln})
contains the nonuniversal microscopic reference length $\at$.
Other reference lengths are expected for other lattice structures,
whereas the amplitude $1/(4\pi)$ is expected to be universal.
The choice of the amplitude of such reference lengths is not unique
but in our case the lattice spacing $\at$ appears to be most natural
for the cubic lattice structure.
(Due to the artifact of the Gaussian model and the $d=2$ Ising model
that $\xi^{-2}\sim t$ and $\xi^{-2}\sim t^2$, respectively, are
analytic functions of $t$, a different choice $c\at$ with $c\neq1$
as a reference length would yield a different decomposition into
singular and nonsingular parts.)

\subsection{Isotropic and anisotropic continuum Hamiltonian}
\label{G.Gauss.cont}

For the purpose of a comparison with the results of
$\vpb^4$ field theory we shall also consider the continuum
version of the Gaussian lattice model (\ref{G.H}) for an $n$-component
vector field $\vpb(\xb)$.
For the choice $2J=1$ the isotropic $\vpb^4$ Hamiltonian reads
\begin{align}
\label{G.H.field}
\Hcal_\text{field}
&=
\int\limits_V d^dx\left[\frac{\rz}{2}
\vpb^2+\frac{1}{2}\sum_{\al=1}^d\left(\frac{\partial \vpb}
{\partial x_\al}\right)^2+u_0(\vpb^2)^2\right],
\end{align}
with some cutoff $\La$ in ${\bf k}$ space.
The field $\vpb(\xb)=\vpb(\yb,z)$ satisfies the various b.c.\ that are
the continuum analogues \cite{KrDi92a} of Eqs.~(\ref{G.bcs}).
In Sec.~\ref{oneloopphi4} our Gaussian results based on $\Hcal$,
(\ref{G.H}), in the limit $\at\to0$ will be renormalized and reformulated
as one-loop contributions of the minimally renormalized $\vpb^4$ field
theory at fixed dimension $2<d<4$ \cite{dohm,Do08a} based on
$\Hcal_\text{field}$, (\ref{G.H.field}), in the limit $\La\to\infty$.
The role played by the $d=3$ RG approach will be to
change the Gaussian critical exponent $\nu=1/2$ to the exact critical
exponent $\nu$ at $d=3$ entering the correlation length $\xi$ which
appears in the scaling argument of the scaling functions of the
renormalized $\vpb^4$ field theory.
This will then justify to compare the resulting one-loop finite-size
scaling functions of the Casimir force in $d=3$ dimensions with MC data
for the three-dimensional Ising model \cite{VaGaMaDi08}, with higher-loop
$\ve=4-d$ expansion results at $\ve=1$ \cite{KrDi92a,GrDi07}, and with the
recent result of an improved $d=3$ perturbation theory \cite{Do08b} in an
${\Lpl^2\times L}$ slab geometry with a finite aspect ratio
$\rho=L/\Lpl=1/4$.

We shall also consider the anisotropic extension of (\ref{G.H.field})
\cite{ChDo04}
\begin{align}
\label{G.H.fieldaniso}
\Hcal_{\begin{smallmatrix}\text{field}\\\text{aniso}\end{smallmatrix}}
&=
\int\limits_V d^dx\left[\frac{\rz}{2}
\vpb^2{+}\!\!\sum_{\al,\be=1}^d
\frac{A_{\al\be}}{2}\frac{\partial \vpb}
{\partial x_\al}\frac{\partial\vpb}{\partial x_\be}
{+}u_0(\vpb^2)^2\right].
\end{align}
The expression for the symmetric anisotropy matrix $\bm{A}=(A_{\al\be})$
in terms of the microscopic couplings $J_{\xb,\xb'}$ of the lattice
Hamiltonian $\Hcal$, (\ref{G.H}), is given by the second moments
\cite{Do06,Do08a}
\begin{align}
\label{G.A}
A_{\al\be}=A_{\be\al}=\f{1}{\Ncal\at^2}
\sum_{\xb,\xb'}(x_{\al}{-}x'_{\al})(x_{\be}{-}x'_{\be})J_{\xb,\xb'}.
\end{align}
In the case of isotropic nearest-neighbor couplings $J$ on a simple-cubic
lattice we have simply $A_{\al\be}=2J\de_{\al\be}$.
In general, $A_{\al\be}$ is non-diagonal and contains $d(d+1)/2$
independent nonuniversal matrix elements.

The relation between the finite-size critical behavior of isotropic and
anisotropic systems was recently discussed in detail for the case of a
{\it finite} rectangular geometry with periodic
b.c.\ \cite{ChDo04,Do06,Do08a}.
It was shown that the relation between the anisotropic and isotropic
critical behavior is brought about by a shear transformation.
In real space, this transformation is described by the matrix product
$\bm{\la}^{-1/2}\bm{U}$,
with an orthogonal matrix $\bm{U}$ that diagonalizes $\bm{A}$ according to
$\bm{\la}=\bm{U}\bm{A}\bm{U}^{-1}$, where $\bm{\la}$ is a diagonal matrix
whose diagonal elements are the eigenvalues of $\bm{A}$.
This transformation causes a nonuniversal distortion of the rectangular
shape to a parallelepipedal shape, of the simple-cubic lattice structure
to a triclinic lattice structure, and of the periodic b.c.\ along the
rectangular symmetry axes to periodic b.c.\ along the corresponding skew
lattice axes of the triclinic lattice.
The general structure of the scaling form of the free energy density is
expressed in terms of the characteristic length $L'= V'^{1/d}$ where
$V'$ is the finite volume of the parallelepiped (see Eqs.~(1.3) and (4.1)
of \cite{Do08a}).
This is, however, not directly applicable to our present model with film
geometry with an {\it infinite} volume and with various b.c..
Furthermore, a significant difference occurs in film geometry due to the
existence of a film transition temperature that is affected by anisotropy
for the cases of antiperiodic, DD, and ND boundary conditions.
Thus anisotropy effects in film geometry deserve a separate discussion.
In particular, we shall compare our results with those of Indekeu
et al.\ \cite{Indekeu}, who studied an anisotropic Ising model on a
two-dimensional infinite strip.

In \cite{ChDo04} it was found that, for $2<d<4$ in the large-$n$ limit of
the $\vpb^4$ theory above bulk $\Tc$ in film geometry with periodic b.c.,
the universal structure (\ref{G.fscal}) is replaced by
\begin{align}
\label{G.fscal-aniso}
f_\text{s,aniso}(t,L;\Ab)
&=
L^{-d}[(\Abarb^{-1})_{dd}]^{-d/2}
\Fcal_\text{iso}((\widetilde{L}/\xi')^{1/\nu}),
\end{align}
where $\Fcal_\text{iso}$ is the scaling function of a film system
described by the isotropic $\vpb^4$ theory with ordinary periodic
b.c., but where the scaling argument contains the transformed length
\begin{align}
\label{G.Ltilde}
\widetilde{L}=[(\Ab^{-1})_{dd}]^{1/2}L,
\end{align}
and where $\xi'$ is the bulk correlation length of the isotropic system.
In (\ref{G.fscal-aniso}), $(\Abarb^{-1})_{dd}$ denotes the $d$th
diagonal element of the inverse of the reduced matrix
$\Abarb=\Ab/(\det\Ab)^{1/d}$.
In \cite{ChDo04} the simplicity of the structure of (\ref{G.fscal-aniso})
was attributed to the large-$n$ limit.
In general one expects that $f_\text{s,aniso}(t,L;\Ab)$ is expressed in
terms of the scaling function of an isotropic system that has
{\it transformed} boundary conditions which are not identical
with those of the original anisotropic system.
For a brief discussion of such boundary conditions see the paragraph
before Eq.~(\ref{G.Zftbc}) in Sec.~\ref{G.Hamiltonian}.

A simplifying feature of film geometry is that the shear transformation
preserves the film geometry except that the original thickness $L$ is
transformed to a different thickness $\bar{L}$.
In general, the length $\widetilde{L}$ appearing in the scaling argument
of $\Fcal_\text{iso}$ in (\ref{G.fscal-aniso}) is not the transformed
thickness $\bar{L}$ but rather the distance between those points on the
opposite surfaces in the transformed film system that are connected via
the periodicity requirement \cite{ChDo04};
this distance is measured along the corresponding skew lattice axis.
The correctness of this geometric interpretation can be seen as follows.
Let $\bm{\hat{x}}_d\equiv\bm{\hat{z}}$ be the unit vector in the
$z$-direction, i.e., orthogonal to the film boundaries.
Then $\widetilde{L}$ is the length of the vector $\bm{\widetilde{L}}$
obtained by transforming the vector $L\bm{\hat{x}}_d$, i.e.,
\begin{align}
\label{G.vectortilde}
\bm{\widetilde{L}}=\bm{\la}^{-1/2}\bm{U}L\bm{\hat{x}}_d,
\end{align}
and therefore
\begin{align}
\label{G.lengthtilde}
\widetilde{L}
&=
|\bm{\la}^{-1/2}\bm{U}L\bm{\hat{x}}_d|
=|\bm{\hat{x}}_d^T\bm{U}^{-1}\bm{\la}^{-1/2}
\bm{\la}^{-1/2}\bm{U}\bm{\hat{x}}_d|^{1/2}L
\nn\\
&=
|\bm{\hat{x}}_d^T\bm{U}^{-1}\bm{\la}^{-1}\bm{U}\bm{\hat{x}}_d|^{1/2}L
=|\bm{\hat{x}}_d^T \bm{A}^{-1}\bm{\hat{x}}_d|^{1/2}L
\nn\\
&=[(\Ab^{-1})_{dd}]^{1/2}L,
\end{align}
in agreement with (\ref{G.Ltilde}).
A corresponding statement holds for antiperiodic b.c..

As we show in Appendix~\ref{G.thickness}, the thickness $\bar{L}$ of the
transformed isotropic film is given by
\begin{align}
\label{G.Lbar}
\bar{L}= (\det{\bm{A}}^{-1}/\det[[{\bm{A}}^{-1}]])^{1/2} \;L,
\end{align}
where the $(d{-}1)\times(d{-}1)$ matrix $[[{\bm{A}}^{-1}]]$ is obtained
by removing the $d$th row and column from ${\bm{A}}^{-1}$.

It is possible to express (\ref{G.fscal-aniso}) in terms of the single
length $\widetilde{L}$ by rewriting
\begin{align}
\label{G.fscal-aniso-iso}
f_\text{s,aniso}(t,L;\Ab)
&=(\det\Ab)^{-1/2} f_\text{s,iso}(t,\widetilde{L}),
\end{align}
\begin{align}
\label{G.fscal-aniso-tilde}
f_\text{s,iso}(t,\widetilde{L})
&=\widetilde{L}^{-d}
\Fcal_\text{iso}((\widetilde{L}/\xi')^{1/\nu}).
\end{align}
Thus, apart from the geometric factor $(\det\Ab)^{-1/2}$ that describes
the change of the volume of the primitive cell under the shear
transformation, $f_\text{s,aniso}$ is given, in the large-$n$ limit, by
the free energy $ f_\text{s,iso}(t,\widetilde{L})$ of an isotropic film
with an effective thickness $\widetilde{L} \neq \bar{L}$ with ordinary
periodic b.c..
We conjecture that the structure of (\ref{G.fscal-aniso-iso}) with
(\ref{G.fscal-aniso-tilde}) is exactly valid also for the Gaussian model
with periodic b.c..
A similar structure is expected to be valid for the Gaussian model with
{\it antiperiodic} b.c.\ except that the scaling argument should be
expressed in terms of $t$ rather than $\xi$ in order to capture the
regime $\Tcfilm\leq T<\Tcbulk$.
Furthermore, the effect of the anisotropy on $\Tcfilm$ needs to be
taken into account (see Sec.~\ref{G.film.crit} below).

A nontrivial situation exists in the case of NN and ND b.c.\ because
Neumann b.c.\ involve a restriction on the spatial derivative
perpendicular to the boundary which, after the transformation, turns into
a derivative in a skew direction not necessarily perpendicular to the
transformed boundary.
Thus the isotropic film system still carries the nonuniversal anisotropy
information of the original system both in its changed thickness and in
the nonuniversal orientation of its transformed boundary conditions.
Thus both nonuniversality and anisotropy are still present at the
boundaries of the transformed system.
The same assertion applies to periodic and antiperiodic b.c..
Clearly, since boundary conditions dominate the finite-size critical
behavior at $\Tc$ where the correlation lengths extend over the entire
thickness of the film system, the above reasoning implies that
universality is {\it not} restored by the shear transformation in spite
of internal isotropy (in the long-wavelength limit) of the transformed
system away from the boundaries.
In other words, even this internal isotropy of a confined system does not
ensure the universality of its critical finite-size properties because
of the nonuniversality contained in the boundary conditions.
In the light of these facts we consider as incorrect the recent assertion
by Diehl and Chamati \cite{diehl2009} that ``the critical properties of
an anisotropic system can be expressed in terms of the universal
properties of the conventional (i.e., isotropic) $ {\bf\varphi}^4$
theory''.

More specifically, even after the shear transformation, the finite-size
effects of the transformed isotropic system still depend, in general, on
$d(d{+}1)/2{+}1$ nonuniversal parameters (see Eqs.~(1.3)--(1.5) of
\cite{Do08a}), contrary to the hypothesis of two-scale factor
universality \cite{PrFi84,priv}.
This {\it multiparameter universality} is fully compatible with the
general framework of the RG theory \cite{Do08a}.
Technically, these parameters enter through the transformed wave vectors
${\bf k}'$ of the isotropic system, thus the dependence of
{\it finite-size} properties on $A_{\alpha \beta}$ cannot be eliminated
by the shear transformation as demonstrated explicitly for the example of
periodic b.c.\ in Eq.~(2.22) of \cite{Do08a}.
We conclude that there is no basis for complying with the traditional
picture of two-scale factor universality according to the suggestion
``to define two-scale factor universality only {\it after} the
transformation to the primed variables (of the isotropic system) has
been made'' \cite{diehl2009}.
This suggestion would be applicable only to {\it bulk} properties of the
transformed system.

A special case is the case of DD b.c.\ (vanishing order-parameter field
$\vpb$ at the boundaries) or free b.c.\ (no condition on the fluctuating
variables at the boundaries) since these b.c.\ are invariant under the
shear transformation and therefore do not violate isotropy.
In particular, these b.c.\ do not introduce any nonuniversal parameter.
Nevertheless, even in this case there is a nontrivial shift of
$\Tcfilm$ of the film critical point of systems in the ordinary
($d,n$) universality classes (for $n=1$, $d>2$, for $n=2$, $d\geq 3$,
and for $n>2$, $d>3$ ) due to anisotropy.
For the special case $d=2$, $n=1$, however, i.e., for a system of the
Ising universality class on an infinite strip of finite width, there is
no separate ``film'' transition and thus no analog of a finite
$\Tcfilm>0$ exists.
This conceptually simplest case was studied by Indekeu
et al.\ \cite{Indekeu} as will be further discussed below.
One may conjecture that for DD b.c.\ the structure of
(\ref{G.fscal-aniso}) is valid also for the $d$-dimensional Gaussian
model where, however, the length $\widetilde{L}$ in (\ref{G.fscal-aniso})
is to be replaced by $\bar{L}$, (\ref{G.Lbar}).

An open question remains as to what extent the structure of
(\ref{G.fscal-aniso}) with (\ref{G.Ltilde}) (and correspondingly of
(\ref{G.fscal-anisospec-neu2}) below) is valid even for the full $\vpb^4$
model with finite $n$ in $d$ dimensions and even for real film systems.
It would be interesting to explore this problem theoretically as well as
by means of MC simulations for a variety of anisotropic spin models in
film geometry with various b.c.\ and various anisotropies.

The situation becomes particularly simple if the matrix $\Ab$ is diagonal
in which case the original simple-cubic lattice of the anisotropic system
is distorted only to an orthorhombic lattice of the isotropic system that
still has a rectangular structure.
Then we have $\bar{L}=\widetilde{L}=A_{dd}^{-1/2}L$, see
App.~\ref{G.thickness}.
In the following we confine ourselves to this simple case.

We consider only two different nearest-neighbor interactions $\Jpl$ and
$\Jpp$ in the ``horizontal'' and ``vertical'' directions.
This corresponds to replacing Eqs.~(\ref{G.Jk}) by
\bse
\label{G.Jplpp}
\begin{align}
\label{G.Jpl}
J_{\pb,d-1}
&\equiv
\f{4\Jpl}{\at^2}\sum_{i=1}^{d-1}\left(1-\cos p_i\at\right),
\\
\label{G.Jpp}
J_{q}
&\equiv
\f{4\Jpp}{\at^2}\left(1-\cos q\at\right),
\end{align}
\ese
in which case $\Ab$ is given in three dimensions by
\begin{align}
\label{G.A3}
\Ab
&=
2\left(\begin{array}{ccc}
\Jpl & 0 & 0 \\
0 & \Jpl & 0 \\
0 & 0 & \Jpp \\
\end{array}\right).
\end{align}
In this case we must distinguish two different correlation lengths
$\xi_\parallel$ and $\xi_\perp$.
For the Gaussian model they are given by
\bse
\label{G.corr}
\begin{align}
\label{G.corr.parall}
\xi_\parallel&=\xi_{0,\parallel} t^{-\nu}, &
\xi_{0,\parallel}&=\left(2\Jpl/\az\right)^\nu, &
\nu&=1/2,
\\
\label{G.corr.perp}
\xi_\perp&=\xi_{0,\perp} t^{-\nu}, &
\xi_{0,\perp}&=\left(2\Jpp/\az\right)^\nu, &
\nu&=1/2.
\end{align}
\ese
The existence of two different correlation lengths implies the absence
of two-scale factor universality \cite{ChDo04,Do06,Do08a}.
As a consequence, all bulk relations involving correlation lengths have
to be modified \cite{Do08a,Do06} and all finite-size scaling functions
are predicted \cite{ChDo04} to become nonuniversal as they depend
explicitly on the ratio $\Jpp/\Jpl$.

For the example (\ref{G.A3}), we obtain
\begin{align}
\label{G.A3barinvers}
\Abarb^{-1}
=
\left(\begin{array}{ccc}
(\Jpp/\Jpl)^{1/3} & 0 & 0 \\
0 & (\Jpp/\Jpl)^{1/3} & 0 \\
0 & 0 & (\Jpp/\Jpl)^{-2/3} \\
\end{array}\right)
\end{align}
for $d=3$ and $(\Abarb^{-1})_{dd}=(\Jpp/\Jpl)^{(1-d)/d}$
and $(\Ab^{-1})_{dd}=(2\Jpp)^{-1}$ for general $d$.
For the isotropic Gaussian model, we have simply \mbox{$\xi'=\rz^{-1/2}$}
(compare Eq.~(B16) of \cite{Do08a}).
Then the scaling form (\ref{G.fscal-aniso}) becomes
\begin{align}
\label{G.fscal-anisospec}
f_\text{s,aniso}&(t,L; \Jpl,\Jpp)
\nn\\
&=
L^{-d}(\Jpp/\Jpl)^{(d-1)/2}
\Fcal_\text{iso}(t(L/\xi_{0,\perp})^{1/\nu}).
\end{align}
Since
\begin{align}
\label{G.corr-ratioGauss}
(\Jpp/\Jpl)^{1/2}=\xi_{0,\perp}/ \xi_{0,\parallel}
\end{align}
according
to (\ref{G.corr}), this relation can be written as
\begin{align}
\label{G.fscal-anisospec-neu1}
f_\text{s,aniso}(t,L;\Jpl,\Jpp)
=L^{-d}\Fcal_\text{aniso}(t(L/\xi_{0,\perp})^{1/\nu};\Jpl,\Jpp),
\end{align}
where the finite-size scaling function $\Fcal_\text{aniso}$ of the
anisotropic system, considered as a function of the single scaling
variable $t(L/\xi_{0,\perp})^{1/\nu}$,  is  {\it nonuniversal},
\begin{align}
\label{G.fscal-anisospec-neu2}
\Fcal_\text{aniso}&(t(L/\xi_{0,\perp})^{1/\nu};\Jpl,\Jpp)
\nn\\
&=
(\xi_{0,\perp}/\xi_{0,\parallel})^{d-1}
\Fcal_\text{iso}(t(L/\xi_{0,\perp})^{1/\nu}),
\end{align}
as it depends on the nonuniversal ratio $\Jpp/\Jpl$ through the factor $(\xi_{0,\perp}/\xi_{0,\parallel})^{d-1}$.
(For the $d=2$ Ising model (see (\ref{G.corr-ratioIsing}) and  Sec. VI B),
this factor depends on $\Jpp$ and $\Jpl$ separately.)
As a consequence, also other thermodynamic quantities have a corresponding
finite-size scaling structure.
This was recently confirmed for the case of {\it antiperiodic} b.c.\ in
the large-$n$ limit in $2<d<4$ dimensions \cite{dan09}.
So far no explicit verification of (\ref{G.fscal-anisospec}) has been
given for systems with surface contributions.
In App.~\ref{G.Ap.free.energy} we shall verify that
(\ref{G.fscal-anisospec}) holds within the Gaussian model for all b.c.\
in $2<d<4$ dimensions, including those involving surface terms, in the
temperature range where finite-size scaling holds.
The consequences for the Casimir force scaling functions will be discussed
in Sec.~\ref{G.Casimir.force}.
Eq.~(\ref{G.fscal-anisospec}) is not directly valid for the free energy
density in $d=2$ dimension since the bulk part has a logarithmic
structure, see (\ref{G.fbs.ln}), but we have verified that it is valid
for the {\it excess} free energy density and for the Casimir force scaling
form  of the  $d=2$  anisotropic Gaussian model for all
b.c.\ (see Secs.~\ref{G.Casimir.force} and \ref{G.2d.Ising}).

The issue of nonuniversality of finite-size amplitudes of the free energy
with respect to coupling anisotropy was studied earlier in the work by
Indekeu et al.\ \cite{Indekeu}.
In this paper an anisotropic Ising model on an infinitely long
two-dimensional strip with free b.c.\ in the vertical direction was
considered.
This corresponds to our geometry for the special case $d=2$ with
DD b.c..
As noted above, this is a particularly simple case as no distortions of
the b.c.\ arise even if the anisotropic couplings correspond to a
nondiagonal anisotropy matrix.
Furthermore, there exists no analog to a ``film'' transition at finite
width of the infinite strip below the two-dimensional ``bulk'' critical
temperature $\Tc$ since there exists no singularity in an effectively
one-dimensional system with short-range interactions.
For the present case of interest, i.e., for the case of two different
nearest-neighbor couplings in the horizontal and vertical directions,
the Ising Hamiltonian (divided by $k_BT)$ of Indekeu
et al.\ \cite{Indekeu} contains ferromagnetic nearest-neighbor couplings
denoted by $K_1$ and $K_2$ which in our notation correspond to
$2\be\Jpl$ and $2\be\Jpp$, respectively, with a lattice spacing $\at=1$.

The authors derived an exact relation between the amplitudes
$\Delta_\text{aniso}$ and $\Delta_\text{iso}$ of the free energies at
criticality of the anisotropic and isotropic Ising strips of the form
\begin{align}
\label{G.Delta-sinh}
\Delta_\text{aniso}
=\left[\frac{\sinh(4\be_c\Jpp)}{\sinh(4\be_c\Jpl)}\right]^{1/2}
\Delta_\text{iso}.
\end{align}
Since
\begin{align}
\label{G.corr-ratioIsing}
[\sinh(4\be_c\Jpp)/\sinh(4\be_c\Jpl)]^{1/2}
=\xi_{0,\perp}/\xi_{0,\parallel}
\end{align}
is the ratio of the amplitudes of the correlation lengths perpendicular
and parallel to the Ising strip \cite{Indekeu}, Eq.~(\ref{G.Delta-sinh})
can be written as
\begin{align}
\label{G.Delta-neu}
\Delta_\text{aniso}
=(\xi_{0,\perp}/\xi_{0,\parallel})\;\;\Delta_\text{iso}.
\end{align}
This is the same structure as given in (\ref{G.fscal-anisospec-neu2})
for $d=2$.

It was also shown that the ratio $\xi_{0,\perp}/\xi_{0,\parallel}$ can be
interpreted as a geometrical factor that arises in a transformation of
lengths such that isotropy is restored \cite{Indekeu}.
This is in complete agreement with the analysis presented here and in
Refs.~\cite{Do08a} and \cite{Do06}.
Nevertheless, in spite of the exact relation (\ref{G.Delta-sinh}), it is
clear that restoring isotropy does not imply ``restoring universality''
\cite{Indekeu} since the finite-size amplitude $\Delta_\text{aniso}$
of the original anisotropic lattice model depends explicitly on the
microscopic couplings $\Jpl$ and $\Jpp$ .

We note that the dependence of $\xi_{0,\perp}/\xi_{0,\parallel}$ on the
nearest-neighbor couplings $\Jpl$ and $\Jpp$ is a nonuniversal property
that has a different form in $d=2$ dimensions for the Gaussian model on
the one hand (see (\ref{G.corr-ratioGauss})) and for the Ising model on
the other hand (see (\ref{G.corr-ratioIsing})).
The latter is not captured by heuristic arguments based on a mapping of
a $d=2$ lattice spin model on a continuum model as seen from Eq.~(6.5)
of Ref.~\cite{diehl2009}.

\section{Film critical behavior}
\label{G.film.crit}
In the following we briefly discuss the film critical behavior of the
$d$-dimensional Gaussian model which we need to refer to in
Sec.~\ref{G.free.energy}.
Here we confine ourselves to $2\leq d<4$.

First we consider the isotropic case.
For finite $L$, the film critical point is determined by $\rz=\rzcf(L)$
with $\rzcf(L)=0$ for periodic and NN b.c., whereas
\begin{align}
\label{G.r0c.film}
\rzcf(L)=-(4J/\at^2)[1-\cos(q_0\at)]<0,
\end{align}
with $q_0=\pi/L$ for antiperiodic b.c., $q_0=\pi/(L+\at)$ for DD b.c.,
and $q_0=\pi/(2L+\at)$ for ND b.c., respectively.
For large $L/\at$, $\rzcf(L)=-2J \pi^2/L^2$ for antiperiodic and
DD b.c.\ and $\rzcf(L)=-2J\pi^2/(4L^2)$ for ND b.c..
Correspondingly, the film critical lines are described, for large $L$, by
\begin{align}
\label{G.r0cDD.film}
\tcfilm(L)\equiv [\Tcfilm(L)-\Tc]/\Tc=-\pi^2(\xiz/L)^{1/\nu}
\end{align}
for antiperiodic and DD b.c.\ and by
\begin{align}
\label{G.tcND.film}
\tcfilm(L)=-(\pi/2)^2(\xiz/L)^{1/\nu}
\end{align}
for ND b.c., in agreement with finite-size scaling.
For the shape of the film critical lines see Figs.~\ref{G.NN.DD.scaling}
and \ref{G.NDscaling.3} below.

Near $\Tcfilm(L)$ there exist long-range correlations parallel
to the boundaries.
A corresponding second-moment correlation length $\xif(\rz,L)$
may be defined by
\begin{align}
\label{G.xi.film}
\xif(\rz,L)^2=\frac{1}{2(d-1)}
\frac{\sum_{\yb,z,\yb',z'}(\yb-\yb')^2\langle S_{\yb,z}S_{\yb',z'}\rangle}%
{\sum_{\yb,z,\yb',z'}\langle S_{\yb,z}S_{\yb',z'}\rangle}.
\end{align}
(The summation over all $z$, $z'$ corresponds to a kind of averaging over
all horizontal layers.)

Define a length
\begin{align}
\label{G.ell}
\ell(\rz,L)= \left(\frac{2J}{\rz-\rzcf(L)}\right)^{1/2}.
\end{align}
For periodic and NN b.c., where $\rzcf(L)=0$, we obtain just as in
the bulk case (\ref{G.xi0.nu}) the exact relationship
$\xif=\ell=\xi=\left(2J/\rz\right)^{1/2}$, which is independent of $L$.
For antiperiodic, DD, and ND b.c.\ and arbitrary $L/\at$, we obtain
$\xif=\ell$ only in the limit where $\ell\gg L$.

At finite $L$, the free energy per unit area divided by $\kB T$
is defined as
\begin{align}
\label{G.ffilm}
\ffilm(\rz,L)&= L f(t,L).
\end{align}
We expect that for $\ell\gg L$ (which is equivalent to the condition
$\xif\gg L$ mentioned in \cite{KrDi92a}),
the film critical behavior corresponds to that of a bulk system in $d-1$
dimensions.
Taking into account (\ref{G.fbs}), this would imply that the singular
part $\ffilms$ has the temperature dependence for $2\leq d<4$, $d\neq3$,
\begin{align}
\label{G.ffs.2-4}
\ffilms(t)&=Y_{d-1}\xif^{-(d-1)},
& d\neq3,
\end{align}
where the dimensionless universal amplitude $Y_{d-1}$ is defined by
(\ref{G.fbs}).
We indeed confirm this expectation for all b.c.\ except for
{\it antiperiodic} b.c.\ whose lowest mode has a two-fold
degeneracy as noted already in Sec.~\ref{G.Hamiltonian} above.
This causes a factor of $2$ in the
corresponding relation
\begin{align}
\label{G.ffs.a.2-4}
\ffilms^\abc(t)&=2Y_{d-1}\xif^{-(d-1)},
& d\neq3,~\text{antiperiodic b.c..}
\end{align}

For $d=3$, the expected structure of $\ffilms$ is less
obvious because of the logarithmic dependence of the corresponding bulk
quantity (\ref{G.fbs.2}) in $d=2$ dimensions.
For $\ell\gg L$ we obtain
\begin{align}
\label{G.f.film.crit}
\ffilm(\rz,L)&=\ffilm(\rzcf,L)+\ffilms+O(\rz-\rzcf),
\end{align}
where it is necessary to specify the singular part $\ffilms$ separately
for the various b.c.,
\bse
\label{G.f.film.crit.s.2}
\begin{align}
\label{G.f.film.crit.p.s.2}
\ffilms^\pbc
&=
\frac{1}{4\pi}\xif^{-2}\ln( \xif/L),
\\
\label{G.f.film.crit.a.s.2}
\ffilms^\abc
&=
\frac{1}{2\pi}\xif^{-2}\ln(\xif/L),
\\
\label{G.f.film.crit.NN.s.2}
\ffilms^\NNbc
&=
\frac{1}{4\pi}\xif^{-2}\ln( \xif/\sqrt{L\at}),
\\
\label{G.f.film.crit.DD.s.2}
\ffilms^\DDbc
&=
\frac{1}{4\pi}\xif^{-2}\ln(\xif\at^{1/2}/L^{3/2}),
\\
\label{G.f.film.crit.ND.s.2}
\ffilms^\NDbc
&=
\frac{1}{4\pi}\xif^{-2}\ln(\xif/L),
\end{align}
\ese
as will be derived in Secs.~\ref{G.F.pa.2-4}, \ref{G.F.NN.DD.3},
and \ref{G.F.ND.2-4} below.

Both microscopic and macroscopic reference lengths may appear
in the logarithmic arguments depending on the b.c..
(Our decomposition is such that no logarithmic dependencies on
$\at$ or $L$ appear in the nonsingular part of $\ffilm$ proportional
to $\rz-\rzcf$.)
By contrast, the amplitude $1/(4\pi)$ appears to have a universal
character, in agreement with (\ref{G.fbs.ln}), except for the factor
of $2$ for antiperiodic b.c..
The expressions for $\ffilm(\rzcf,L)$ at the critical line
$T=\Tcfilm(L)$ are nonuniversal and depend on the b.c..

For the anisotropic case, we consider only two different nearest-neighbor
interactions $\Jpl$ and $\Jpp$ as described by (\ref{G.Jplpp}) and
corresponding different bulk correlation lengths $\xi_\parallel$ and
$\xi_\perp$ as described by (\ref{G.corr}).
In the paragraph containing Eqs.~(\ref{G.r0c.film})--(\ref{G.tcND.film})
the only necessary changes are a replacement of $J$ by $\Jpp$ and of
$\xiz$ by $\xi_{0,\perp}$.

Define the film correlation length, now called $\xi_\filmaniso$, by
(\ref{G.xi.film}) and lengths $\ell_{\perp}$ and $\ell_{\parallel}$ by
\begin{align}
\label{G.ell.aniso}
\ell_{\perp(\parallel)}(\rz,L)
=\left(\frac{2J_{\perp(\parallel)}}{\rz-\rzcf(L)}\right)^{1/2}.
\end{align}
For periodic and NN b.c., we obtain, in close analogy to the isotropic
case, the exact relationship
$\xifa=\ell_\parallel=\xi_\parallel=\left(2J_\parallel/\rz\right)^{1/2}$,
which is again independent of $L$.
For antiperiodic, DD, and ND b.c.\ and arbitrary $L/\at$, we obtain
$\xifa=\ell_\parallel$ only in the limit where $\ell_\perp\gg L$.

The considerations of the paragraphs containing
Eqs.~(\ref{G.ffilm})--(\ref{G.f.film.crit.s.2})
translate one to one to the anisotropic case if $\xif$ is replaced by
$\xifa$ and the condition $\ell\gg L$ is replaced by $\ell_\perp\gg L$.

\section{\boldmath Free energy in $1<d<4$ dimensions}
\label{G.free.energy}

In the following we present exact results for the asymptotic structure
of the finite-size critical behavior of the Gaussian free energy density
near the bulk transition temperature for large $L/\at\gg1$ in the
isotropic case.
These results include both the bulk critical behavior
(\ref{G.fbs})--(\ref{G.fbs.ln}) for $L\to\infty$ at fixed $t>0$ and the
film critical behavior (\ref{G.ffs.2-4})--(\ref{G.f.film.crit.s.2}) for
$T\to \Tcfilm(L)$ at fixed finite $L$.
Thus our results provide an exact description of the dimensional crossover
from the $d$-dimensional finite-size critical behavior near bulk $\Tc$ to
the $(d{-}1)$-dimensional critical behavior near $\Tcfilm$ of (the
isotropic subclass of) the Gaussian universality class.
Our scaling functions ${\cal F}$ are analytic at bulk $\Tc$ for
antiperiodic, DD, and ND b.c., in agreement with the general discussion
given in Sec.~VII of Ref.~\cite{KrDi92a}.
Our Gaussian results go beyond the corresponding
one-loop results of Ref.~\cite{KrDi92a} in the following respects:
(i) Our exact calculation includes nonnegligible logarithmic non-scaling
lattice effects in $d=3$ dimensions for the case of NN and DD
b.c., whereas these effects are not captured by the method of dimensional
regularization used in Ref.~\cite{KrDi92a}.
(ii) For the case of ND b.c., a strong power-law violation of scaling is
found in general dimensions $1<d<4$
that has an important impact on the
scaling structure of the free energy density in a large part of the
$L^{-1/\nu}$--$t$ plane of the Gaussian model and that is expected to
imply unusually large corrections to scaling in the $\varphi^4 $ theory.
(iii) Our representation of the scaling functions is directly applicable
to the region $\Tcfilm(L)\leq T \leq \Tc$ for antiperiodic, DD and ND
b.c., whereas the representation of Ref.~\cite{KrDi92a} is applicable
only to $T>\Tc$, apart from a few results in Sec.~VII of
Ref.~\cite{KrDi92a}.
(iv) We study the approach to the critical behavior near $\Tcfilm$ and
compare it with the critical behavior of a $(d{-}1)$-dimensional bulk
system; this comparison confirms the unexpected factor of two of the
leading universal amplitude for the case of antiperiodic b.c.\ that was
presented in our Sec.~\ref{G.film.crit} as a consequence of the twofold
degeneracy of the lowest mode.
(v) Our analysis includes, for all b.c., the exponential {\it nonscaling}
part of the excess free energy due to the lattice-dependent nonuniversal
exponential bulk correlation length (\ref {G.xie}) that was not taken into
account in \cite{KrDi92a}.
(vi) Our analysis also includes the $d=2$ scaling functions of the
finite-size part of the free energy that provide the basis for the
Casimir force scaling functions in $d=2$ dimensions to be discussed in
Sec.~\ref{G.Casimir.force} and \ref{G.2d.Ising} (it is only the $d=2$ bulk
part of the free energy that exhibits a logarithmic deviation
$\sim\ln(\xi/\at)$ from scaling, see Sec.~\ref{G.bulk.film.crit});
the case $d=2$ was not discussed in \cite{KrDi92a}.

Because of the special role played by the borderline dimension $d^*=3$
for the surface properties of the Gaussian model it is necessary to
distinguish the cases {\it without} surface contributions (periodic and
antiperiodic b.c.) from those with surface contributions (NN, DD, and
ND b.c.).

\subsection{Periodic and antiperiodic b.c.}
\label{G.F.pa.2-4}

For periodic and antiperiodic b.c., the finite-size scaling structure
of (\ref{G.fscal}) and (\ref{G.fexdef}) is confirmed.
For $d\neq2$ we find (see Appendix~\ref{G.Ap.free.energy}) the finite-size
scaling functions
\bse
\label{G.Fcal.pa}
\begin{align}
\label{G.Fcal.p}
\!\!\!\Fcal^\pbc(\xt)&=\Ic{d}^\pbc(\xt)+Y_d\xt^{d/2},
& \xt&\geq0,
\\
\label{G.Fcal.a}
\!\!\!\Fcal^\abc(\xt)&=\Ic{d}^\abc(\xt{+}\pi^2){+}Y_d (\xt{+}\pi^2)^{d/2},
& \xt&\geq -\pi^2,
\end{align}
\ese
with the universal bulk amplitude from (\ref{G.Yd}), where, for
$\yv\geq0$,
\bse
\label{G.Itd.pa}
\begin{align}
\label{G.Itd.p}
&\Ic{d}^\pbc(\yv)
=
-\f{1}{2\pi}\int_0^\infty d\zi\,(\pi/\zi)^{(d+1)/2}
\times
\nn\\
&~~~~~~~~~~~~~~~~~~
e^{-\zi\yv/(2\pi)^2}\left[K(\zi)-\sqrt{\pi/\zi}\right],
\\
\label{G.Itd.a}
&\Ic{d}^\abc(\yv)
=
-\f{1}{2\pi}\int_0^\infty d\zi\,(\pi/\zi)^{(d+1)/2}
\times
\nn\\
&
\left\{e^{-\zi\yv/(2\pi)^2}
\left[e^{\zi/4}[K(\zi/4)-K(\zi)]-\sqrt{\pi/\zi}\right]
-\f{\sqrt{\pi\zi}}{4}\right\}.
\end{align}
\ese
Eq.~(\ref{G.Itd.a}) is valid for $2<d<4$, while for $1<d<2$ the
subtraction of the term $\f{1}{4}\sqrt{\pi\zi}$ inside the curly
brackets has to be omitted.
The function $K(z)$ is defined by
$K(z)\equiv\sum_{n=-\infty}^{+\infty}\exp(-n^2 z)$,
which converges rapidly for large $z$.
It may be expressed in terms of the third elliptic theta function
$\vt_3(u,e^{-z})$ \cite{GrRy94} via $K(z)=\vt_3(0,e^{-z})$.
It satisfies the relation $K(z)=\sqrt{\pi/z}K(\pi^2/z)$
with the expansion
$K(z)=\sqrt{\pi/z}\sum_{n=-\infty}^{+\infty}\exp(-n^2\pi^2/z)$,
which converges rapidly for small $z$.

The function $\Fcal^\abc(\xt)$ is regular at $\xt=0$ in agreement with
general analyticity requirements, whereas $\Fcal^\pbc(\xt)$ is nonanalytic
at $\xt=0$ due to the film critical point.

Eqs.~(\ref{G.Fcal.pa}) include the singular parts of both the bulk
critical behavior ($\xt\to\infty$) and the film critical
behavior ($\xt=L^2/\xi_\film^2\to0$ for periodic b.c.\ and
\mbox{$\xt+\pi^2=L^2/\xi_\film^2\to0$} for antiperiodic b.c.).
The latter is obtained from the singular parts of the small-$\yv$
expansions for $\yv>0$
\bse
\label{G.Itd.pa.smallx}
\begin{align}
\label{G.Itd.p.smallx}
\Ic{d}^\pbc(\yv)
&=
\Ic{d}^\pbc(0)+Y_{d-1}\yv^{(d-1)/2}+O(\yv,\yv^{d/2}),
\\
\label{G.Itd.a.smallx}
\Ic{d}^\abc(\yv)
&=
\Ic{d}^\abc(0)+2Y_{d-1}\yv^{(d-1)/2}+O(\yv,\yv^{d/2}),
\end{align}
\ese
for $d\neq3$, while for $d=3$
\bse
\label{G.It3.pa.smallx}
\begin{align}
\label{G.It3.p.smallx}
\Ic{3}^\pbc(\yv)
&=
-\f{\ze(3)}{2\pi}-\f{1}{8\pi}\yv\left(\ln\yv-1\right)+O(\yv^{3/2}),
\\
\label{G.It3.a.smallx}
\Ic{3}^\abc(\yv)
&=
-\f{\ze(3)}{2\pi}-\f{1}{4\pi}\yv\left[\ln\yv-1-\ln(2\pi)\right]
+O(\yv^{3/2}).
\end{align}
\ese
Contrary to the naive expectation based on universality,
the amplitudes of the leading singular $\yv^{(d-1)/2}$ and
$\yv\ln\yv$ terms of (\ref{G.Itd.pa.smallx})
and (\ref{G.It3.pa.smallx}), respectively, differ by a factor of two for
periodic and antiperiodic b.c.\ as already mentioned in
Sec.~\ref{G.film.crit}.
These terms yield the right hand sides of (\ref{G.ffs.2-4}),
(\ref{G.ffs.a.2-4}), (\ref{G.f.film.crit.p.s.2}), and
(\ref{G.f.film.crit.a.s.2}).

Comparison of (\ref{G.fexdef}) and (\ref{G.Fcal.pa}) leads to the
finite-size parts for $\xt\geq0$
\bse
\label{G.Gcal.pa}
\begin{align}
\label{G.Gcal.p}
\Gcal^\pbc(\xt)&=\Ic{d}^\pbc(\xt),
\\
\label{G.Gcal.a}
\Gcal^\abc(\xt)&=2^{1-d}\Gcal^\pbc(4\xt)-\Gcal^\pbc(\xt),
\end{align}
\ese
where (\ref{G.Gcal.a}) follows from (\ref{G.BN.a}).
Eqs.~(\ref{G.Gcal.pa}) remain valid for $d=2$.
For $d=3$, Eq.~(\ref{G.Gcal.a}) agrees with Eqs.~(9.3) for $N=1$ of
Ref.~\cite{KrDi92a}.

The representation of our results differs from that of \cite{KrDi92a},
where Eqs.~(6.8) provide an integral representation of $\Gcal^\pbc(\xt)$
and $\Gcal^\abc(\xt)$.
Both representations have the same expansions in terms of modified Bessel
functions, see Appendix~\ref{G.Ap.Galternative}, which suggests that, for
$\xt>0$, indeed $\Gcal^\pbc(\xt)=\Th_{+\text{per}}^{(1)}(y_+)$ and
$\Gcal^\abc(\xt)=\Th_{+\text{aper}}^{(1)}(y_+)$, with the identification
$y_+=\sqrt{\xt}$.
Our representation of $\Fcal^\pbc(\xt)$, $\Gcal^\pbc(\xt)$, and
$\Gcal^\abc(\xt)$ in terms of $\Ic{d}^\pbc$ has the advantage that it is
directly applicable to the bulk critical point at $\xt=0$, whereas the
integral representation of $\Th_{+\text{per}}^{(1)}(y_+)$ and
$\Th_{+\text{aper}}^{(1)}(y_+)$ given in Eqs.~(6.8) of \cite{KrDi92a}
require an extra small-$y_+$ treatment of the divergent integrals so that
after multiplication with the prefactor $y_+^d$ finite results are
obtained.
More importantly, the representation of $\Fcal^\abc(\xt)$ in terms of
$\Ic{d}^\abc$ has the advantage that it is valid also for $\xt\leq0$
including the film critical point at $\xt=-\pi^2$, whereas the integral
representation of $\Th_{+\text{aper}}^{(1)}(y_+)$ in Eqs.~(6.8) of
\cite{KrDi92a} is not suitable for an analytic continuation to the
region $\xt<0$.

A representation of $\Fcal^\abc(\xt)$ valid for all $\xt\geq-\pi^2$
may also be extracted from the result (3.26) with (3.27) in
Ref.~\cite{dan09} for the singular part of the excess free energy of
the mean spherical model with antiperiodic b.c.\ in film geometry.
After omitting the term $\propto x_t$, restoring the bulk contribution
by removing the term $\propto y_\text{b}^{d/2}$, and replacing
$y\to\xt+\pi^2$, the last two terms within the curly brackets of
Eq.~(3.26) in Ref.~\cite{dan09} may be shown to be equivalent to the
integral representation (\ref{G.Fcal.a}) with (\ref{G.Itd.a}) of
$\Fcal^\abc(\xt)$.

The universal finite-size amplitudes at $\Tc$ are
\bse
\label{G.Fcal.pa.0}
\begin{align}
\label{G.Fcal.p.0}
\Fcal^\pbc(0)=\Gcal^\pbc(0) &= -\pi^{-d/2}\Ga(d/2)\ze(d),
\\
\label{G.Fcal.a.0}
\Fcal^\abc(0)=\Gcal^\abc(0) &=(1-2^{1-d})\pi^{-d/2}\Ga(d/2)\ze(d),
\end{align}
\ese
which agree with the corresponding $N=1$ amplitudes $\De_\text{per}^{(1)}$
and $\De_\text{aper}^{(1)}$, respectively, in Eq.~(5.7) of \cite{KrDi92a}
(up to a sign misprint there for periodic b.c.).

At fixed $t>0$ the results for $\Gcal^\pbc$ and $\Gcal^\abc$ yield the
large-$L$ approach to the bulk critical behavior
\begin{align}
\label{G.f.pa.large.L}
f(t,L)-\fb(t)
&=
\mp\f{1}{L^d}\f{\xt^{(d-1)/4}}{(2\pi)^{(d-1)/2}}e^{-\sqrt{\xt}},
& \xt&\gg1,
\end{align}
where the upper (lower) sign refers to periodic (antiperiodic) b.c.,
see the paragraph around (\ref{G.Gp.large.xt}).
For sufficiently large $L$ at fixed $t>0$, however, the exponential
scaling form (\ref{G.f.pa.large.L}) must be replaced by an exponential
{\it nonscaling} form \cite{Do08a} which is obtained from
(\ref{G.f.pa.large.L}) by replacing the exponential argument
$-\sqrt{\xt}$ by $-L/\xie$, where $\xie$ is the exponential correlation
length (\ref{G.xie}).

In $d=3$ dimensions the scaling functions (\ref{G.Fcal.pa}) can be
expressed as
\bse
\label{G.Fcal.pa.3}
\begin{align}
\label{G.Fcal.p.3}
\Fcal^\pbc(\xt)
&=
\Gcal^\pbc(\xt)
-\f{1}{12\pi}\xt^{3/2},
& \xt&\geq0,
\\
\label{G.Fcal.a.3}
\Fcal^\abc(\xt)
&=
\Gcal^\abc(\xt)
-\f{1}{12\pi}\xt^{3/2},
& \xt&\geq-\pi^2,
\end{align}
\ese
with the finite-size parts
\bse
\label{G.Gcal.pa.3}
\begin{align}
\label{G.Gcal.p.3}
\hspace{-6pt}
\Gcal^\pbc(\xt)
&{=}
{-}\f{1}{2\pi}\left[\Li_3(e^{-\sqrt{\xt}})
+\sqrt{\xt}\Li_2(e^{-\sqrt{\xt}})\right],
\\
\label{G.Gcal.a.3}
\hspace{-6pt}
\Gcal^\abc(\xt)
&{=}
{-}\f{1}{2\pi}\left[\Li_3(-e^{-\sqrt{\xt}})
+\sqrt{\xt}\Li_2(-e^{-\sqrt{\xt}})\right],
\end{align}
\ese
where $\Li_n(\zi)$ are polylogarithmic functions (see
Appendix~\ref{G.Ap.functions}).
With the identification $y_+=\xt$ we find that $\Gcal^\pbc(\xt)$
and $\Gcal^\abc(\xt)$ agree with $\Th_{+\text{per}}(y_+)$
and $\Th_{+\text{aper}}(y_+)$ in Eqs.~(9.3) of \cite{KrDi92a},
respectively, but that the representation of
$\Th_{+\text{aper}}(y_+)$ is more elaborate than that of
$\Gcal^\abc(\xt)$.
It is understood that in (\ref{G.Fcal.a.3}) for \mbox{$-\pi^2<\xt<0$},
the function $\Gcal^\abc(\xt)$ means the analytic continuation
of (\ref{G.Gcal.a.3}) to $\xt<0$ which is complex; together with
the complex term $-\xt^{3/2}/(12\pi)$, however, the right-hand
side of (\ref{G.Fcal.a.3}) becomes real and analytic for all
$\xt>-\pi^2$ with a finite real value
\begin{align}
\label{G.Fcal.a.cfilm.d3}
\Fcal^\abc(-\pi^2)=-\f{\ze(3)}{2\pi}\approx-0.191313,
\end{align}
see Appendix~\ref{G.Ap.analytic.properties}.
For $\xt\geq0$, the $d=3$ scaling functions $\Gcal^\pbc(\xt)$ and
$\Gcal^\abc(\xt)$ will be shown in Sec.~\ref{G.Casimir.force} together
with the corresponding scaling functions $X^\pbc(\xt)$ and $X^\abc(\xt)$
of the Casimir force.

\begin{figure}
\begin{center}
\includegraphics[width=7cm,angle=0]{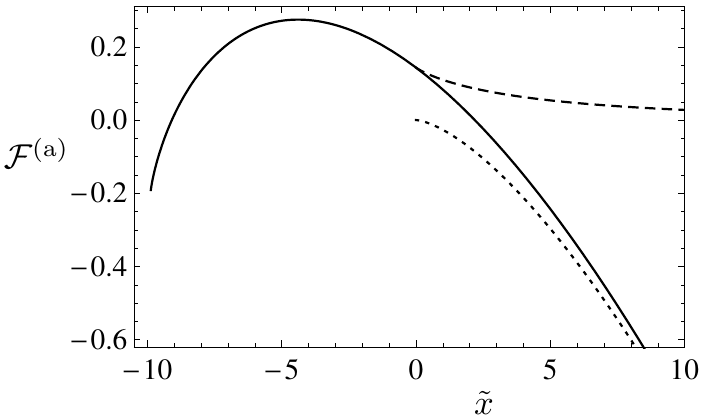}

\end{center}
\caption{\label{G.FbGa3}
Scaling function $\Fcal^\abc(\xt)$, (\ref{G.Fcal.a.3}), of the free
energy of the Gaussian model in three dimensions with antiperiodic
b.c.\ for $\xt\geq-\pi^2$ (solid line).
For $\xt\geq0$ are also shown the bulk part $Y_3\xt^{3/2}$ (dotted)
and finite-size part $\Gcal^\abc(\xt)$ (dashed).}
\end{figure}
In Fig.~\ref{G.FbGa3} we show the scaling function $\Fcal^\abc(\xt)$,
(\ref{G.Fcal.a.3}), of the Gaussian free energy density in three
dimensions for antiperiodic b.c.\ including the range for negative $\xt$
down to the film transition at $\xt=-\pi^2$.
It would be interesting to compare this result with the corresponding
$\ve$ expansion result at $\ve=1$ which, however, is not
available in the literature so far.

\subsection{\boldmath NN and DD b.c.\ in $d\neq3$ dimensions}
\label{G.F.NN.DD.2-3-4}

For NN and DD b.c.\ there exist well-defined surface free energy
densities for $t>0$ in $d>1$ dimensions.
They are given by (see Appendix~\ref{G.Ap.free.energy})
\bse
\label{G.fsf}
\begin{align}
\label{G.fsf.N}
&\fsf^\Nbc(t)
\nn\\
&~~=
\f{1}{8\at^{d-1}}\int_0^\infty\f{d\yi}{\yi}\B(\yi)^{d-1}
\left[e^{-2\yi}-1\right]e^{-\yi\rzt/2}
\nn\\
&~~=
\f{1}{8\at^{d-1}}
\left[\Wt{d-1}(\rzt)-\Wt{d-1}(4+\rzt)\right],
\\
\label{G.fsf.D}
&\fsf^\Dbc(t)
\nn\\
&~~=
\f{1}{8\at^{d-1}}\int_0^\infty\f{d\yi}{\yi}\B(\yi)^{d-1}
\left[e^{-2\yi}+1-2\B(\yi)\right]e^{-\yi\rzt/2}
\nn\\
&~~=
\f{1}{8\at^{d-1}}
\left[-\Wt{d-1}(\rzt)-\Wt{d-1}(4+\rzt)
+2\Wt{d}(\rzt)\right],
\end{align}
\ese
with $\rzt$ defined after Eq.~(\ref{G.fbulk}).
The result for $\fsf^\Dbc$ agrees with $f_\text{surface}$ in Eq.~(67)
of \cite{ChDo03}.
For $d\neq3$ and small $t>0$, the singular parts are
\bse
\label{G.fsf.s.exp}
\begin{align}
\label{G.fsf.s.N.exp}
\fsfs^\Nbc(t)
&=
\f{\Asf^\Nbc}{\xi^{d-1}}
+O(\xi^{-(d+1)}),
\\
\label{G.fsf.s.D.exp}
\fsfs^\Dbc(t)
&=
\f{\Asf^\Dbc}{\xi^{d-1}}
-\f{\Ga(-\f{d}{2})}{4(4\pi)^{d/2}}\f{\at}{\xi^d}
+O( \xi^{-(d+1)}),
\end{align}
\ese
with the universal surface amplitudes
\begin{align}
\label{G.Asf.ND}
\Asf^\Nbc
&=
-\Asf^\Dbc
=\f{1}{4}Y_{d-1},
& 1&<d<5, & d\neq3,
\end{align}
in agreement with Eq.~(6.3) in \cite{KrDi92a} and the remark about the
surface contribution in the last paragraph on page 1910 of \cite{KrDi92a}
as well as with Eqs.~(76) and (88) in \cite{ChDo03}.
The nonsingular parts are for $\tau=\text{N},\text{D}$
\begin{align}
\label{G.fsf.ns.exp}
\fsfns^\tbc(t)
&=
\fsf^\tbc(0)
-\f{\btd^\tbc}{\at^{d-1}}\rzt+O(\rzt^2),
\end{align}
with the nonuniversal constants
\bse
\label{G.btd.N.D}
\begin{align}
\label{G.btd.N.1d3}
\!\!\btd^\Nbc
&\equiv
\f{1}{16}\int_0^\infty\!\!\!d\yi
\left[\B(\yi)^{d-1}(e^{-2\yi}{-}1)
{+}(2\pi\yi)^{(1-d)/2}\right]\,{>}\,0,
\\
\label{G.btd.D.1d2}
\!\!\btd^\Dbc
&\equiv
\f{1}{16}\int_0^\infty\!\!\!d\yi
\Big\{\B(\yi)^{d-1}[e^{-2\yi}+1-2\B(\yi)]
\nn\\
&~~~~~~~~~~
-(2\pi\yi)^{(1-d)/2}
+2(2\pi\yi)^{-d/2}
\Big\}>0.
\end{align}
\ese
Eq.~(\ref{G.btd.N.1d3}) holds for $1\,{<}\,d\,{<}\,3$, while for
$3\,{<}\,d\,{<}\,5$
(where $\btd^\Nbc\,{<}\,0$) the addition of $(2\pi\yi)^{(1-d)/2}$ inside
the curly brackets has to be omitted.
These integral expressions for $\btd^\Nbc$ are connected by analytical
continuation in $d$.
Eq.~(\ref{G.btd.D.1d2}) holds for $1\,{<}\,d\,{<}\,2$, while for
$2\,{<}\,d\,{<}\,3$ (where $\btd^\Nbc\,{<}\,0$) the addition of
$2(2\pi\yi)^{-d/2}$ inside the curly brackets has to be omitted.
For $3\,{<}\,d\,{<}\,5$ (where $\btd^\Dbc\,{>}\,0$ again), additionally
the subtraction of $(2\pi\yi)^{(1-d)/2}$ has to be omitted.
These integral expressions for $\btd^\Dbc$ are connected by analytical
continuation in $d$, as already noted in \cite{ChDo03}, where closely
related integral representations of $\btd^\Dbc$ were given in Eqs.~(77)
and (89), valid for $2\,{<}\,d\,{<}\,3$ and $3\,{<}\,d\,{<}\,5$,
respectively.
For $d\to3$, both the amplitudes $\Asf^\tbc$ and the coefficients
$\btd^\tbc$ diverge, while the nonuniversal constants
$\fsf^\Nbc(0)\,{<}\,0$ and $\fsf^\Dbc(0)\,{>}\,0$ remain finite,
see Sec.~\ref{G.F.NN.DD.3} below.
For $d\to2$, $\btd^\Dbc$ diverges but the corresponding term in
(\ref{G.fsf.ns.exp}) combines with the subleading term in
(\ref{G.fsf.s.D.exp}) to give a finite contribution to $\fsf^\Dbc$.

The $\xi^{1-d}$ terms in (\ref{G.fsf.s.exp}) agree with the corresponding
contributions in Eq.~(6.3) and Appendix~C of \cite{KrDi92a}.
The sum of (\ref{G.fsf.s.D.exp}) and (\ref{G.fsf.ns.exp}) for
$\tau=\text{D}$ b.c.\ with (\ref{G.Asf.ND}), and $\btd^\Dbc$ as given
in Ref.~\cite{ChDo03} agrees with Eqs.~(75)--(77) and (87)--(91) of
\cite{ChDo03}.
In (\ref{G.fsf.s.D.exp}) we have included a singular term of order
$\xi^{-d}$.
Such a term does not exist in (\ref{G.fsf.s.N.exp}).
Although this term is subleading compared to the leading singular
$\xi^{1-d}$ term, it becomes a leading singular term for ND b.c.\ (to
be discussed in Sec.~\ref{G.F.ND.2-4} below), where the terms $\xi^{1-d}$
of (\ref{G.fsf.s.N.exp}) and (\ref{G.fsf.s.D.exp}) cancel because of
(\ref{G.Asf.ND}).

For NN and DD b.c.\ the finite-size scaling structure of (\ref{G.fscal})
and (\ref{G.fexdef}) is confirmed for $d\neq3$.
We find the finite-size scaling functions
(see Appendix~\ref{G.Ap.free.energy})
\begin{align}
\label{G.Fcal.NN.DD}
\Fcal^\ttbc&(\xt)
=
\Ic{d}^\ttbc(\xt+c\pi^2)+Y_d(\xt+c\pi^2)^{d/2}
\hspace{-35pt}
\nn\\
&
+2\Asf^\tbc(\xt+c\pi^2)^{(d-1)/2},
& \xt&\geq -c\pi^2,
\end{align}
with $c=0$ for $\tau=\text{N}$ and $c=1$ for $\tau=\text{D}$,
with the universal bulk amplitude $Y_d$ from (\ref{G.Yd}), and where
\bse
\begin{align}
\label{G.Itd.NN}
&\Ic{d}^\NNbc(\yv)
=
2^{-d}\Ic{d}^\pbc(4\yv),
\\
\label{G.Itd.DD.3d4}
&\Ic{d}^\DDbc(\yv)
=
-\f{1}{2^{d+1}\pi}
{\ds\int_0^\infty}d\zi\,(\pi/\zi)^{(d+1)/2}
\times
\nn\\
&~
\left\{e^{-\zi\yv/\pi^2}
\left[e^{\zi}[K(\zi)-1]
-\sqrt{\pi/\zi}+1\right]-\sqrt{\pi\zi}+\zi\right\},
\end{align}
\ese
with $\Ic{d}^\pbc$ from (\ref{G.Itd.p}).
Eq.~(\ref{G.Itd.DD.3d4}) is valid for $3\,{<}\,d\,{<}\,4$, while for
$2\,{<}\,d\,{<}\,3$ ($1\,{<}\,d\,{<}\,2$), the addition of $\zi$
(the addition of $\zi$ and the subtraction of $\sqrt{\pi\zi}$) inside
the curly brackets has to be omitted.
The function $\Fcal^\DDbc(\xt)$ is regular at $\xt=0$ in agreement with
general analyticity requirements, whereas $\Fcal^\NNbc(\xt)$ is
nonanalytic at $\xt=0$ due to the film critical point.

Eqs.~(\ref{G.Fcal.NN.DD}) include the singular parts
of both the bulk critical behavior (\ref{G.fbs}) ($\xt\to\infty$) and the
film critical behavior (\ref{G.ffs.2-4}) ($\xt=L^2/\xi_\film^2\to0$
for NN b.c.\ and \mbox{$\xt+\pi^2=L^2/\xi_\film^2\to0$} for DD b.c.).
The latter is obtained from the surface terms of Eqs.~(\ref{G.Fcal.NN.DD})
and from singular parts of the small-$\yv$ expansions for $\yv>0$,
\bse
\label{G.Itd.NN.DD.smallx}
\begin{alignat}{2}
\label{G.Itd.NN.smallx}
\Ic{d}^\NNbc(\yv)
&=
\Ic{d}^\NNbc(0)+\fr{1}{2}Y_{d-1}\yv^{(d-1)/2}
&&+O(\yv,\yv^{d/2}),
\nn\\
&&&
d\neq3,
\\
\label{G.Itd.DD.smallx}
\Ic{d}^\DDbc(\yv)
&=
\Ic{d}^\DDbc(0)+\fr{3}{2}Y_{d-1}\yv^{(d-1)/2}
&&+O(\yv,\yv^{d/2}),
\nn\\
&&&
d\neq2,3.
\end{alignat}
\ese
We note that, according to (\ref{G.Asf.ND}), the surface amplitudes
$\Asf$ of the $d$-dimensional film system have the same
$d$-dependence as the bulk amplitude $Y_{d-1}$ of the
$(d{-}1)$-dimensional bulk system, apart from a constant factor of
$\pm1/4$.
This implies
\begin{align}
\label{G.A-Y}
2\Asf^\Nbc+\fr{1}{2}Y_{d-1}
=2\Asf^\Dbc+\fr{3}{2}Y_{d-1}
=Y_{d-1},
\end{align}
which explains how the $\yv^{(d-1)/2}$ terms on the right hand sides of
(\ref{G.Itd.NN.DD.smallx}) and the terms in (\ref{G.Fcal.NN.DD}) involving
the surface amplitudes (\ref{G.Asf.ND}) lead to identical amplitudes
$Y_{d-1}$ for the film free energy in (\ref{G.ffs.2-4}) for both NN and DD
b.c., in agreement with the expectation based on universality.

For the finite-size contribution $\sim L^{-d}$ in (\ref{G.fexdef}) we
find the scaling functions for $\xt\geq0$
\begin{align}
\label{G.Gcal.NN.DD}
\Gcal^\NNbc(\xt)
&=
\Gcal^\DDbc(\xt)=2^{-d}\Gcal^\pbc(4\xt)
\nn\\
&=
\Ic{d}^\NNbc(\xt)=2^{-d}\Ic{d}^\pbc(4\xt),
\end{align}
where $\Gcal^\DDbc(\xt)$
agrees with Eq.~(71) in \cite{ChDo03} with \mbox{$x=\sqrt{\xt}\geq0$}.

The representation of our results differs from that of \cite{KrDi92a},
where Eqs.~(6.8) and (6.6) provide an integral representation of
$\Gcal^\NNbc$ and $\Gcal^\DDbc$, respectively.
Both representations have the same expansions in terms of modified Bessel
functions, see Appendix~\ref{G.Ap.Galternative}, which suggests that, for
$\xt>0$, indeed $\Gcal^\NNbc(\xt)=\Th_{+\text{SB,SB}}^{(1)}(y_+)$ and $\Gcal^\DDbc(\xt)=\Th_{+\text{O,O}}^{(1)}(y_+)$,
with the identification $y_+=\sqrt{\xt}$.
Our representation of $\Fcal^\NNbc(\xt)$, $\Gcal^\NNbc(\xt)$, and
$\Gcal^\DDbc(\xt)$ in terms of $\Ic{d}^\NNbc$ has the advantage that it
is directly applicable to the bulk critical point at $\xt=0$, whereas the
integral representation of $\Th_{+\text{SB,SB}}^{(1)}(y_+)$ and
$\Th_{+\text{O,O}}^{(1)}(y_+)$ given in Eqs.~(6.8) and (6.6),
respectively, of \cite{KrDi92a} require an extra small-$y_+$ treatment
of the divergent integrals so that after multiplication with the
prefactor $y_+^d$ finite results are obtained.
More importantly, the representation of $\Fcal^\DDbc(\xt)$ in terms of
$\Ic{d}^\DDbc$ has the advantage that it is valid also for $\xt\leq0$
including the film critical point at $\xt=-\pi^2$, whereas the integral
representation of $\Th_{+\text{O,O}}^{(1)}(y_+)$ in Eq.~(6.6) of
\cite{KrDi92a} is not suitable for an analytic continuation to the
region $\xt<0$.

The universal finite-size amplitudes at $\Tc$ are
\begin{align}
\Fcal^\NNbc(0)
&=
\Gcal^\NNbc(0)=\Fcal^\DDbc(0)=\Gcal^\DDbc(0)
\nn\\
&=-(4\pi)^{-d/2}\Ga(d/2)\ze(d),
\end{align}
which agree with the corresponding $N=1$ amplitudes
$\De_\text{SB,SB}^{(1)}$ for NN b.c.\ and
$\De_\text{O,O}^{(1)}$ for DD b.c.\ in Eqs.~(5.7) and (5.6)
of \cite{KrDi92a}, respectively.

At fixed $t>0$ the results for $\Gcal^\NNbc$ and $\Gcal^\DDbc$
yield the same large-$L$ approach to the bulk critical behavior
\begin{align}
\label{G.f.NN.DD.large.L}
f(t,L)-\fb(t)
=
\frac{2\fsf(t)}{L}
-\f{1}{L^d}\f{\xt^{(d-1)/4}}{2(4\pi)^{(d-1)/2}}&e^{-2\sqrt{\xt}},
\nn\\
\xt\gg1,&
\end{align}
see the paragraph around (\ref{G.Gp.large.xt}).
Eq.~(\ref{G.f.NN.DD.large.L}) is in agreement with the result Eq.~(72) in
\cite{ChDo03} for free (DD) b.c..
For sufficiently large $L$ at fixed $t>0$, the exponential part of the
scaling form (\ref{G.f.NN.DD.large.L}) must be replaced by an exponential
{\it nonscaling} form \cite{Do08a} which is obtained from
(\ref{G.f.NN.DD.large.L}) by replacing the exponential argument
$-2\sqrt{\xt}$ by $-2L/\xie$, where $\xie$ is the exponential correlation
length (\ref{G.xie}).
The same remark applies to the exponential parts contained in the scaling
functions that are presented in Eqs.~(\ref{G.f.s.NN.DD.new}),
(\ref{G.fsf.s.N.exp2}), and (\ref{G.f.ND.large.L}) below.

\subsection{\boldmath NN and DD b.c.\ in $d=3$ dimensions}
\label{G.F.NN.DD.3}

For NN and DD b.c.\ in $d=3$ dimensions, the vanishing of the critical
exponent (\ref{G.surface.exponent}) of the surface energy density
\cite{ChDo03} causes logarithmic deviations $\sim\ln(\xi/\at)$ from the
scaling structure of (\ref{G.fscal}).
From (\ref{G.fsf.s.exp}), (\ref{G.fsf.ns.exp}), and (\ref{G.btd.N.D}),
we obtain for $d\to3$ the singular and nonsingular parts of the surface
free energy density for small $t>0$ as
\bse
\label{G.fsf.N.D.3.exp}
\begin{align}
\label{G.fsf.N.3.exp}
\fsf^\Nbc(t)
&=
\fsf^\Nbc(0)+\f{\ln(\xi/\at)}{16\pi\xi^2}-\f{\bt^\Nbc}{\at^2}\rzt
\nn\\
&\ph{=}
+O(\rzt^2,\xi^{-4}\ln\xi),
\\
\label{G.fsf.D.3.exp}
\fsf^\Dbc(t)
&=
\fsf^\Dbc(0)-\f{\ln(\xi/\at)}{16\pi\xi^2}
-\f{1}{24\pi}\f{\at}{\xi^3}-\f{\bt^\Dbc}{\at^2}\rzt
\nn\\
&\ph{=}
+O(\rzt^2,\xi^{-4}\ln\xi),
\end{align}
\ese
with the nonuniversal constants
\bse
\label{G.bt.N.D.3}
\begin{align}
\label{G.bt.N.3}
&\bt^\Nbc
\equiv
\lim_{d\to3}\left(\btd^\Nbc-\Asf^\Nbc\right)
\nn\\
&=
\f{1}{16}\int_0^\infty d\yi\left\{\B(\yi)^2
\left[e^{-2\yi}-1\right]+\f{1-e^{-\yi/2}}{2\pi\yi}\right\}
-\f{1}{32\pi}
\nn\\
&=
\f{1}{8}\W{2}(4)-\f{5\ln2+1}{32\pi}
\approx
-0.027653,
\\
\label{G.bt.D.3}
&\bt^\Dbc
\equiv
\lim_{d\to3}\left(\btd^\Dbc-\Asf^\Dbc\right)
\nn\\
&=
\f{1}{16}\int_0^\infty d\yi\left\{\B(\yi)^2
\left[e^{-2\yi}+1-2\B(\yi)\right]-\f{1-e^{-\yi/2}}{2\pi\yi}\right\}
\nn\\
&\ph{=}
+\f{1}{32\pi}
\nn\\
&=
\f{1}{8}\left[\W{2}(4)-2\W{3}(0)\right]+\f{5\ln2+1}{32\pi}
\approx-0.00199279.
\end{align}
\ese
The limit $d\to3$ in (\ref{G.bt.N.D.3}) is independent of whether it
is taken as $d\to3_-$ or $d\to3_+$, in agreement with Eqs.~(79) and (92)
of \cite{ChDo03} for the case of Dirichlet b.c..
The structure of the leading singular terms of (\ref{G.fsf.N.3.exp})
and (\ref{G.fsf.D.3.exp}) agrees with that of the two-dimensional
result (\ref{G.fbs.2}) but the amplitudes are different.
For the same reason as in (\ref{G.fsf.s.D.exp}), we have included the
subleading $\xi^{-3}$ term in (\ref{G.fsf.D.3.exp}).
Eq.~(\ref{G.fsf.D.3.exp}) with (\ref{G.bt.D.3}) agrees with
Eqs.~(80)--(82) of \cite{ChDo03} but here we give a simplified expression
of $\bt^\Dbc$ as compared to Eqs.~(81) and (82) in \cite{ChDo03}.

The singular surface contributions (\ref{G.fsf.N.3.exp}) and
(\ref{G.fsf.D.3.exp}) appear also in the resulting singular parts
of the free energy densities for $t\geq0$
(see Appendix~\ref{G.Ap.free.energy}),
\begin{align}
\label{G.f.s.NN.DD.new}
\fs^\tbc(t,L)
&=
\f{Y_3}{\xi^3}\pm2\f{\ln(\xi/\at)}{16\pi\xi^2L}
+\f{\Gcal^\tbc(\xt)}{L^3},
\end{align}
with ``$+$'' for $\tau=\text{NN}$ and ``$-$'' for $\tau=\text{DD}$,
and where
\begin{align}
\label{G.Gcal.NN.DD.3}
&\Gcal^\NNbc(\xt)
=\Gcal^\DDbc(\xt)=2^{-3}\Gcal^\pbc(4\xt)
\nn\\
&~~=
-\f{1}{16\pi}\left[\Li_3(e^{-2\sqrt{\xt}})
+2\sqrt{\xt}\Li_2(e^{-2\sqrt{\xt}})\right].
\end{align}
Eq.~(\ref{G.f.s.NN.DD.new}) for DD b.c.\ agrees with Eq.~(86) in
\cite{ChDo03} (there is a sign misprint in Eq.~(85) of \cite{ChDo03}).
With $y_+=\sqrt{\xt}$, Eq.~(\ref{G.Gcal.NN.DD.3}) agrees with Eqs.~(9.3)
for $N=1$ in \cite{KrDi92a}.
\begin{figure}[t]
\begin{center}
\includegraphics[width=7cm,angle=0]{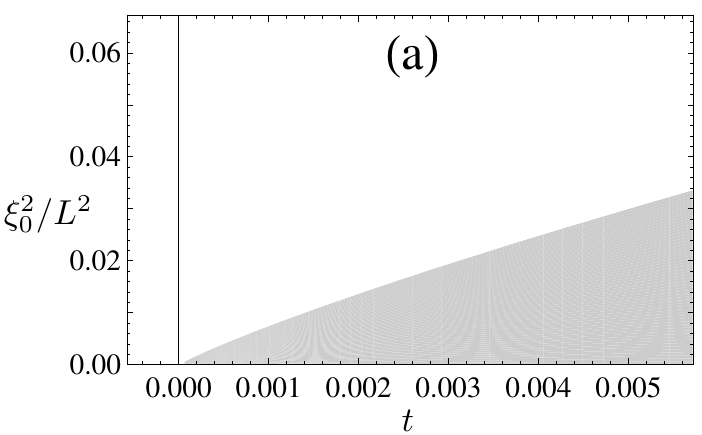}
\includegraphics[width=7cm,angle=0]{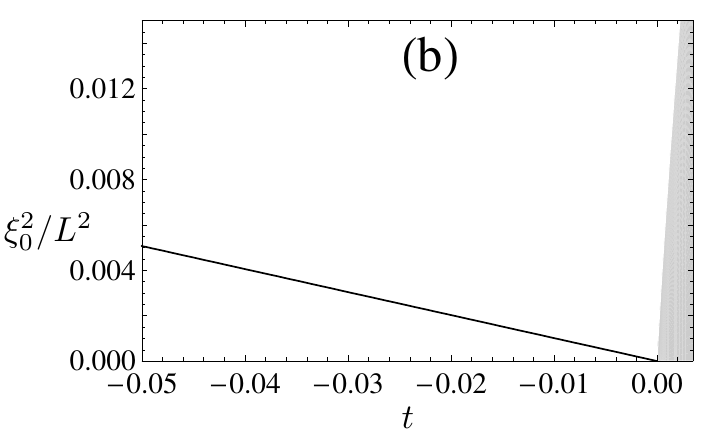}
\end{center}
\caption{\label{G.NN.DD.scaling}
Asymptotic part of the $(\xiz/L)^{1/\nu}$--$t$ plane of the Gaussian
model in three dimensions with isotropic short-range interaction in
film geometry with NN (a) and DD (b) b.c..
The solid lines indicate the film critical temperatures $\Tcfilm(L)$
at finite $L$:
(a) vertical line at $t=0$ for NN b.c.,
(b) Eq.~(\ref{G.r0cDD.film}) for DD b.c..
No low-temperature phases exist for $T<\Tcfilm(L)$.
Finite-size scaling is valid between the film critical lines and the
shaded areas.
The shaded areas (as defined in the text) are nonscaling regions that
depend on $\at/\xiz$.
Their shapes are shown here for the example $\at/\xiz=1$.
These shapes start at the origin with infinite slope.
The film critical lines in (a) and (b) have the same form as for the
cases of periodic and antiperiodic b.c., respectively.
In these cases surface terms are absent and only very small nonscaling
regions exist due to the nonscaling exponential parts
$\sim\exp(-L/\xie)$ mentioned after Eq.~(\ref{G.f.pa.large.L}).}
\end{figure}

Because of the dependence of $\ln(\xi/\at)$ on the nonuniversal lattice
spacing $\at$, no finite-size scaling functions of $\fs^\NNbc(t,L)$ and
$\fs^\DDbc(t,L)$ can be defined.
The nonuniversal surface terms $\sim L^{-1}$ constitute the leading
deviations from the bulk critical behavior for large $L$ at fixed $t>0$.
One may define ``nonscaling regions'' in the $(\xiz/L)^{1/\nu}$--$t$
planes (see Figs.~\ref{G.NN.DD.scaling}(a) and (b)) by requiring that
these logarithmic terms are comparable to or larger than the scaling
terms $L^{-3}\Gcal(\xt)$.
These nonscaling regions depend on $\at/\xiz$ and are shown for the
example $\at/\xiz=1$ as the shaded regions in
Figs.~\ref{G.NN.DD.scaling}(a) and (b).

The logarithmic deviations from scaling are not present right at bulk
$\Tc$, where $\fs(0,L)= L^{-3}\Gcal(0)$ with the universal critical
amplitudes $\Gcal^\NNbc(0)=\Gcal^\DDbc(0)=-\ze(3)/(16\pi)$, in agreement
with Eq.~(9.2) for $N=1$ in \cite{KrDi92a}.

For the remaining part of the discussion we need to distinguish the cases
of NN and DD b.c..
For NN b.c., the function $\fs^\NNbc(t,L)$ is not regular at $t=0$ as it
includes the film critical behavior (\ref{G.f.film.crit}) with
(\ref{G.f.film.crit.NN.s.2}) for $t\to0$ at fixed $L$.
To derive this behavior we use the small-$\xt$ expansion for $\xt\geq0$
\begin{align}
\label{G.G.NN.DD.smalls}
\Gcal^\NNbc(\xt)
&=-\f{\ze(3)}{16\pi}-\f{\xt(\ln\xt{+}2\ln2{-}1)}{16\pi}+\f{\xt^{3/2}}{12\pi}
+O(\xt^2),
\end{align}
which implies
\begin{align}
\fs^\NNbc(t,L)
=\fs^\NNbc(0,L)-\f{1}{L^3}\left[\frac{\xt\ln(\xt\at/L)}{8\pi}
+O(\xt)\right].
\end{align}
The second term yields (\ref{G.f.film.crit.NN.s.2}) because of
$\xt=L^2/\xi_\film^2$ for NN b.c..

By contrast, the film critical point for DD b.c.\ is located at
$\tcfilm<0$, see (\ref{G.r0cDD.film}), thus no singularity exists
at $t=0$ for finite $L$ for DD b.c., which implies that $\fs^\DDbc(t,L)$
should be regular at $t=0$.
This is indeed the case as shown in the following.
The first three terms of $\fs^\DDbc(t,L)$ from (\ref{G.f.s.NN.DD.new})
can be rewritten as
\begin{align}
\label{G.f.s.DD.new.3.terms}
&\fs^\DDbc(t,L)
\nn\\
&=\f{1}{L^3}\left[\Gcal^\DDbc(\xt)
-\f{\xt^{3/2}}{12\pi}+\f{\xt\ln\xt}{16\pi}-\f{\xt\ln(L/\at)}{8\pi}\right],
\end{align}
where now the logarithmic deviation from scaling appears in the form of
$\ln(L/\at)$ but the temperature dependence through $\xt\sim t$ is
regular at $t=0$ since
\begin{align}
\label{G.f.s.DD.regular.new}
\Gcal^\DDbc(\xt)-\f{\xt^{3/2}}{12\pi}+\f{\xt\ln\xt}{16\pi}
\end{align}
is regular at $\xt=0$
(see Appendix~\ref{G.Ap.analytic.properties}).
It is understood that in (\ref{G.f.s.DD.regular.new}) for $-\pi^2<\xt<0$,
the function $\Gcal^\DDbc(\xt)$ means the analytic
continuation to $\xt<0$ as given by (\ref{G.Gcal.NN.DD.3}), which is
complex; together with the complex terms
$-\xt^{3/2}/(12\pi)+\xt\ln\xt/(16\pi)$, however,
(\ref{G.f.s.DD.regular.new}) becomes real and analytic for $\xt>-\pi^2$
with a finite real value at $\xt=-\pi^2$, see
Appendix~\ref{G.Ap.analytic.properties}.
The representation (\ref{G.f.s.DD.new.3.terms}) has the advantage that it
is valid down to $\xt=-\pi^2$ corresponding to the film critical point.
For $\xt\to-\pi^2$ it includes the film critical behavior
(\ref{G.f.film.crit}) with (\ref{G.f.film.crit.DD.s.2}).
To derive this behavior we use an expansion around $\xt=-\pi^2$ for
$\xt>-\pi^2$ (see Appendix~\ref{G.Ap.analytic.properties}),
\begin{align}
\label{G.It3.NN.smallsplusi2}
&\Gcal^\DDbc(\xt)-\f{\xt^{3/2}}{12\pi}+\f{\xt\ln\xt}{16\pi}
\nn\\
&\!=-\f{\ze(3)+2\pi^2\ln\pi}{16\pi}
-\f{(\xt+\pi^2)[2\ln(\xt+\pi^2)-4\ln\pi-3]}{16\pi}
\nn\\
&\ph{=}
+O((\xt+\pi^2)^2),
\end{align}
which implies
\begin{align}
&\fs^\DDbc(t,L)
=\fs^\DDbc(\tcfilm,L)
\nn\\
&\ph{=}
+\f{1}{L^3}\left\{
-\f{(\xt+\pi^2)\ln[(\xt+\pi^2)L/\at)]}{8\pi}
+O(\xt+\pi^2)\right\}.
\end{align}
The second term yields (\ref{G.f.film.crit.DD.s.2}) because of
\mbox{$\xt+\pi^2=L^2/\xi_\film^2$} for DD b.c..

For $\xt\geq0$, the $d=3$ scaling functions $\Gcal^\NNbc(\xt)$ and
$\Gcal^\DDbc(\xt)$ will be shown in Sec.~\ref{G.Casimir.force}
together with the corresponding scaling functions $X^\NNbc(\xt)$
and $X^\DDbc(\xt)$ of the Casimir force.

\subsection{\boldmath ND b.c.\ in $1<d<4$ dimensions}
\label{G.F.ND.2-4}

For ND b.c., the leading terms of the singular parts of the surface
free energies, i.e., the $O(\xi^{1-d})$ terms in
(\ref {G.fsf.s.N.exp}) and (\ref {G.fsf.s.D.exp}) and the logarithmic
terms in (\ref{G.fsf.N.3.exp}) and (\ref{G.fsf.D.3.exp}), cancel.
Then the leading term of the singular part of the total surface free
energy density for $t>0$,
\begin{align}
\label{G.fsf.s.N.exp1}
\fsfs^\NDbc =\fsfs^\Nbc(t)+\fsfs^\Dbc(t)
=-\f{\Ga(-\f{d}{2})}{4(4\pi)^{d/2}}\f{\at}{\xi^d}
+O( \xi^{-(d+1)}),
\end{align}
does not have the universal scaling form (\ref{G.Asurface}), but depends
explicitly on $\at$.
The cancelation of the leading surface terms $\sim\xi^{1-d}$ for
ND b.c.\ was already noted in Appendix~C of \cite{KrDi92a}, where,
however, the next-to-leading surface term $\sim\xi^{-d}$,
(\ref{G.fsf.s.N.exp1}), was not taken into account.
In contrast to the weak logarithmic deviations from scaling in $d=3$
dimensions according to (\ref{G.fsf.N.D.3.exp}),
Eq.~(\ref{G.fsf.s.N.exp1}) constitutes a strong power-law violation of
scaling (within the Gaussian model) that has an important impact on the
scaling structure of the free energy density in a large part of the
$L^{-1/\nu}$--$t$ plane.
The resulting singular and nonsingular parts of the free energy density
for ND b.c.\ read for $t\geq0$
\begin{align}
\label{G.fsf.s.N.exp2}
&\fs^\NDbc(t,L)
=
\fbs(t)+\f{1}{L}
\left[-\f{\Ga(-\f{d}{2})}{4(4\pi)^{d/2}}\f{\at}{\xi^d}\right]
+\f{\Gcal^\NDbc(\xt)}{L^d},
\\
\label{G.fsf.ns.N.exp2}
&\fns^\NDbc(t,L)
=
\fbns(t)
\nn\\
&\ph{=}
+\f{1}{L}\left[\fsf^\Nbc(0)+\fsf^\Dbc(0)
-\f{\btd^\Nbc+\btd^\Dbc}{\at^{d-1}}\rzt
+O(\rzt^2)\right],
\end{align}
where $\btd^\Nbc+\btd^\Dbc<0$ and, for $\xt\geq0$,
\begin{align}
\label{G.Gcal.ND}
\Gcal^\NDbc(\xt)=
2^{-d}\Gcal^\abc(4\xt),
\end{align}
with $\Gcal^\abc$ from (\ref{G.Gcal.a})
(see Appendix~\ref{G.Ap.free.energy}).
This result remains valid for $d\to3$, where
$\lim_{d\to3}(\btd^\Nbc+\btd^\Dbc)=\bt^\Nbc+\bt^\Dbc$ with $\bt^\Nbc$
and $\bt^\Dbc$ given by (\ref{G.bt.N.D.3}).
For $d\to2$, the divergent terms of (\ref{G.fsf.s.N.exp2}) and
(\ref{G.fsf.ns.N.exp2}) combine to give a finite result, while
$\Gcal^\NDbc(\xt)$ remains finite and continues to provide the scaling
function of the finite-size contribution to the free energy.

The representation of our results differs from that of
Ref.~\cite{KrDi92a}, where Eqs.~(6.8) provide an integral representation
of $\Gcal^\NDbc$.
Both representations have the same expansions in terms of modified Bessel
functions, see Appendix~\ref{G.Ap.Galternative}, which suggests that, for
$\xt>0$, $\Gcal^\NDbc(\xt)=\Th_{+\text{O,SB}}^{(1)}(y_+)$,
with the identification $y_+=\sqrt{\xt}$.
Our representation of $\Gcal^\NDbc(\xt)$ in terms of $\Gcal^\abc(\xt)$
and thus $\Ic{d}^\pbc$ has the advantage that it is directly applicable
to the bulk critical point at $\xt=0$, whereas the integral representation
of $\Th_{+\text{O,SB}}^{(1)}(y_+)$ given in Eqs.~(6.8) of \cite{KrDi92a}
requires an extra small-$y_+$ treatment of the divergent integral so that
after multiplication with the prefactor $y_+^d$ a finite result is
obtained.
More importantly, the related representation of $\Fcal^\NDbc(\xt)$ in
terms of $\Ic{d}^\NDbc$ provided in Eq.~(\ref{G.Fcal.ND}) below has the
advantage that it is valid also for $\xt\leq0$ including the film
critical point at $\xt=-(\pi/2)^2$, whereas the integral representation
of $\Th_{+\text{O,SB}}^{(1)}(y_+)$ in Eqs.~(6.8) of \cite{KrDi92a} is not
suitable for an analytic continuation to the region $\xt<0$.

\begin{figure}[t]
\begin{center}
\includegraphics[width=7cm,angle=0]{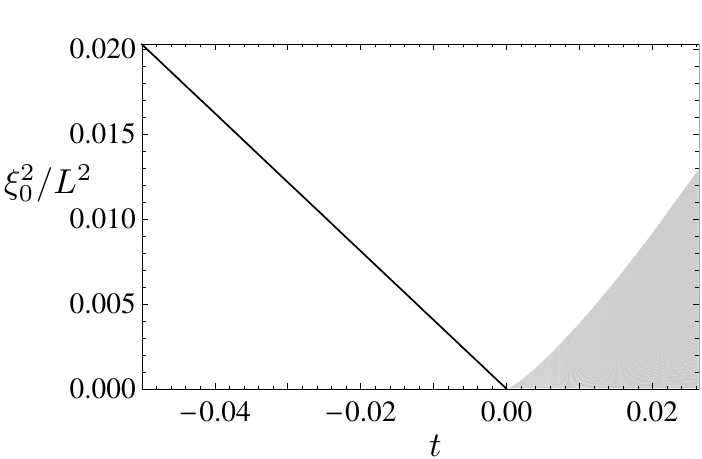}
\end{center}
\caption{\label{G.NDscaling.3}
Asymptotic part of the $(\xiz/L)^{1/\nu}$--$t$ plane of the Gaussian
model in three dimensions with isotropic short-range interaction in film
geometry with ND boundary conditions.
The solid line indicates the film critical temperature
$\Tcfilm(L)$ at finite $L$ according to Eq.~(\ref{G.tcND.film}).
No low-temperature phase exists for $T<\Tcfilm$.
Finite-size scaling is valid between the film critical line and the
shaded area.
The shaded area is a nonscaling region that depends on $\at/\xiz$.
Its shape (shown here for the example $\at/\xiz=1$) is determined
by Eq.~(\ref{G.fsf.s.N.exp3}).
The shape of the shaded area starts at the origin with zero slope.}
\end{figure}

In (\ref{G.fsf.s.N.exp2}) the nonscaling structure of the surface term
$\sim L^{-1}$ destroys the finite-size scaling form of $\fs^\NDbc(t,L)$
above $\Tc$ in the regime where the surface term is comparable to or
larger than the finite-size term $L^{-d}\Gcal^\NDbc(\xt)$, i.e., in
the regime
\begin{align}
\label{G.fsf.s.N.exp3}
\left(\frac{L}{\xi}\right)^{d-1}\;\f{\at}{\xi}
\;\agt\;
\left|\f{4(4\pi)^{d/2}}{\Ga(-d/2)}
\Gcal^\NDbc(\xt)\right|,
\end{align}
with $\xt=(L/\xi)^{1/\nu}$ for $\xt\geq0$ (this is valid only
for $d\neq2$; for $d=2$ the logarithm in the singular part
of the bulk free energy causes deviations from scaling in any case).
For $d=3$ this regime is indicated by the shaded area in
Fig.~\ref{G.NDscaling.3}.
This violation of finite-size scaling is significantly more important
than that due to the exponential correlation length $\xie$,
(\ref{G.xie}), which happens only for considerably larger
$L\agt 24\xi^3/\at^2$.
At fixed $t>0$, the result (\ref{G.fsf.s.N.exp2}) yields the large--$L$
approach to the bulk critical behavior
\begin{align}
\label{G.f.ND.large.L}
\fs^\NDbc(t,L)&-\fbs(t)
=L^{-1}\left[-\f{\Ga(-\f{d}{2})}{4(4\pi)^{d/2}}\f{\at}{\xi^d}\right]
\hspace{-50pt}
\nn\\
&
+L^{-d}\f{\xt^{(d-1)/4}}{2(4\pi)^{(d-1)/2}}e^{-2\sqrt{\xt}},
& \xt&\gg1,
\end{align}
see the paragraph around (\ref{G.Gp.large.xt}).
Eq.~(\ref{G.f.ND.large.L}) implies that for $\xi/\at\gg1$ the estimate
(\ref{G.fsf.s.N.exp3}) for the nonscaling region in
Fig.~\ref{G.NDscaling.3} can be replaced by
$L\agt\fr{1}{2}\xi\ln(\xi/\at)$.

The cancelation of the leading surface scaling terms in the Gaussian model
does not persist in the $\varphi^4$ theory at $O(u^*)$ (two-loop order)
as can be seen from Eq.~(D11) of \cite{KrDi92a}.
Two-loop terms, however, are typically smaller than one-loop terms,
therefore it is expected that the two-loop contributions to the scaling
part are less important than ordinary 1-loop scaling contributions.
This means that now corrections to scaling are expected to become
considerably more important compared to the scaling part.
This would imply a shrinking of the asymptotic region for the case of
ND b.c..
Thus we expect that a more careful analysis of future experiments or of
MC simulations is required for systems with ND b.c.\ because of unusually
large corrections to scaling.

Above the shaded area in Fig.~\ref{G.NDscaling.3} the nonscaling surface
term is negligible and the leading $L$ dependence of $\fs^\NDbc(t,L)$
is, for $d\neq2$, described by the scaling form
\begin{align}
\label{G.fs.ND}
\fs^\NDbc(t,L)&=L^{-d}\Fcal^\NDbc(\xt),
& \xt\geq-(\pi/2)^2,
\end{align}
where
\begin{align}
\label{G.Fcal.ND}
\Fcal^\NDbc(\xt)
&=
\Ic{d}^\NDbc(\xt+(\pi/2)^2)+Y_d[\xt+(\pi/2)^2]^{d/2}
\nn\\
&=
2^{-d}\Fcal^\abc(4\xt)
\end{align}
and
\begin{align}
\label{G.Itd.ND}
\Ic{d}^\NDbc(\yv)\equiv 2^{-d}\Ic{d}^\abc(4\yv),
\end{align}
with $\Fcal^\abc$ and $\Ic{d}^\abc$ from (\ref{G.Fcal.a}) and
(\ref{G.Itd.a}), respectively.
This result includes the film
critical behavior (\ref{G.ffs.2-4}) and (\ref{G.f.film.crit.ND.s.2})
for $\xt\to-(\pi/2)^2$ at fixed finite $L$.
To derive this behavior we use an expansion around $\xt=-(\pi/2)^2$,
\begin{multline}
\label{G.Itd.ND.smallx}
\Ic{d}^\NDbc(\yv)
=\Ic{d}^\NDbc(0)+Y_{d-1}\yv^{(d-1)/2}+O(\yv,\yv^{d/2}),
\\
d\neq3,
\end{multline}
while for $d=3$
\begin{align}
\label{G.It3.ND.smallx}
\Ic{3}^\NDbc(\yv)
&=
-\f{\ze(3)}{16\pi}
-\f{\yv[\ln(2\yv/\pi)-1]}{8\pi}+O(\yv^{3/2}).
\end{align}
The second terms on the right hand sides of (\ref{G.Itd.ND.smallx})
and (\ref{G.It3.ND.smallx}), respectively, yield (\ref{G.ffs.2-4}) and
(\ref{G.f.film.crit.ND.s.2}) because of $\xt+(\pi/2)^2=L^2/\xi_\film^2$.

The universal finite-size amplitude at $\Tc$ is
\begin{align}
\Fcal^\NDbc(0)
&=
\Gcal^\NDbc(0)=(1-2^{1-d})(4\pi)^{-d/2}\Ga(d/2)\ze(d),
\end{align}
which agrees with the corresponding $N=1$ amplitude
$\De_\text{O,SB}^{(1)}$ in Eq.~(5.7) of \cite{KrDi92a}.

For $d=3$ we combine (\ref{G.Fcal.a.3}) and (\ref{G.Fcal.ND}) to obtain
\begin{align}
\label{G.Fcal.ND.3}
\Fcal^\NDbc(\xt)
&=
\Gcal^\NDbc(\xt)-\f{1}{12\pi}\xt^{3/2},
\end{align}
\begin{align}
\label{G.Gcal.ND.3}
\Gcal^\NDbc(\xt)
&=
-\f{1}{16\pi}\left[\Li_3(-e^{-2\sqrt{\xt}})
+2\sqrt{\xt}\Li_2(-e^{-2\sqrt{\xt}})\right].
\end{align}
With the identification $y_+=\sqrt{\xt}$ we find that $\Gcal^\NDbc(\xt)$
agrees with the more elaborate representation of $\Th_{+\text{O,SB}}(y_+)$
provided by Eq.~(9.3) in Ref.~\cite{KrDi92a}.
Because of the relations (\ref{G.Gcal.ND}) and (\ref{G.Fcal.ND}), the
situation is similar to that explained after (\ref{G.Fcal.pa.3}) and
(\ref{G.Gcal.pa.3}), thus $\Fcal^\NDbc(\xt)$ is real for
$\xt\geq-(\pi/2)^2$ and an analytic function for $\xt>-(\pi/2)^2$, even
though the analytic continuation of $\Gcal^\NDbc(\xt)$ to negative
$\xt$ becomes complex.
For $\xt\geq0$, the $d=3$ scaling function $\Gcal^\NDbc(\xt)$
will be shown in Sec.~\ref{G.Casimir.force} together with the
corresponding scaling function $X^\NDbc(\xt)$ of the Casimir force.

In the region where finite-size scaling is valid
(see Fig.~\ref{G.NDscaling.3}) there exists a scaling function
$\Fcal^\NDbc(\xt)$ of the free energy density for $\xt\geq-(\pi/2)^2$.
Due to Eqs.~(\ref{G.Gcal.ND}) and (\ref{G.Fcal.ND}), a plot of this
function in three dimensions can be obtained from the solid curve in
Fig.~\ref{G.FbGa3} with appropriately rescaled axes (the same holds for
$\Gcal^\NDbc(\xt)$ and the bulk part $Y_3\xt^{3/2}$).

\section{Casimir force}
\label{G.Casimir.force}

The excess free energy density per component divided by $\kB T$ is
defined by
\begin{align}
\label{G.fex}
\fex(t,L)=f(t,L)-\fb(t),
\end{align}
where $\fb(t)$, (\ref{G.fb}), is the bulk free energy density.
The latter exists only for $t\geq 0$.
Thus, as a shortcoming of the Gaussian model, $\fex(t,L)$ can be defined
only for $t\geq0$ although $f(t,L)$, (\ref{G.ftbc}), exists for $t<0$ for
the cases of antiperiodic, DD, and ND boundary conditions.

The Casimir force $\FCas$ per component and per unit area divided by
$\kB T$ is related to $\fex$ by
\begin{align}
\FCas(t,L)/(\kB T)=-\f{\ptl[L\fex(t,L)]}{\ptl L}.
\end{align}
For the subclass of isotropic systems, the asymptotic scaling form of
its singular contribution is
\begin{align}
\label{G.Cas.scal}
\FCass(t,L)/(\kB T)=L^{-d}X(\xt),
\end{align}
where, for $\xt\geq0$, the universal scaling function $X(\xt)$ is
determined by the universal scaling function $\Gcal(\xt)$ of the
finite-size contribution to the free energy defined by (\ref{G.fexdef})
according to
\begin{align}
\label{G.X}
X(\xt)
&=
(d-1)\Gcal(\xt)-\nu^{-1}\xt \frac{d \Gcal(\xt)}{d \xt}.
\end{align}
The surface contributions to the free energy density do not contribute
to $X(\xt)$.
As an important consequence, finite-size scaling is found to be valid for
the Casimir force for all b.c.\ in $1<d<4$ dimensions, i.e., no scaling
violations exist for the Casimir force in the three-dimensional Gaussian
model with NN, DD, and ND b.c., in contrast to the free energy density
itself.

As a consequence of (\ref{G.fscal-anisospec}), the asymptotic scaling
form of the singular part of the Casimir force becomes {\it nonuniversal}
in the case of the anisotropic couplings (\ref{G.Jplpp}).
Then (\ref{G.Cas.scal}) is replaced by
\begin{align}
\label{G.Cas.scal.aniso}
\FCass&(t,L)/(\kB T)
\nn\\
&=L^{-d}(\Jpp/\Jpl)^{(d-1)/2}
X(t(L/\xi_{0,\perp})^{1/\nu}),
\end{align}
where $\xi_{0,\perp}$ is the amplitude of the correlation length
(\ref{G.corr.perp}).
Thus the Casimir force depends explicitly on the ratio of the microscopic
couplings $\Jpp$ and $\Jpl$ for all b.c., in agreement with earlier
results for periodic \cite{ChDo04,Do08a} and antiperiodic
\cite{dan09} b.c..
In the following we primarily discuss the isotropic case.

For $\xt\geq0$ follow from (\ref{G.X}) with (\ref{G.Gcal.NN.DD}) and
(\ref{G.Gcal.ND})
\bse
\label{G.X.other}
\begin{align}
\label{G.X.NN.DD}
X^\NNbc(\xt)=X^\DDbc(\xt)
&=
2^{-d}X^\pbc(4\xt),
\\
\label{G.X.ND}
X^\NDbc(\xt)
&=
2^{-d}X^\abc(4\xt),
\end{align}
\ese
and with (\ref{G.Gcal.a})
\begin{align}
\label{G.X.ap}
X^\abc(\xt)=2^{1-d}X^\pbc(4\xt)-X^\pbc(\xt).
\end{align}
Thus we only need $X^\pbc$, which we obtain in $1<d<4$ dimensions
by applying (\ref{G.X}) to (\ref{G.Gcal.p}).
In three dimensions we utilize (\ref{G.Gcal.p.3}), its derivative
\begin{align}
\label{G.Gcal.deriv.p}
\frac{d\Gcal^\pbc(\xt)}{d\xt}=
-\f{1}{4\pi}\ln\left(1-e^{-\sqrt{\xt}}\right),
\end{align}
and (\ref{G.X.ap}) to obtain the $d=3$ scaling functions for periodic
and antiperiodic b.c.\ for $\xt\geq0$,
\bse
\label{G.X.pa.3}
\begin{align}
\label{G.X.p.3}
X^\pbc(\xt)
&=
-\f{1}{\pi}\left[\Li_3(e^{-\sqrt{\xt}})
+\sqrt{\xt}\Li_2(e^{-\sqrt{\xt}})\right]
\nn\\
&\ph{=}
+\f{\xt}{2\pi}\ln\left(1-e^{-\sqrt{\xt}}\right)
\nn\\
&=
-\f{\ze(3)}{\pi}+\f{\xt}{4\pi}-\f{\xt^{3/2}}{12\pi}+O(\xt^2),
\\
\label{G.X.a.3}
X^\abc(\xt)
&=
-\f{1}{\pi}\left[\Li_3(-e^{-\sqrt{\xt}})
+\sqrt{\xt}\Li_2(-e^{-\sqrt{\xt}})\right]
\nn\\
&\ph{=}
+\f{\xt}{2\pi}\ln\left(1+e^{-\sqrt{\xt}}\right)
\nn\\
&=
\f{3\ze(3)}{4\pi}-\f{\xt^{3/2}}{12\pi}+O(\xt^2).
\end{align}
\ese
The scaling functions for the other b.c.\ follow by employing
(\ref{G.X.other}).

At $\xt=0$ the critical Casimir amplitude \cite{Kr94}
\begin{align}
\label{G.FSS.Delta}
\De \equiv(d-1)^{-1}X(0)
\end{align}
is obtained for $1<d<4$ as
\bse
\label{G.Casampl}
\begin{align}
\label{G.Casampl.NN.DD.p}
\De^\pbc\!
&=
2^d\De^\NNbc\!=2^d\De^\DDbc
=-\pi^{-d/2}\Ga(\fr{d}{2})\ze(d),
\\
\label{G.Casampl.ND.a}
\De^\abc\!
&=
2^d\De^\NDbc\!
=(2^{1-d}-1)\De^\pbc,
\end{align}
\ese
specializing for $d=3$ to
\bse
\label{G.Casampl.3}
\begin{align}
\label{G.Casampl.NN.DD.p.3}
\hspace{-3pt}
\De^\pbc\!\!
&=
8\De^\NNbc\!=8\De^\DDbc
=-\f{\ze(3)}{2\pi}
\approx-0.191313,
\\
\label{G.Casampl.ND.a.3}
\hspace{-3pt}
\De^\abc\!\!
&=
8\De^\NDbc\!
=\f{3\ze(3)}{8\pi}
\approx0.143485.
\end{align}
\ese
The results (\ref{G.Casampl}) are identical to the results (5.6) and
(5.7) of \cite{KrDi92a} after setting $N=1$ (there is a misprint
concerning the sign of $\De_\text{per}^{(1)}$ in (5.7) of \cite{KrDi92a}).
The results for $\De^\pbc$ are also in agreement with Eq.~(3.42) of
\cite{DaKr04}.

As a consequence of (\ref{G.Cas.scal.aniso}), the Casimir amplitude
$\Delta_\text{aniso}$ of the anisotropic system (with $\Jpp\neq\Jpl$)
is nonuniversal and is related to $\Delta$ of the isotropic system
(with $J=\Jpp=\Jpl$) for all b.c.\ by
\begin{align}
\label{G.FSS.Delta.aniso}
\Delta_\text{aniso}=(\Jpp/\Jpl)^{(d-1)/2}\Delta
=(\xi_{0,\perp}/\xi_{0,\parallel})^{d-1}\Delta,
\end{align}
in agreement with earlier results for periodic \cite{ChDo04} and
antiperiodic \cite{dan09} b.c..
For $d=2$, the right hand side of (5.13) is of the same form as found
in \cite{Indekeu} for the anisotropic Ising model on a two-dimensional
strip with free b.c..

\section{\boldmath Casimir force in $d=2$ dimensions}
\label{G.2d.Ising}
Our exact results for $X_\text{Gauss}$ in $d=2$ dimensions are of
particular interest in view of results for the exact Casimir force scaling
functions $X_\text{Ising}$
  for an Ising model with
{\it isotropic} couplings on a two-dimensional strip of infinite length
and finite width $L$ for free b.c.~\cite{Evans} and for periodic and
antiperiodic b.c.\ in the recent work  by
Rudnick et al. \cite{RuZaShAb10}.
Moreover, there exist earlier results for the Casimir amplitude at
$\Tc$ of the two-dimensional Ising model with free b.c.\ and
{\it anisotropic} couplings by Indekeu et al.~\cite{Indekeu}.
This calls for a comparison with the corresponding Gaussian model results
$X_\text{Gauss}$.

\subsection{Isotropic case}
\label{G.Ising.iso}
For the isotropic case, the two-dimensional Gaussian model results
for periodic, antiperiodic, and DD b.c.\ are obtained from
Eqs.~(\ref{G.Itd.p}), (\ref{G.Gcal.p}), (\ref{G.X}), (\ref{G.X.NN.DD}),
and (\ref{G.X.ap}) for $d=2$.
They read
\bse
\label{G.Gauss.Xp}
\begin{align}
X_\text{Gauss}^\pbc(\xt)
&=
-\f{1}{2\pi}\int_0^\infty d\zi\left(\f{\pi}{\zi}\right)^{3/2}
\left(1+\f{\zi\xt}{2\pi^2}\right)
\times
\nn\\
&~~~~~~
e^{-\zi\xt/(2\pi)^2}\left[K(\zi)-\sqrt{\f{\pi}{\zi}}\right],
\\
X_\text{Gauss}^\abc(\xt)
&=
\fr{1}{2}X_\text{Gauss}^\pbc(4\xt)-X_\text{Gauss}^\pbc(\xt),
\\
X_\text{Gauss}^\DDbc(\xt)
&=
\fr{1}{4}X_\text{Gauss}^\pbc(4\xt),
\end{align}
\ese
with the scaling variable $\xt=(L/\xi_{0})^{1/\nu}t=(L/\xi)^2$,
$\nu=1/2$, $t\,{\geq}\,0$.
Above $\Tc$, the corresponding Ising model results for periodic,
antiperiodic, and free b.c.\ read \cite{Evans,RuZaShAb10}
\bse
\label{G.Ising.Xp}
\begin{align}
X_\text{Ising}^\text{(p)}(x_\text{I})
&=
\f{1}{2\pi}\int_0^\infty\!d\om\,q\left[\tanh(q/2)-1\right],
\\
\label{G.Ising.Xa}
X_\text{Ising}^\text{(a)}(x_\text{I})
&=
\f{1}{2\pi}\int_0^\infty\!d\om\,q\left[\coth(q/2)-1\right],
\\
X_\text{Ising}^\text{(free)}(x_\text{I})
&=
\f{1}{2\pi}\int_0^\infty\!d\om\,q
\left[\f{\left(q{+}x_\text{I}\right)e^{q}-\left(q{-}x_\text{I}\right)e^{-q}}%
{\left(q{+}x_\text{I}\right)e^{q}+\left(q{-}x_\text{I}\right)e^{-q}}{-}1\right],
\end{align}
\ese
with $q\equiv\sqrt{x_\text{I}^2+\om^2}$.
Here our Ising scaling variable
$x_\text{I}=(L/\xi_{0,I})^{1/\nu}t= L/\xi$ with $\nu=1$ and
$\xi_{0,I}=(8\be_cJ)^{-1}$ \cite{exp.corr.length} is related to the
scaling variables $X$ and $x$ used in \cite{Evans} and \cite{RuZaShAb10},
respectively, by $x_\text{I}=2X=2x$ (compare, e.g., with the isotropic limit of
the correlation-length results in Appendix~A\,2 of \cite{Indekeu}; see
also Sec.~\ref{G.Ising.aniso} below).
As an unexpected result, we find the surprising identities
\bse
\begin{align}
\label{G.Ising.1}
X^\pbc_\text{Gauss}((L/\xi)^2)&=-X^\abc_\text{Ising}(L/\xi),
\\
\label{G.Ising.2}
X^\abc_\text{Gauss}((L/\xi)^2)&=-X^\pbc_\text{Ising}(L/\xi),
\end{align}
\ese
whose derivation will be presented elsewhere \cite{KaDo10}.

\begin{figure}
\begin{center}
\includegraphics[width=6.6cm,angle=0]{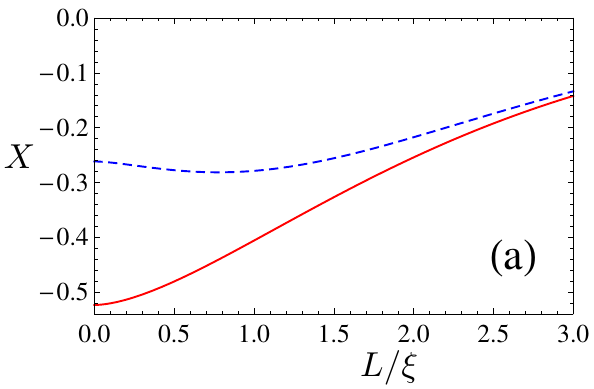}
\includegraphics[width=6.6cm,angle=0]{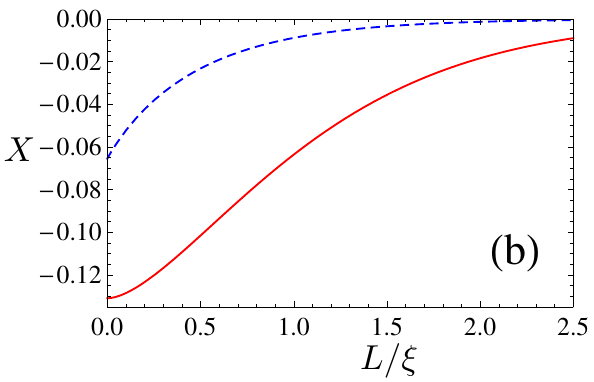}
\end{center}
\caption{\label{G.Gauss_Ising}
(Color online) Casimir force scaling functions (\ref{G.Gauss.Xp}) and
(\ref{G.Ising.Xp}) of the Gaussian and Ising models in $d=2$ dimensions.
(a) Scaling functions
$X^\pbc_\text{Gauss}((L/\xi)^2)=-X^\abc_\text{Ising}(L/\xi)$
(solid line) according to (\ref{G.Ising.1}) and
$X^\pbc_\text{Ising}((L/\xi)^2)=-X^\abc_\text{Gauss}(L/\xi)$
(dashed line) according to (\ref{G.Ising.2}).
(b) Scaling functions $X^\DDbc_\text{Gauss}((L/\xi)^2)$ (solid line) and $X^\freebc_\text{Ising}(L/\xi)$ (dashed line).}
\end{figure}

For a comparison of these results see Fig.~\ref{G.Gauss_Ising}.
In Fig.~\ref{G.Gauss_Ising}(b), we have assumed that (asymptotically
close to criticality) free b.c.\ in the Ising model correspond to
DD b.c.\ in the Gaussian model.
We see some similarity on a qualitative level: both the Gaussian and
the Ising scaling functions are negative for periodic and for DD
(or free) b.c., thus implying an attractive Casimir force whereas for
antiperiodic b.c.\ they are positive implying a repulsive Casimir force.
On a quantitative level, however, the Casimir amplitudes at $\Tc$ differ
significantly, namely by a factor of two according to the exact results
$X^\pbc_\text{Gauss}(0)=2X^\pbc_\text{Ising}(0)=-\pi/6$,
$2X^\abc_\text{Gauss}(0)=X^\abc_\text{Ising}(0)=\pi/6$, and
$X^\DDbc_\text{Gauss}(0)=2X^\freebc_\text{Ising}(0)=-\pi/24$,
obtained from (\ref{G.FSS.Delta}), (\ref{G.Casampl}), and
(\ref{G.Ising.Xp}).
Furthermore we note that, according to the dashed line in
Fig.~\ref{G.Gauss_Ising}(a), $X^\abc_\text{Gauss}$ has a weak maximum
above $\Tc$, and correspondingly $X^\pbc_\text{Ising}$ has a weak minimum
above $\Tc$, in agreement with Fig.~2 of \cite{RuZaShAb10} and Fig.~15
of \cite{VaGaMaDi08}.

These results can be interpreted in terms of the two-dimensional
$\varphi^4$ model which should be in the same universality class as the
two-dimensional Ising model.
In all cases, the scaling functions at $\Tc$ of the Gaussian model differ
by a factor of two from the scaling functions at $\Tc$ of the
two-dimensional $\varphi^4$ model.
This indicates that a low-order perturbation approach in the
{\it two}-dimensional $\varphi^4$ model (in terms of a perturbation
expansion with respect to the four-point coupling) is inappropriate,
in contrast to the situation in {\it three} dimensions to be discussed
in Sec.~\ref{oneloopphi4}.
This is quite plausible since the fixed-point value of the renormalized
four-point coupling in two dimensions is quite large, i.e., far from the
vanishing Gaussian fixed-point value.
This is in line with the known fact that non-Gaussian fluctuations are
generally larger in two than in three dimensions, as seen, e.g., from
the bulk critical exponents.

\subsection{Anisotropic case}
\label{G.Ising.aniso}
We have extended the analysis of the isotropic Ising model by Rudnick
et al.\ \cite{RuZaShAb10} for periodic and antiperiodic b.c.\ to the
anisotropic Ising model on a square lattice with nearest-neighbor
couplings $\Jpl>0$ and $\Jpp>0$.
This corresponds to the ``rectangular lattice'' of Indekeu
et al.\ \cite{Indekeu} with the identifications of the couplings
$K_1=2\be\Jpl$, $K_2=2\be\Jpp$, and $K_3=0$.
The corresponding bulk-correlation-length amplitudes above $\Tc$ follow
from Appendix~A\,2 of \cite{Indekeu} (for $\at=1$) as
\cite{exp.corr.length}
\bse
\begin{align}
\label{xi.perp.0.p}
\xi_{0,\perp}
&=
\frac{1}{4}
\left(\be_cJ_\parallel+\f{\be_cJ_\perp}{\sinh(4\be_cJ_\perp)}\right)^{-1},
\\
\xi_{0,\parallel}
&=
\frac{1}{4}\left(
\be_cJ_\perp+\f{\be_cJ_\parallel}{\sinh(4\be_cJ_\parallel)}
\right)^{-1},
\end{align}
\ese
with the ratio (\ref{G.corr-ratioIsing}), where we have used the condition
$\sinh(4\be_c\Jpl)\sinh(4\be_c\Jpp)=1$ for $d=2$ bulk criticality.
From Sec.~\ref{G.Gauss.cont} we obtain the nonuniversal Casimir force
scaling function of the anisotropic Gaussian model for $d>2$ above $\Tc$
\begin{align}
\label{G.X.aniso.iso}
X_\text{aniso}&(t(L/\xi_{0,\perp})^{1/\nu};J_\parallel,J_\perp)
\nn\\
&=(\xi_{0,\perp}/\xi_{0,\parallel})^{d-1}
X_\text{iso}(t(L/\xi_{0,\perp})^{1/\nu}).
\end{align}
We have verified that this relation holds also for $d=2$ dimensions
(i) for the Gaussian model with periodic, antiperiodic, DD, NN, and
ND b.c.\ and (ii) for the Isingmodel for periodic and antiperiodic b.c..
There is little doubt that it also holds for the two-dimensional Ising
model with corresponding other b.c..
Right at $\Tc$ this was established already in \cite{Indekeu} for the case
of free b.c., as noted in Sec.~\ref{G.section}.C .

\section{\boldmath $\varphi^4$ field theory at $d=3$}
\label{oneloopphi4}
\begin{figure*}
\begin{center}\vspace{-10pt}
\bmp{8cm}
\includegraphics[width=6.6cm,angle=0]{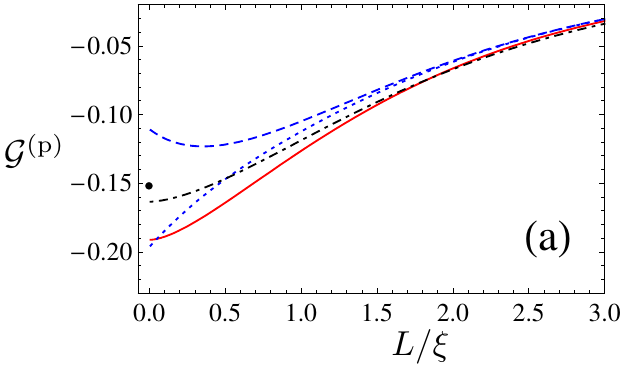}
\emp
\bmp{8cm}
\includegraphics[width=6.6cm,angle=0]{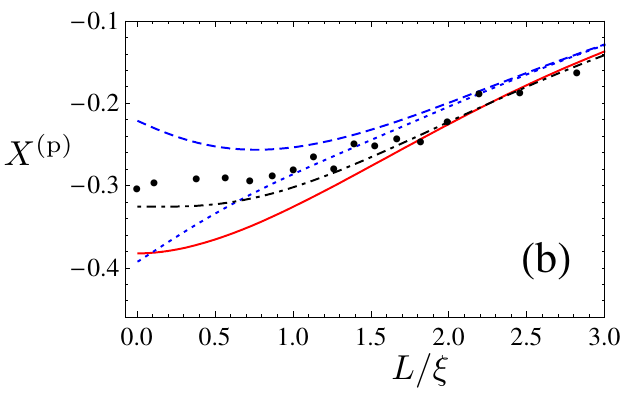}
\emp
\vspace{-5pt}\\
\bmp{8cm}
\includegraphics[width=6.6cm,angle=0]{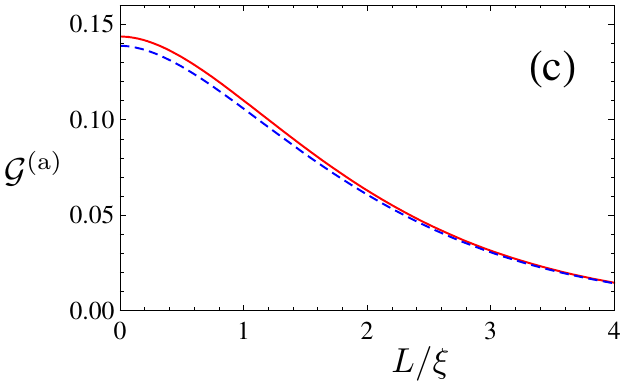}
\emp
\bmp{8cm}
\includegraphics[width=6.6cm,angle=0]{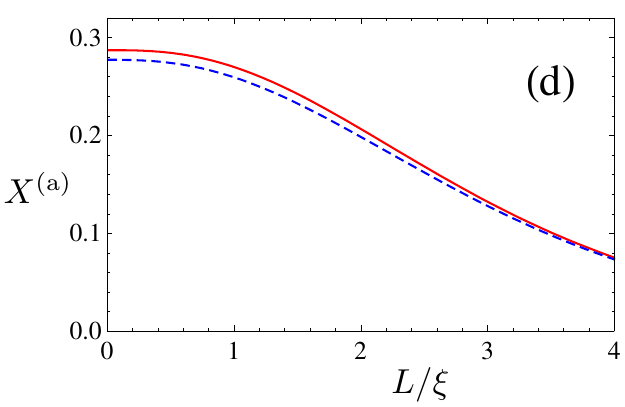}
\emp
\vspace{-5pt}\\
\bmp{8cm}
\includegraphics[width=6.6cm,angle=0]{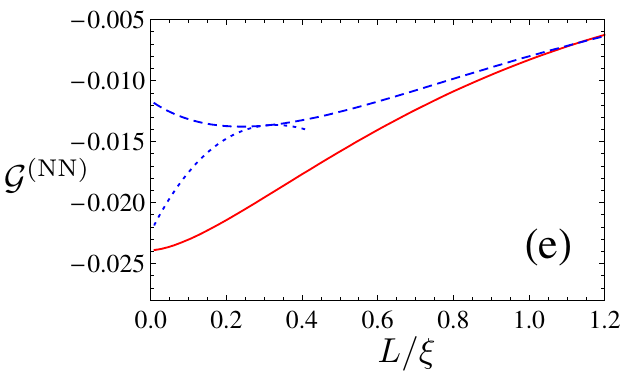}
\emp
\bmp{8cm}
\includegraphics[width=6.6cm,angle=0]{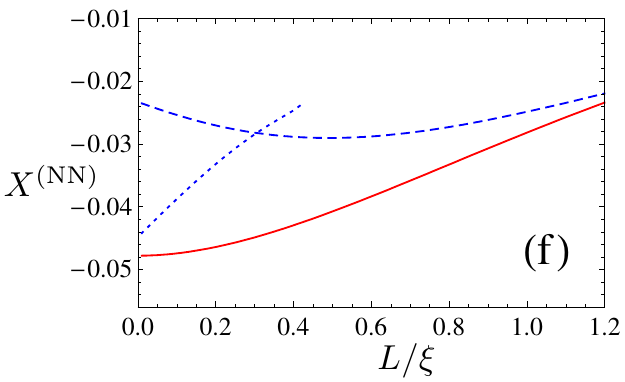}
\emp
\vspace{-5pt}\\
\bmp{8cm}
\includegraphics[width=6.6cm,angle=0]{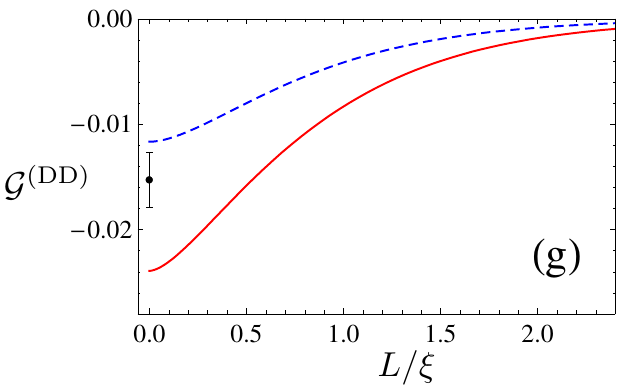}
\emp
\bmp{8cm}
\includegraphics[width=6.6cm,angle=0]{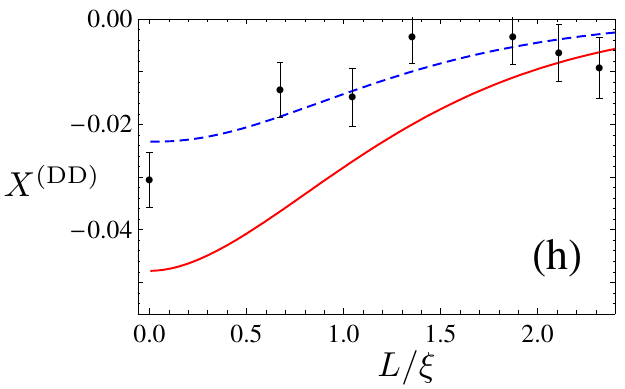}
\emp
\vspace{-5pt}\\
\bmp{8cm}
\includegraphics[width=6.6cm,angle=0]{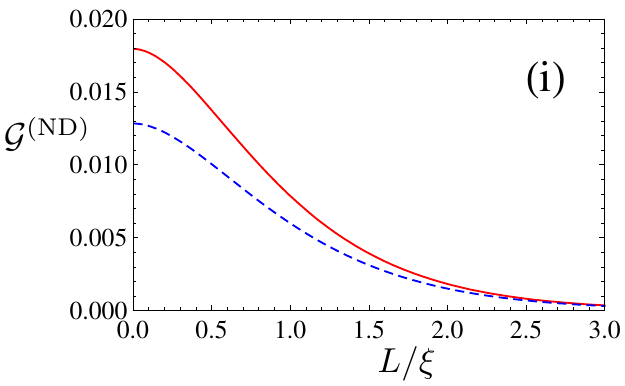}
\emp
\bmp{8cm}
\includegraphics[width=6.6cm,angle=0]{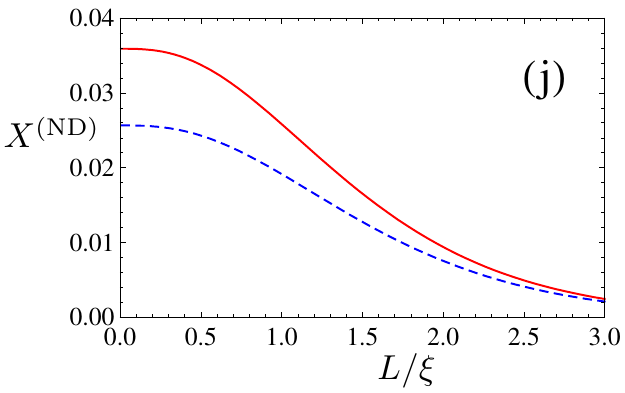}
\emp
\end{center}
\vspace{-17pt}
\caption{\label{G.GcalX}\footnotesize
(Color online) Scaling functions $\Gcal$ and $X$ of the finite-size part
of the free energy density and of the Casimir force, respectively, for
$T\geq\Tc$ in a three-dimensional film of thickness $L$ with isotropic
interactions for the various b.c.\ as a function of the scaling argument
$L/\xi$.
Solid lines:
one-loop results of the fixed $d=3$ $\vp^4$ theory according to
Eqs.~(\ref{G.Gcal.pa.3}), (\ref{G.Gcal.NN.DD.3}), (\ref{G.Gcal.ND.3}),
(\ref{G.fexscal.pa}), and (\ref{G.fexscal.NNDDND}) for $\Gcal((L/\xi)^2)$
and according to Eqs.~(\ref{G.X.other}), (\ref{G.X.pa.3}), and (\ref{G.X})
for $X((L/\xi)^2)$.
Dashed lines in left panels:
two-loop $\ve$ expansion results $\Th_{+\text{per}}(L/\xi)$,
$\Th_{+\text{aper}}(L/\xi)$, $\Th_{+\text{SB,SB}}(L/\xi)$,
$\Th_{+\text{O,O}}(L/\xi)$, and $\Th_{+\text{O,SB}}(L/\xi)$ at
$\ve=1$ for $n=1$ for periodic, antiperiodic, NN, DD, and ND b.c.,
respectively, according to Eqs.~(6.12) and (6.13) in \cite{KrDi92a}.
Dashed lines in right panels:
two-loop $\ve$ expansion results
$\vt_{+\text{per}}(L/\xi)$, $\vt_{+\text{aper}}(L/\xi)$,
$\vt_{+\text{SB,SB}}(L/\xi)$, $\vt_{+\text{O,O}}(L/\xi)$,
and $\vt_{+\text{O,SB}}(L/\xi)$ at $\ve=1$ for $n=1$, obtained
through Eq.~(3.9) in \cite{KrDi92b}.
Dotted line in (a):
improved $\ve$ expansion result $\Th^{(\text{per})}(L/\xi)$ at
$\ve=1$ for $n=1$ according to Eq.~(4.57) in \cite{GrDi07}, compare
Fig.~6 of \cite{GrDi07}.
Dot-dashed line in (a):
$F^{\text{ex}}((L/\xi)^{1/\nu},\rho=1/4)$ in fixed $d=3$ according to
Eq.~(17) in \cite{Do08b}.
Data point in (a): From MC result $\vt_\text{P}(0)=-0.3040(4)$
in \cite{VaGaMaDi08}.
Dotted line in (b):
improved $\ve$ expansion results $\Xi^{(\text{per})}(L/\xi)$ for $n=1$,
obtained through
Eq.~(1.7) in \cite{GrDi07}.
Dot-dashed line in (b):
$X((L/\xi)^{1/\nu},\rho=1/4)$ in fixed $d=3$ according to Eq.~(19)
in \cite{Do08b}.
The $L/\xi=0$ data point in (b) is twice the $L/\xi=0$ data point
displayed in (a).
Other data points in (b): MC results from Fig.~15 in
\cite{VaGaMaDi08} for $L=20$ and $\rho=1/6$.
Dotted line in (e):
improved $\ve$ expansion result $\Th^{(\text{sp,sp})}(L/\xi)$ at
$\ve=1$ for $n=1$ according to Eqs.~(4.47), (4.50), (4.53), and (4.56)
in \cite{GrDi07}, compare Fig.~8 of \cite{GrDi07}
(where the variable on the abscissa should read
$\sqrt{L/\xi_\infty}=\sqrt{\sf L}$).
Dotted line in (f):
improved $\ve$ expansion results $\Xi^{(\text{sp,sp})}(L/\xi)$ at
$\ve=1$ for $n=1$, obtained through Eq.~(1.7) in \cite{GrDi07}.
Data point in (g): MC result for $L=20$ and $\rho=1/6$ by dividing the
$L/\xi=0$--result displayed in (h) by $2$.
Data points in (h): MC results for $L=20$ and $\rho=1/6$ from the inset
of Fig.~13 in \cite{VaGaMaDi08}
(where the $\ve$ expansion line is misrepresented).
}
\end{figure*}

In this section we present the first (one-loop) step within the
framework of the minimally renormalized $\varphi^4$ field theory at fixed
dimensions $2<d<4$ \cite{dohm,Do08a} for the calculation of the Casimir
force scaling function in film geometry in the regime $T \geq \Tc$ for all
five b.c.\ defined in Sec.~\ref{G.section}.
As noted in Sec.~\ref{G.Gauss.cont}, our Gaussian results for the various
scaling functions $\Gcal(\xt)$ and $X(\xt)$ can be incorporated in such
a theory based on the isotropic $\vp^4$ Hamiltonian (\ref{G.H.field}).
We emphasize, however, that we do not set $u_0$ equal to zero from the
outset and that our one-loop treatment goes beyond the simple
Gaussian model in that it includes the effect of the renormalized
four-point coupling $u$ via the exact {\it exponent} function $\zeta_r(u)$
(see Eq.~(\ref{G.rvonl}) below), whose fixed-point value determines the
exact (non-Gaussian) critical exponent $\nu$.
The one-loop approximation manifests itself only in neglecting the
(two-loop) $O(u^*)$ contribution to the {\it amplitude} function of the
free energy density.
Such a treatment has recently been presented in Sec.~X\,A of \cite{Do08a}
for the case of cubic geometry with periodic b.c..
For the specific heat in film geometry with Dirichlet b.c.\ a
corresponding treatment was given in \cite{sutter}.
As suggested by the earlier successes \cite{sutter,Do08a,Do08b}, the
minimally renormalized $\varphi^4$ theory at fixed $d$ is expected to
constitute an important alternative in the determination of the Casimir
force scaling function in comparison to the earlier $\ve$
expansion approach \cite{KrDi92a,KrDi92b,GrDi07}.
Our quantitative results to be presented in Fig.~\ref{G.GcalX} below will
support this expectation.
Other fixed-$d$ renormalization schemes are, of course, conceivable which
would lead to the same one-loop results at $d=3$ as obtained in our
approach.
We believe, however, that the fixed-$d$ minimal subtraction scheme has
considerable advantages in extending the finite-size theory to two-loop
order and to the temperature regime below $\Tc$ \cite{Do08a,Do08b}.

As a temperature variable we use the shifted parameter
$r_0-r_\text{0c}=a_0t$, where
$r_\text{0c}=-4(n+2)u_0\int_{\bm{k}}^{(d)}\bm{k}^{-2}$ is the critical
value of $r_0$ up to $O(u_0)$.
In the following we sketch the relevant steps of calculating the singular
part of the minimally renormalized free energy density in one-loop order
for film geometry with periodic or antiperiodic b.c..
After subtracting the regular bulk part up to linear order in $r_0-r_{0c}$
and performing the limit $\La\to\infty$ at fixed $r_0-r_{0c}$ we
obtain the bare one-loop expression of the remaining part $\delta f$ of
the free energy density per component divided by $k_B T$ in $2<d<4$
dimensions as
\begin{align}
\label{G.delta-f-bare}
\delta f(r_0-r_{0c},&u_0,L)
=-\frac{A_d}{d\ve}\;(r_0-r_{0c})^{d/2}
\nn\\
&
+L^{-d}\Gcal((r_0-r_{0c})L^2)+O(u_0)
\end{align}
for periodic and antiperiodic b.c.
where $\Gcal^\pbc(y)$ and $\Gcal^\abc(y)$ are given by
(\ref{G.Gcal.p}) and (\ref{G.Gcal.a}).
The renormalized parameters $r$ and $u$ are defined in the standard way
\cite{Do08a} as $r=Z_r^{-1}(r_0-r_{0c})$ and
$u=\mu^{-\ve}A_dZ_{u}^{-1}Z_{\varphi}^2u_0$
with an inverse reference length $\mu=\xi_0^{-1}$, where $\xi_0$ is the
correlation-length amplitude above $\Tc$.
The additively renormalized counterpart $f_R$ of $\delta f$ is defined
as \cite{Do08a}
\begin{align}
\label{G.renorm}
f_R(r, u, L,\mu)
&=
\delta f(Z_rr,\mu^{\ve}Z_{u}Z_{\varphi}^{-2}A_d^{-1}u,L)
\nn\\
&\ph{=}
-\frac{1}{8}\mu^{-\ve}n^{-1}r^2A_dA(u,\ve),
\end{align}
where $A(u,\ve)=-2n/\ve+O(u)$ is the additive
renormalization constant of the minimal renormalization scheme.
After integration of the renomalization-group equation (see Eqs.~(5.6)
and (5.7) of \cite{Do08a}) and with the choice $\mu^2 l^2=r(l)$ of the
flow parameter $l$, the finite-size part of $f_R$ is then given by
\begin{align}
\label{G.frenorm}f_R(r,u,L,\mu)-f_R(r,u,\infty,\mu)
&=
L^{-d}\Gcal(r(l)L^2)+O(u(l)),
\end{align}
with the effective temperature
variable \cite{dohm}
\begin{align}
\label{G.rvonl}
r(l)
&=
r\exp\left(\int_{1}^{l}\zeta_r(u(l'))\frac{dl'}{l'}\right).
\end{align}
This variable contains the field-theoretic function
$\zeta_r(u)=\mu(\partial_{\mu}\ln Z_r^{-1})_0$ \cite{dohm} with the fixed
point value $\zeta_r(u^*)=2-\nu^{-1}$.
Asymptotically ($l\to0$), this leads to the scaling form of the singular
part of the excess free energy per component in one-loop order
\begin{align}
\label{G.fexscal.pa}\fexs(t,L)&=L^{-d}\Gcal(L^2/\xi^2)+O(u^*)
\end{align}
for periodic and antiperiodic b.c., with $\Gcal^\pbc(y)$ and
$\Gcal^\abc(y)$ given by (\ref{G.Gcal.p}) and (\ref{G.Gcal.a}) where now
$\xi= \xi_0 t^{-\nu}$ is the correlation length above $\Tc$ with the exact
critical exponent $\nu$ of the $(d,n)$ universality class.

A complication appears to arise at $d=3$ for NN and DD
b.c.\ because in these cases the limit $\La\to\infty$ at fixed
$r_0-r_{0c}$ does not exist which in the dimensionally regularized
form of the free energy density shows up as pole terms at $d=3$.
These divergent contributions, however, are restricted only to the surface
part $L^{-1}\fsf$ which does not contribute to the Casimir force.
Thus $\fex(t,L)-L^{-1}\fsf(t)$ is well behaved at $d=3$ in the limit
$\La\to\infty$ at fixed $r_0-r_{0c}$.
Consequently there exists no problem in the calculation of the finite-size
part scaling function $\Gcal$ and of the Casimir force scaling function
$X$ in the framework of the minimally renormalized theory at fixed $d=3$
for NN and DD b.c..
This holds also for ND b.c..
The resulting asymptotic scaling form in one-loop order is
\begin{align}
\label{G.fexscal.NNDDND}\fexs(t,L)-L^{-1}\fsfs(t)
&= L^{-d}\Gcal(L^2/\xi^2)+O(u^*),
\end{align}
for NN, DD, and ND b.c., where $\Gcal^\NNbc(y)$, $\Gcal^\DDbc(y)$, and
$\Gcal^\NDbc(y)$ are given by (\ref{G.Gcal.NN.DD}) with (\ref{G.Gcal.p}),
and by (\ref{G.Gcal.ND}) with (\ref{G.Gcal.a}), respectively.
Again, the correlation length $\xi$ in (\ref{G.fexscal.NNDDND}) contains
the exact critical exponent $\nu$ of the $(d,n)$ universality class.
The results (\ref{G.fexscal.pa}) and (\ref{G.fexscal.NNDDND}) can be
applied directly to $d=3$ dimensions.
The Casimir force scaling functions
$X(L^2/\xi^2)$ follow from (\ref{G.X}).
Thus we are in the position to perform a reasonable comparison both with
MC data for the three-dimensional Ising model \cite{VaGaMaDi08} and with
earlier $\ve=4-d$ expansion results \cite{KrDi92a,GrDi07} of the $\vp^4$
theory evaluated at $\ve=1$ as well as with the very recent result of an
improved $d=3$ perturbation theory \cite{Do08b} (in a ${\Lpl^2\times L}$
slab geometry with a finite aspect ratio $\rho=L/\Lpl=1/4$) for periodic
b.c..
This comparison is one of the central results of this paper.

The comparison is shown in Figs.~\ref{G.GcalX}(a)--(j) as a function of
the variable $L/\xi$.
The solid lines represent our one-loop results.
The $\ve$ expansion results are represented by the dashed lines (two-loop
$\ve$ expansion \cite{KrDi92a}) and by the dotted lines (improved $\ve$
expansion \cite{GrDi07}), respectively, and the result of the improved
$d=3$ perturbation theory \cite{Do08b} is represented by dot-dashed lines
in Figs.~\ref{G.GcalX}(a),(b).
In the large--$L/\xi$ regime (not shown in Fig.~\ref{G.GcalX}), the solid
and dashed lines have an exponential approach to zero and differ very
little from each other in all cases.
This statement holds also for the dotted and dot-dashed lines for the case
of periodic b.c.\ (Figs.~\ref{G.GcalX} (a),(b)) but not for the dotted
lines for the case of NN b.c.\ (Figs.~\ref{G.GcalX} (e),(f)), where the
$\ve$ expansion result of \cite{GrDi07} breaks down in the large--$L/\xi$
regime.
Also shown are recent MC data \cite{VaGaMaDi08} for the
three-dimensional Ising model (in a ${\Lpl^2\times L}$ slab geometry with
the aspect ratio $\rho=L/\Lpl=1/6$ and with $L=20$) for the cases of
periodic and DD b.c.\ in Figs.~\ref{G.GcalX} (a),(b),(g),(h),
respectively.

For periodic b.c., our one-loop result for $\Gcal^\pbc(L^2/\xi^2)$ and
$X^\pbc(L^2/\xi^2)$ (solid lines in Figs.~\ref{G.GcalX}(a) and (b)) is in
remarkable agreement with the improved $\ve$ expansion result of
\cite{GrDi07} (dotted lines in Figs.~\ref{G.GcalX}(a) and (b)) at $\Tc$
(i.e., $L/\xi=0$) and in the large--$L/\xi$ regime.
The slope of our one-loop result at $\Tc$ is in better agreement with the
slope of the MC data than the slopes of the $\ve$
expansion results at $\Tc$.
In particular, there is no artifact of the one-loop result of the fixed
$d=3$ theory such as the minimum of the two-loop $\ve$ expansion result
above $\Tc$ shown by the dashed line in Figs.~\ref{G.GcalX}(a) and (b).
This suggests that, for periodic b.c., the $d=3$ approach is a better
starting point of perturbation theory than the $\ve$ expansion around
$d=4$ dimensions.
This is consistent with recent findings of finite-size effects in cubic
geometry (Fig.~5 in \cite{Do08a}) and in finite-slab geometry
(Fig.~4 in \cite{Do08b}).
For antiperiodic b.c.\ (see Figs.~\ref{G.GcalX}(c) and (d)), there is
surprisingly good agreement between our one-loop $d=3$ result for
$\Gcal^\abc(L^2/\xi^2)$ and $X^\abc(L^2/\xi^2)$ and the two-loop $\ve$
expansion result \cite{KrDi92a}.
This is quite remarkable in view of the fact that the
computational effort in obtaining $d=3$ one-loop RG results is
considerably smaller than that for deriving two-loop $\ve$ expansion
results.

For NN, DD, and ND b.c., there are considerable differences between our
one-loop results of the fixed $d=3$ $\vp^4$ theory and the two-loop
$\ve$ expansion results as shown in Figs.~\ref{G.GcalX}(e)--(j).
For NN b.c., however, the improved $\ve$ expansion result of \cite{GrDi07}
at $\Tc$ is not far from our one-loop result, but for $L/\xi >0$ our $d=3$
result does not agree with the strong increase of the $\ve$ expansion
result of \cite{GrDi07}.

In summary, the fixed $d=3$ theory yields reasonable results already in
one-loop order.
It would be a rewarding task to perform a two-loop calculation of the
fixed $d$ theory for all b.c.\ as well as to proceed to higher-orders
of the $\ve$ expansion for DD and ND b.c..
Also MC data for the cases of antiperiodic, NN, and ND b.c.\ are highly
desirable for a comparison with the predictions shown in
Fig.~\ref{G.GcalX} in order to clarify the reliability of the different
perturbative approaches.

\section{Dimensional crossover: Specific heat}
\label{G.specific.heat}

In the following, we present an explicit study of the dimensional
crossover in the Gaussian model from the finite-size critical behavior
near the $d$-dimensional bulk transition at $\Tc$ to the
$(d{-}1)$-dimensional critical behavior near the film transition at the
critical temperature $\Tcfilm(L)\leq \Tc$ of the film of finite
thickness $L$.
(The equality sign holds only for periodic and NN b.c., whereas
$\Tcfilm(L)<\Tc$ for antiperiodic, DD, and ND b.c..)
The most interesting candidate for this study is the divergent
specific heat $C(t,L)$, whose $d$-dependent critical exponent $\al$
changes from $\al_\bulk=(4-d)/2$ near bulk $\Tc$ (see Eq.~(\ref{G.alpha}))
to $\al_\film=[4-(d-1)]/2$ near $\Tcfilm(L)$ in $ d<4$ dimensions.
We shall verify that the film
critical behavior of a $d$-dimensional system corresponds to that of a
bulk system in $d-1$ dimensions for all b.c.\ except for antiperiodic
b.c., where an unexpected factor of two appears due to a two-fold
degeneracy of the lowest mode as noted already in Sec.~\ref{G.Hamiltonian}
above.
It remains to be seen which effect this feature may have in non-Gaussian
models and in the $\varphi^4$ theory.

The dimensional crossover behavior is particularly simple in the Gaussian
model because the correlation-length exponent $\nu=1/2$ is independent of
the dimension $d$ and the correlation-length amplitude is the same both
for the bulk and the film critical point.
To the best of our knowledge, this crossover behavior has not been
presented in the literature so far.
Nevertheless it is worthwhile to study the exact Gaussian crossover
behavior for various b.c.\ as it provides part of the mathematical basis
also for the crossover behavior in the more complicated mean spherical
model in film geometry in $3<d<4$ dimensions that we shall study in a
separate paper \cite{KaDo10}.

According to (\ref{G.ftbc}) and (\ref{G.spec-heat.u}), the expression for
the specific heat reads
\begin{align}
\label{G.spec-heat}
C(t,L)
&=
\frac{T^2a^2_0}{2\Tc^2L}\sum_q\int_{\pb}^{(d-1)}
\f{1}{(\rz+J_{\pb,d-1}+J_q)^2}
\nn\\
&\ph{=}
-\frac{Ta_0}{\Tc L}\sum_q\int_{\pb}^{(d-1)}\f{1}{\rz+J_{\pb,d-1}+J_q}.
\end{align}
For small $t$ and large $L/\at$, the specific heat can be decomposed
into singular and nonsingular parts as
\begin{align}
\label{G.C.s.C.ns}
C(t,L)=C_\text{s}(t,L)+C_\text{ns}(t,L).
\end{align}
The first term on the right hand side of (\ref{G.spec-heat}) provides the
leading singular contribution to $C_\text{s}$, whereas the second term
yields only subleading corrections.
The non-scaling structures of the type discussed in the preceding sections
(for NN, DD, and ND b.c.\ in $d=3$ dimensions) appear only in the
subleading corrections, whereas the leading part of the first term of
(\ref{G.spec-heat}) is in full agreement with the finite-size scaling form
(for the subclass of isotropic systems)
\begin{align}
\label{G.C-scal}
C_\text{s}(t,L)=
\xiz^{-2/\nu}L^{\al/\nu}\Ccal(\xt),
\rule[-5pt]{0pt}{0pt}
\end{align}
as noted already in \cite{ChDo03} for the case of free (DD) b.c..
For the Gaussian model, the scaling structure (\ref{G.C-scal}),
together with the critical exponents (\ref{G.xi0.nu}) and (\ref{G.alpha}),
holds in $1<d<4$ dimensions for all boundary conditions.
If the finite-size scaling function $\Fcal(\xt)$ of the free energy
density, (\ref{G.fscal}), exists, the universal scaling function
$\Ccal(\xt)$ is related to it by
\begin{align}
\label{G.C-F}
\Ccal(\xt)=-\f{d^2\Fcal(\xt)}{d\xt^2}.
\end{align}
For the simplest anisotropic case, discussed at the end of
Sec.~\ref{G.Gauss.cont}, the corresponding nonuniversal result can be
derived from (\ref{G.fscal-anisospec}).
$\Ccal(\xt)$ exists also in $d=3$ dimensions for the cases of NN, DD, and
ND b.c., where $\Fcal(\xt)$ does not exist for all $\xt$.

From (\ref{G.fbs}) follows the bulk singular part
\begin{align}
\label{G.bulk.Cs}
C_\text{b,s}(t)
&=
Y_{C,d}\xiz^{-d}t^{-\al}
=Y_{C,d}\xiz^{-2/\nu}\xi^{\al/\nu},
\end{align}
with the universal bulk amplitude
\begin{align}
\label{G.bulk.YCd}
Y_{C,d}
=-\frac{d(d-2)}{4}Y_d
=\frac{\Ga({\ts\f{4-d}{2}})}{2(4\pi)^{d/2}},
\end{align}
which is valid for $d>0$, $d\neq4,6,8,\ldots$ dimensions, even though
(\ref{G.fbs}) does not hold for $d=2$.
This implies in two and three dimensions
\bse
\label{G.Cs.d.23}
\begin{numcases}{C_\text{b,s}(t)=}
\label{G.Cs.d.2}
\f{1}{8\pi}\xiz^{-2}t^{-1}
=\f{1}{8\pi}\xiz^{-4}\xi^2,
& $d=2$,~~~~~~~~
\\
\label{G.Cs.d.3}
\f{1}{16\pi}\xiz^{-3} t^{-1/2}
=\f{1}{16\pi}\xiz^{-4}\xi,
& $d=3$.~~~~~~~~
\end{numcases}
\ese
The finite-size scaling functions read
\bwt
\vspace{-20pt}
\bse
\label{G.Ccalneu}
\begin{align}
\label{G.Ccal.p}
\Ccal^\pbc(\xt)
&=
-\Kc{d}^\pbc(\xt)+Y_{C,d}\xt^{(d-4)/2},
& \xt&>0,
\\
\label{G.Ccal.a}
\Ccal^\abc(\xt)
&=
-\Kc{d}^\abc(\xt+\pi^2)+Y_{C,d}(\xt+\pi^2)^{(d-4)/2},
& \xt&>-\pi^2,
\\
\label{G.Ccal.NN}
\Ccal^\NNbc(\xt)
&=
-\Kc{d}^\NNbc(\xt)+Y_{C,d}\xt^{(d-4)/2}+2\ACsf^\Nbc\xt^{(d-5)/2},
& \xt&>0,
\\
\label{G.Ccal.DD}
\Ccal^\DDbc(\xt)
&=
-\Kc{d}^\DDbc(\xt+\pi^2)+Y_{C,d}(\xt+\pi^2)^{(d-4)/2}
+2\ACsf^\Dbc(\xt+\pi^2)^{(d-5)/2},
& \xt&>-\pi^2,
\\
\label{G.Ccal.ND}
\Ccal^\NDbc(\xt)
&=
-\Kc{d}^\NDbc(\xt+(\pi/2)^2)+Y_{C,d}[\xt+(\pi/2)^2]^{(d-4)/2},
& \xt&>-(\pi/2)^2,
\end{align}
\ese
\ewt
with
\begin{align}
\label{G.ACsfND}
\ACsf^\Nbc
&=
-\ACsf^\Dbc
=-\f{(d-1)(d-3)}{4}\Asf^\Nbc
=\f{\Ga(\f{5-d}{2})}{8(4\pi)^{(d-1)/2}},
\end{align}
as follows from (\ref{G.C-F}),(\ref{G.Fcal.pa}),
(\ref{G.Fcal.NN.DD}), and (\ref{G.Fcal.ND}).
The functions $\Kc{d}(\yv)$ are listed in Appendix~\ref{G.Ap.functions}.
They are given by the second derivatives of the functions $\Ic{d}$
defined in Sec.~\ref{G.free.energy}, i.e., $\Kc{d}(\yv)=\Ic{d}''(\yv)$.
Eqs.~(\ref{G.Ccalneu}) are valid in $1<d<4$ dimensions.
Eq.~(\ref{G.ACsfND}) agrees with Eqs.~(125) and (126) of \cite{ChDo03}
for the case of DD b.c.\ (apart from a factor two due to a different
definition of $\ACsf$).

The scaling functions (\ref{G.Ccalneu}) are valid not only near $\xt=0$,
but also near the film transition at $\Tcfilm(L)<\Tc$ for antiperiodic,
DD, and ND b.c., i.e., near $\xt=-\pi^2$ or $\xt=-\pi^2/4$, respectively.
This means that they provide an exact description of the crossover from
the \mbox{$d$-dimensional} to the $(d{-}1)$-dimensional critical behavior
of the specific heat.
The result (\ref{G.Ccal.DD}) for $\Ccal^\DDbc(\xt)$ agrees with the
result for $\Ccal(\yv,0)$ of Eqs.~(124)--(126) in \cite{ChDo03} with
the identification $\xt=\yv^2$.

Similar to (\ref{G.ffilm}), at finite $L$, we define the film specific
heat $C_\film$ (heat capacity per unit area divided by $\kB$) as
\begin{align}
\label{G.Cfilm}
C_\film(\rz,L) = L C(t,L).
\end{align}
One expects that, asymptotically ($\xif\gg L$), the film
critical behavior of a $d$-dimensional system corresponds to that of a
bulk system in $d-1$ dimensions.
We indeed obtain from (\ref{G.Ccalneu}) and (\ref{G.C-scal}) for small
\mbox{$[\rz-\rzcf(L)]/J\ll L^{-2}$} the singular part of the film specific
heat for finite $L$ for all b.c.\
\bse
\label{G.C-film}
\begin{flalign*}
C_\text{film,s}(t_\film,L)
&=&
\end{flalign*}
\vspace{-18pt}
\begin{numcases}{\hspace{0pt}}
\label{G.C-film-a}
\f{Y_{C,d-1}}{\xiz^{d-1}}t_\film^{(d-5)/2}
=\f{Y_{C,d-1}}{\xiz^4}\xi_\film^{5-d},
& $\begin{smallmatrix}\text{\normalsize periodic, NN,}\\
\text{\normalsize DD, ND b.c.,}\end{smallmatrix}$
\nn\\
\\
\label{G.C-film-b}
\f{2Y_{C,d-1}}{\xiz^{d-1}}t_\film^{(d-5)/2}
=\f{2Y_{C,d-1}}{\xiz^4}\xi_\film^{5-d},
& \text{antiperiodic b.c.,}
\nn\\
\end{numcases}
\ese
with $t_\film\equiv t-\tcfilm(L)$, in agreement with the bulk critical
behavior in $d{-}1$ dimensions (compare (\ref{G.bulk.Cs}) and
(\ref{G.Cs.d.23}), as expected on the basis of universality.
An exception is the additional factor of $2$ for antiperiodic b.c.\ which
is a consequence of the two-fold degeneracy of the lowest mode, as
already noted in Sec.~\ref{G.film.crit} (see Eq.~(\ref{G.ffs.a.2-4})).

\begin{figure}[t]
\begin{center}
\includegraphics[width=7cm,angle=0]{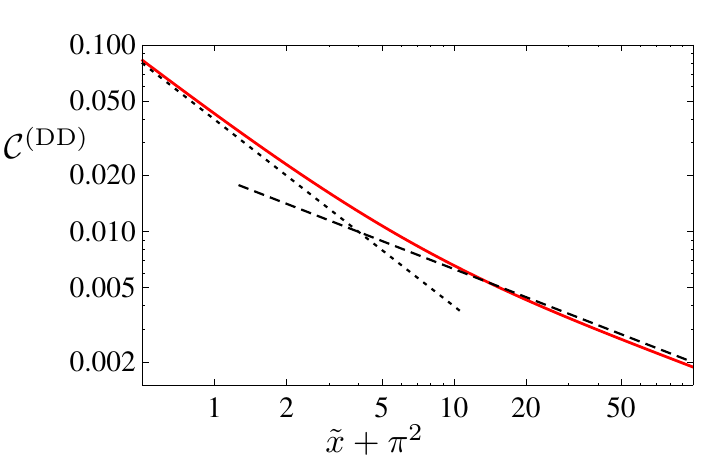}\vspace{-10pt}
\end{center}
\caption{\label{G.specificheat.3}
(Color online) Solid line: Double-logarithmic plot of the specific-heat
scaling function $\Ccal^\DDbc(\xt)$, (\ref{G.Ccal.3.DD}), of the Gaussian
model with DD b.c.\ in three dimensions as a function of $\xt+\pi^2$,
with $\xt=t(L/\xiz)^2$, $t=(T-\Tc)/\Tc$.
Dotted: asymptotic behavior at small $\xt+\pi^2>0$ according to
(\ref{G.Ccal.3.DD.asymp.a}) and (\ref{G.C(t,L).3.DD.asymp.a}),
displaying the divergent critical behavior near
\mbox{$\Tcfilm(L)<\Tc$} with an exponent $\al=1$.
Dashed: asymptotic behavior at large $\xt$ according to
(\ref{G.Ccal.3.DD.asymp.c}) and (\ref{G.C(t,L).3.DD.asymp.c}),
displaying the three-dimensional bulk critical behavior above bulk
$\Tc$ with an exponent $\al=1/2$.
Near $\xt=0$ corresponding to bulk $\Tc$, $\Ccal^\DDbc(\xt)$ is an
analytic function with the finite amplitude $\Ccal^\DDbc(0)=1/(48\pi)$.}
\end{figure}

Since the crossover behavior is qualitatively similar in all dimensions
$1<d<4$ we confine ourselves to illustrating only the example of
DD b.c.\ in $d=3$ dimensions which is obtained from (\ref{G.Ccal.DD}) as
\begin{align}
\label{G.Ccal.3.DD}
\Ccal^\DDbc(\xt)
&=
\f{1}{16\pi}\left(\f{\coth\sqrt{\xt}}{\sqrt{\xt}}-\f{1}{\xt}\right),
& \xt&>-\pi^2.
\end{align}
(Eq.~(\ref{G.Ccal.3.DD}) follows also from (\ref{G.spec-heat})
together with (\ref{G.f.s.DD.new.3.terms}) and (\ref{G.Gcal.NN.DD.3}) or
(\ref{G.Gcal.DD}).)
Its asymptotic behavior is
\pagebreak
\bse
\label{G.Ccal.3.DD.asymp}
\begin{flalign*}
\Ccal^\DDbc(\xt)
&=&
\end{flalign*}
\vspace{-18pt}
\begin{numcases}{\hspace{-5pt}}
\label{G.Ccal.3.DD.asymp.a}
\f{1}{8\pi(\xt+\pi^2)}+O((\xt+\pi^2)^0),
&
$0<\xt+\pi^2\ll1$,
\nn\\\\
\label{G.Ccal.3.DD.asymp.b}
\f{1}{48\pi}-\f{\xt}{720\pi}+O(\xt^2),
&
$|\xt|\ll1$,
\\
\label{G.Ccal.3.DD.asymp.c}
\f{1}{16\pi\sqrt{\xt}}-\f{1}{16\pi\xt}
+O(e^{-\sqrt{\xt}}/\sqrt{\xt}),
&
$\xt\gg1$,
\end{numcases}
\ese
which implies the corresponding asymptotic behavior
\bse
\label{G.C(t,L).3.DD.asymp}
\begin{flalign*}
C_\text{s}^\DDbc(t,L)
&=&
\end{flalign*}
\vspace{-18pt}
\begin{numcases}{\hspace{-15pt}}
\label{G.C(t,L).3.DD.asymp.a}
\f{1}{8\pi}\xiz^{-2}L^{-1}t_\film^{-1}, & $0<L/\xi_\film\ll1$,
\\
\label{G.C(t,L).3.DD.asymp.b}
\f{1}{48\pi}\xiz^{-4}L, & $T=\Tc$, $L/\at\gg1$,
\\
\label{G.C(t,L).3.DD.asymp.c}
\f{1}{16\pi}\xiz^{-3}t^{-1/2}, & $L/\xi\gg1$.
\end{numcases}
\ese
This indeed represents a two-dimensional critical behavior near
$\Tcfilm$ with the exponent $\al=1$ (compare (\ref{G.Cs.d.2})),
a three-dimensional finite-size critical behavior at $\Tc$ with
$\nu=1/2$, $\al=1/2$ (compare Eq.~(\ref{G.C-scal})), and a
three-dimensional bulk critical behavior above $\Tc$ with the
exponent $\al=1/2$ (compare Eq.~(\ref{G.Cs.d.3})), respectively.
The crossover is illustrated in Fig.~\ref{G.specificheat.3}.
Similar illustrations can be given for the other b.c..
\vspace{-10pt}

\section*{Acknowledgments}
\vspace{-5pt}
We are grateful to S.~Dietrich and A.~Gambassi for providing the
MC data of Ref.~\cite{VaGaMaDi08} in numerical form.
We also acknowledge financial support by DLR (German Aerospace Center)
under grant number 50WM0443.

\appendix

\section{Film Thickness}
\label{G.thickness}
Here we derive the expression (\ref{G.Lbar}) for the thickness $\bar{L}$
of the transformed isotropic film.
This thickness is given by $L$ times the height of the $d$-dimensional
parallelepiped spanned by the transforms
$\bm{\hat{x}}_i'\equiv\bm{\la}^{-1/2}\bm{U}\bm{\hat{x}}_i$
of the orthogonal unit vectors $\bm{\hat{x}}_i$, $i=1,\ldots,d$ over
the surface given by the $(d{-}1)$-dimensional parallelepiped spanned by
$\bm{\hat{x}}_i'$, $i=1,\ldots,d{-}1$.
Since the volume of a parallelepiped is given by the product of
one of its surface areas and the corresponding height, we may write
\begin{align}
\label{G.LVdVdmo}
\bar{L}=(V_d/V_{d-1})L,
\end{align}
where the volume of the $d$-dimensional parallelepiped has been denoted
by $V_d$ and the surface area is given by the volume $V_{d-1}$ of the
$(d{-}1)$-dimensional parallelepiped.
The $m$-dimensional volume $V_m$ of a
parallelepiped spanned by the $m$ vectors $\vec{v}_1$,\ldots,$\vec{v}_m$
in an $n$-dimensional space with $n\geq m$ is given by
$V_m=\left(\det\bm{V}^T\bm{V}\right)^{1/2}$,
where $\bm{V}$ is the $n\times m$ matrix whose columns are the vectors
$\vec{v}_1$,\ldots,$\vec{v}_m$.
Thus we obtain
\begin{align}
\label{G.Vd}
V_d
&=
\left[\det(\bm{\la}^{-1/2}\bm{U})^T(\bm{\la}^{-1/2}\bm{U})\right]^{1/2}
=\left(\det{\bm{\la}}^{-1}\right)^{1/2}
\nn\\
&=
\prod_{i=1}^d\la_i^{-1/2}
=\left(\det{\bm{A}}^{-1}\right)^{1/2},
\end{align}
where ${\bm{A}}^{-1}$ may also be written as
\begin{align}
{\bm{A}}^{-1}
=\left(\ba{ccc}
\bm{\hat{x}}_1'\cdot\bm{\hat{x}}_1'
&\cdots&
\bm{\hat{x}}_1'\cdot\bm{\hat{x}}_d'\\
\vdots&\ddots&\vdots\\
\bm{\hat{x}}_d'\cdot\bm{\hat{x}}_1'
&\cdots&
\bm{\hat{x}}_d'\cdot\bm{\hat{x}}_d'
\ea\right),
\end{align}
and
\begin{align}
\label{G.Vdmo}
&V_{d-1}
=
\left(\det[\bm{\la}^{-1/2}\bm{U}]^T
[\bm{\la}^{-1/2}\bm{U}]\right)^{1/2}
\nn\\
&~=
\left(
\det
[\bm{A}^{-1/2}]^T
[\bm{A}^{-1/2}]\right)^{1/2}
=\left(\det[[\bm{A}^{-1}]]\right)^{1/2},
\end{align}
where $[\bm{\la}^{-1/2}\bm{U}]$ and $[\bm{A}^{-1/2}]$
are the $d\times(d{-}1)$ matrices that result from removing the last
column from the matrices $\bm{\la}^{-1/2}\bm{U}$ and $\bm{A}^{-1/2}$,
respectively, and where $[[\bm{A}^{-1}]]$ is the
$(d{-}1)\times(d{-}1)$ left upper part of ${\bm{A}}^{-1}$.
Combining (\ref{G.LVdVdmo}), (\ref{G.Vd}), and (\ref{G.Vdmo}) gives
(\ref{G.Lbar}).

For the special case where $\bm{A}$ is diagonal, we obtain
$V_d=\prod_{i=1}^dA_{ii}^{-1/2}$ and
$V_{d-1}=\prod_{i=1}^{d-1}A_{ii}^{-1/2}$
and thus $V_d/V_{d-1}=A_{dd}^{-1/2}$.

\section{Free energy}
\label{G.Ap.free.energy}

Here we derive $f_s(t,L)$ of the Gaussian lattice model in \mbox{$1<d<4$}
dimensions for film geometry with the various b.c.\ for both the isotropic
case and the anisotropic case.

While we follow in spirit the derivation given for DD b.c.\ in
\cite{ChDo03}, two simplifications arise:
(i) In \cite{ChDo03} a slab geometry was investigated, of which the film
geometry is only a limiting case;
(ii) we perform an exact separation of the surface contributions at
an early stage of the calculation and reduce the remaining computations
for all b.c.\ to the case of periodic b.c..

First we consider the isotropic case.
We start from (\ref{G.ftbc}) and (\ref{G.fb}) and use
$\ln z=\int_0^\infty d\yi\yi^{-1}\left(e^{-\yi}-e^{-\yi z}\right)$
to write the excess free energy as
\begin{align}
\label{G.fex.int}
\fex(t,L)
&=
\f{1}{2\at^d}\int_0^\infty\f{d\yi}{\yi}e^{-\yi\rzt/2}
\B(\yi)^{d-1}\De\B_N(\yi),
\end{align}
with $\rzt$ defined after Eq.~(\ref{G.fbulk}),
\begin{align}
\label{G.DeBN}
\De\B_N(\yi)\equiv\B(\yi)-\B_N(\yi),
\end{align}
\begin{align}
\label{G.By.2pi}
\B(\yi)
&=
\f{1}{2\pi}\int_0^{2\pi}d\vp\,\exp[-\yi(1-\cos\vp)],
\end{align}
\begin{align}
\label{G.BN.def}
\B_N(\yi)
&=
\f{1}{N}\sum_{q_m}\exp[-\yi(1-\cos q_m\at)],
\end{align}
where the sum $\sum_{q_m}$ runs over the wave numbers given in
(\ref{G.ul}).
The quantity $\B(\yi)\equiv\lim_{N\to\infty}\B_N(\yi)$ in
(\ref{G.By.2pi}) is identical to $\B(\yi)$ from (\ref{G.Bfunc}).
By rearranging the sums it is possible to express $\B_N^\abc(\yi)$,
$\B_N^\NNbc(\yi)$, $\B_N^\DDbc(\yi)$, and $\B_N^\NDbc(\yi)$
in terms of $\B_N^\pbc(\yi)$.
For example, for DD b.c.\
\begin{align}
\label{G.DD.from.p.bc}
\lefteqn{\B_N^\DDbc(\yi)}
\nn\\
&=
\f{1}{2N}\left(\sum_{m=0}^{N-1}{+}\sum_{m=N+1}^{2N}\right)
\exp\left[-\yi\left(1{-}\cos\f{\pi(m{+}1)}{N{+}1}\right)\right]
\nn\\
&=
-\f{1+e^{-2\yi}}{2N}+\f{N+1}{N}
\B_{2(N+1)}^\pbc(\yi),
\end{align}
where in the first step we have exploited the symmetry of the cosine
about $\pi$, while in the second step $m=-1$ and $m=N$ terms have been
added to and subtracted from the sum and subsequently $m$ has been
renamed $m-1$.
Similar rearrangements can be performed for the other nonperiodic
b.c.\ and we obtain the exact relations
\bse
\begin{align}
\label{G.BN.a}
\De\B_N^\abc(\yi)
&=
2\De\B_{2N}^\pbc(\yi)-\De\B_N^\pbc(\yi),
\\
\label{G.BN.NN}
\De\B_N^\NNbc(\yi)
&=
\f{e^{-2\yi}-1}{2N}+\De\B_{2N}^\pbc(\yi),
\\
\label{G.BN.DD}
\De\B_N^\DDbc(\yi)
&=
\f{1{+}e^{-2\yi}{-}2\B(\yi)}{2N}
+\left(1{+}\f{1}{N}\right)\De\B_{2(N+1)}^\pbc(\yi),
\\
\label{G.BN.ND}
\De\B_N^\NDbc(\yi)
&=
\f{e^{-2\yi}-\B(\yi)}{2N}+\left(1+\f{1}{2N}\right)\De\B_{2N+1}^\abc(\yi).
\end{align}
\ese
This leads to the exact representation
\pagebreak
\begin{flalign*}
\fex(t,L)&=\f{2\fsf(t)}{L}
+\f{1}{2\at^d}\int_0^\infty\f{d\yi}{\yi}e^{-\yi\rzt/2}\B(\yi)^{d-1}
\times&
\end{flalign*}
\vspace{-15pt}
\bse
\begin{numcases}{\hspace{-15pt}}
\label{G.DeBN.p}
\De\B_N^\pbc(\yi), & \text{periodic b.c.,}
\\
\label{G.DeBN.a}
\De\B_N^\abc(\yi), & \text{antiperiodic b.c.,}
\\
\label{G.DeBN.NN}
\De\B_{2N}^\pbc(\yi), & \text{NN b.c.,}
\\
\label{G.DeBN.DD}
\left(1+\fr{1}{N}\right)\De\B_{2(N+1)}^\pbc(\yi), & \text{DD b.c.,}
\\
\label{G.DeBN.ND}
\left(1+\fr{1}{2N}\right)\De\B_{2N+1}^\abc(\yi), & \text{ND b.c.,}
\end{numcases}
\ese
for arbitrary \mbox{$L=N\at$} with $\De\B_N^\abc$ from (\ref{G.BN.a}).
For NN, DD, and ND b.c., the surface contribution $2\fsf(t)/L$
originates from the first term on the right hand sides of
(\ref{G.BN.NN}), (\ref{G.BN.DD}), and (\ref{G.BN.ND}),
respectively.
It is given by $\fsf^\Nbc(t)$, $\fsf^\Dbc(t)$, and $\fsf^\NDbc(t)$
provided in (\ref{G.fsf}) and (\ref{G.fsf.ND}).

The remaining tasks are (i) to determine the large-$N$ behavior of
$\B_N^\pbc(\yi)$ for periodic b.c.\ and (ii) to translate the result
to the other b.c..
For the first task it is useful to distinguish the regimes
$0\leq\yi\alt\yi_0$ and $\yi\agt\yi_0$ in the integral
(\ref{G.fex.int}), with $\yi_0$ chosen such that $1\ll\yi_0\ll N^2$.
Thus, for periodic b.c., we separate
\begin{align}
\label{G.fextL.12}
\fex(t,L)=\f{1}{2\at^d}(f_1+f_2),
\end{align}
\begin{align}
f_{1,2}
=
\int_{1,2}\f{d\yi}{\yi}e^{-\yi\rzt/2}\B(\yi)^{d-1}
\De\B_N^\pbc(\yi),
\end{align}
with $\int_1\equiv\int_0^{\yi_0}$ and $\int_2\equiv\int_{\yi_0}^\infty$,
corresponding to (A10) and (A11) of \cite{ChDo03}.
As shown in (A12)--(A17) of \cite{ChDo03}, the large-$N$ dependence
of $\De\B_N^\pbc(\yi)$ in the regime $0\leq\yi\alt\yi_0$ is of
$O(e^{-N})$, thus $f_1$ yields only exponentially small contributions.

Now consider $\yi\agt\yi_0$ with $\yi_0\gg1$.
Rewrite the sum over $q_m$ in (\ref{G.BN.def}) for periodic b.c.\ by
letting $m$ run over $m=-N/2,\ldots,N/2-1$ for even $N$ and
\mbox{$m=-(N-1)/2,\ldots,(N-1)/2$} for odd $N$.
Then only $|q_m\at|\ll1$ can lead to contributions to $\B_N^\pbc(\yi)$
in (\ref{G.BN.def}) that are not exponentially small and we may expand
\begin{align}
\label{G.ig.exp}
\exp&[-\yi(1-\cos q_m\at)]
\nn\\
&=\exp\left\{-\fr{1}{2}\yi(q_m\at)^2
\left[1+O((q_m\at)^2)\right]\right\}.
\end{align}
Correspondingly, we obtain in the regime $\yi\agt\yi_0$ , apart from
exponentially small corrections,
\begin{align}
\label{G.BNgpexp}
\B_N^\pbc(\yi)
&\approx
\f{1}{N}\sum_{m=-\infty}^{+\infty}
\exp\left[-2\yi\left(\fr{\pi m}{N}\right)^2\right]
=\f{1}{N}K\left(2\yi(\fr{\pi}{N})^2\right).
\end{align}

For the evaluation of $f_2$, we use (\ref{G.BNgpexp}) and keep only the
first term of
\begin{align}
\label{G.Byg}
\B(\yi)=\f{1}{\sqrt{2\pi\yi}}\left[1+O(\yi^{-1})\right].
\end{align}
Extending the lower integration limit in $f_2$ to $0$ leads only to
exponentially small corrections.
Changing the integration variable according to $\zi=2\yi(\pi/N)^2$
gives for periodic b.c.\ the result (\ref{G.fexdef}) with
$\Gcal^\pbc(\xt)$ in (\ref{G.Gcal.p}).
Due to (\ref{G.DeBN.a})--(\ref{G.DeBN.ND}), we confirm (\ref{G.fexdef})
also for the other b.c.\ under consideration here (the surface terms are
absent also for antiperiodic b.c.), with the $\Gcal(\xt)$ provided in
(\ref{G.Gcal.a}), (\ref{G.Gcal.NN.DD}), and (\ref{G.Gcal.ND}), with
specializations to $d=3$ in (\ref{G.Gcal.pa.3}), (\ref{G.Gcal.NN.DD.3}),
and (\ref{G.Gcal.ND.3}).

Now add to $L^{-d}\Gcal(\xt)$ the bulk singular part of the free energy
$\fbs$ from (\ref{G.fbs}) and, for NN or DD b.c.\ and $d\neq3$,
the surface singular part from (\ref{G.fsf.s.exp}) (the corresponding
part for ND b.c.\ vanishes, see Sec.~\ref{G.F.ND.2-4}).
Observing (\ref{G.fscal}) with $C_1$ given after (\ref{G.scalvar}) leads
to the scaling functions (\ref{G.Fcal.pa}), (\ref{G.Fcal.NN.DD}), and
(\ref{G.Fcal.ND}).

For $d=3$ and NN or DD b.c., we add to $L^{-3}\Gcal(\xt)$ with
$\Gcal(\xt)$ from (\ref{G.Gcal.NN.DD.3}) the bulk singular part of the
free energy $\fbs$ from (\ref{G.fbs}) with $Y_3$ from after (\ref{G.Yd})
and the surface singular part from (\ref{G.fsf.N.D.3.exp}) to obtain
the results (\ref{G.f.s.NN.DD.new}).

Inserting the small-$z$ expansion of $K(z)$ into (\ref{G.Gcal.p}) with
(\ref{G.Itd.p}) gives
\begin{align}
\label{G.Gp.large.xt}
&\Gcal^\pbc(\xt)
\nn\\
&=
-\int_0^\infty \frac{d\zi}{\pi}
\left(\f{\pi}{\zi}\right)^{(d+2)/2}
e^{-\zi\xt/(2\pi)^2}\sum_{n=1}^\infty e^{-n^2\pi^2/z}.
\end{align}
For large $\xt$, the right hand side is dominated by the first term
of the sum.
The remaining integral may be evaluated in a saddle point approximation.
This yields, together with (\ref{G.Gcal.a}),
(\ref{G.Gcal.NN.DD}), and (\ref{G.Gcal.ND}), the exponential decay of the
non-surface terms in (\ref{G.f.pa.large.L}), (\ref{G.f.NN.DD.large.L}),
and (\ref{G.f.ND.large.L}).

For the anisotropic case, we consider two different couplings $\Jpp$
and $\Jpl$, see (\ref{G.Jplpp}).
Then (\ref{G.fex.int}) is replaced by
\begin{align}
\label{G.fex.int.aniso}
&\hspace{-5pt}\fex(t,L)
\nn\\
&\hspace{-5pt}=
\f{1}{2\at^d}\int_0^\infty\f{d\yi}{\yi}e^{-\yi\rztp/2}
\B((\Jpl/\Jpp)\yi)^{d-1}\De\B_N(\yi),
\end{align}
with $\rztp\equiv\rz\at^2/(2\Jpp)$.
For the leading singular finite-size terms only the leading large-$\yi$
behavior of $\B(\yi)$ given in (\ref{G.Byg}) matters, see the
derivation of (\ref{G.Gcal.p}) for periodic b.c.\ above and its
translation through (\ref{G.DeBN.a})--(\ref{G.DeBN.ND}) to the other b.c.,
manifested in (\ref{G.Gcal.a}), (\ref{G.Gcal.NN.DD}), and
(\ref{G.Gcal.ND}).
The same is true for the leading singular surface terms as may be
inferred from the derivation of (\ref{G.fsf.s.exp}) for $d\neq3$.
Thus the factor $\Jpl/\Jpp$ in the argument of $B$ in
(\ref{G.fex.int.aniso}) leads to an additional factor
$(\Jpp/\Jpl)^{(d-1)/2}$ in front of the leading singular contributions
to the excess free energy $\fex(t,L)$.
Eqs.~(\ref{G.fbulk})--(\ref{G.Wtdz}) are replaced
by
\begin{align}
\fb(t)=\f{1}{2\at^d}\left[\ln\f{\Jpp}{\pi}
+\Wt{d}(\rztp,\Jpl/\Jpp)\right],
\end{align}
\begin{align}
\label{G.Wtdz.aniso}
\Wt{d}(z,w)\equiv
\int_0^\infty\f{d\yi}{\yi}\left[e^{-\yi/2}
-e^{-z\yi/2}\B(\yi)\B(w\yi)^{d-1}\right].
\end{align}
Thus, because of the argument $(\Jpl/\Jpp)\yi$ of $B$, the same factor
$(\Jpp/\Jpl)^{(d-1)/2}$ appears in front of the leading singular bulk
part.
Since the temperature dependence enters only through the parameter
$\rztp=\at^2/\xi_\perp^2$, where $\xi_\perp$ is the
correlation length (\ref{G.corr.perp}), it is straightforward to confirm
(\ref{G.fscal-anisospec}) for all b.c.\ on the basis of these properties.

\section{Comparison with Ref.~\cite{KrDi92a}}
\label{G.Ap.Galternative}

In Sec.~\ref{G.free.energy} we stated the identity of $\Gcal(\xt)$ with
the functions $\Th_+^{(1)}(y_+)$ used in \cite{KrDi92a} (as noted for
periodic and antiperiodic b.c.\ after Eq.~(\ref{G.Gcal.pa}), for NN and DD
b.c.\ after Eq.~(\ref{G.Gcal.NN.DD}), and for ND b.c.\
after Eq.~(\ref{G.Gcal.ND})).
For periodic b.c., this equivalence follows from using the expansion
(3.38) in \cite{DaKr04} in terms of Bessel functions, which provides
another representation of $\Gcal^\pbc$.
With $\xt=y_+^2$, we obtain
\begin{align}
\label{G.G.pid}
&\Gcal^\pbc(y_+^2)
\nn\\
&\equiv
-\f{1}{2\pi}\int_0^\infty d\zi\,(\pi/\zi)^{(d+1)/2}
e^{-\zi y_+^2/(2\pi)^2}
\left[K(\zi)-\sqrt{\pi/\zi}\right]
\nn\\
&=
-\f{1}{\pi}\sum_{n=1}^\infty\int_0^\infty d\zi(\pi/\zi)^{d/2+1}
e^{-\zi y_+^2/(2\pi)^2}e^{-n^2\pi^2/z}
\nn\\
&=
-2y_+^{d/2}\sum_{n=1}^\infty\f{K_{d/2}(ny_+)}{(2\pi n)^{d/2}}
\nn\\
&=
-\f{y_+^d}{(4\pi)^{(d-1)/2}\Ga(\f{d+1}{2})}
\sum_{n=1}^\infty
\int_1^\infty d\zi(\zi^2-1)^{(d-1)/2}e^{-n\zi y_+}
\nn\\
&=
-\f{y_+^d}{(4\pi)^{(d-1)/2}\Ga(\f{d+1}{2})}
\int_1^\infty d\zi\f{(\zi^2-1)^{(d-1)/2}}{e^{\zi y_+}-1}
\nn\\
&\equiv
\Th_{+\text{per}}^{(1)}(y_+),
\end{align}
valid for $y_+>0$.
The other identities between $\Gcal(\xt)$ and $\Th_+^{(1)}(y_+)$
follow similarly.
They may also be derived by showing the identities
(\ref{G.Gcal.a}), (\ref{G.Gcal.NN.DD}), and (\ref{G.Gcal.ND})
for the functions $\Th_+^{(1)}$ as represented in \cite{KrDi92a}.

\section{Functions}
\label{G.Ap.functions}

For $|z|<1$, the polylogarithms are defined by
$\Li_\nu(z)=\sum_{k=1}^\infty z^k/k^\nu$ and for $|z|\geq1$
by their analytic continuation.
They are analytic in the complex plane except at $z=1$ and except
for a branch cut that we take along $z\in]1,\infty[$.
We need $\Li_\nu(z)$ for $\nu=1,2,3$.
Well known relations are $\Li_\nu(1)=\ze(\nu)$ for $\Rp\nu>1$,
$\Li_\nu(-1)=(2^{1-\nu}-1)\ze(\nu)$,
$\Li_1(z)=-\ln(1-z)$, and
$\Li_\nu'(z)=\Li_{\nu-1}(z)/z$ for $z\notin[1,\infty[$, where
$\ze(\nu)\equiv\sum_{k=1}^\infty k^{-\nu}$ is Riemann's zeta function.
Combining them we may write for $z\geq0$
\begin{align}
\label{G.Li2Li3}
\Li_3(\pm e^{-z})&+z\Li_2(\pm e^{-z})
\nn\\
&=
\f{1}{8}(1\pm7)\ze(3)+\int_0^zdx\,x\ln(1\mp e^{-x}),
\end{align}
which is needed for App.~\ref{G.Ap.analytic.properties}.

The various functions $\Kc{d}^\tbc$ appearing in (\ref{G.Ccalneu}) read
\bse
\label{G.KtdpaNNN.DDD}
\begin{align}
\label{G.Ktdp}
\Kc{d}^\pbc(\yv)
&=
-\f{1}{32\pi^3}\int_0^\infty d\zi
\left(\pi/\zi\right)^{(d-3)/2}
e^{-\zi\yv/(2\pi)^2}
\times
\nn\\
&\ph{=}
\left[K(\zi)-\sqrt{\pi/\zi}\right],
\\
\label{G.Ktda}
\Kc{d}^\abc(\yv)
&=
-\f{1}{32\pi^3}\int_0^\infty d\zi
\left(\pi/\zi\right)^{(d-3)/2}
e^{-\zi\yv/(2\pi)^2}
\times
\nn\\
&\ph{=}
\left\{e^{\zi/4}\left[K(\zi/4)-K(\zi)\right]
-\sqrt{\pi/\zi}\right\},
\\
\label{G.KtdDD}
\Kc{d}^\DDbc(\yv)
&=
-\f{1}{2^{d+1}\pi^3}
\int_0^\infty d\zi\left(\pi/\zi\right)^{(d-3)/2}
e^{-\zi\yv/\pi^2}
\times
\nn\\
&\ph{=}
\left\{e^{\zi}[K(\zi)-1]-\sqrt{\pi/\zi}+1\right\}.
\end{align}
\ese
and $\Kc{d}^\NNbc(\yv)=2^{4-d}\Kc{d}^\pbc(4\yv)$,
$\Kc{d}^\NDbc(\yv)=2^{4-d}\Kc{d}^\abc(4\yv)$.

\section{\boldmath Analyticity properties}
\label{G.Ap.analytic.properties}
Here we show that the $d=3$ expressions (\ref{G.Fcal.a.3}) for
$\Fcal^\abc(\xt)$ and (\ref{G.f.s.DD.regular.new}) for
$L^3\fs^\DDbc(t,L)+(8\pi)^{-1}\xt\ln(L/\at)$ are analytic for
$\xt>-\pi^2$ and are finite and real at $\xt=-\pi^2$.
Combining (\ref{G.Gcal.a.3}) and (\ref{G.Li2Li3}), we obtain
for $\xt\geq0$
\begin{align}
\label{G.Gcal.a.int}
&\Gcal^\abc(\xt)
=
-\f{1}{2\pi}\left[-\f{3}{4}\ze(3)
+\int_0^{\sqrt{\xt}}dzz\ln(1+e^{-z})\right]
\nn\\
&=
\f{1}{4\pi}
\left\{\f{3}{2}\ze(3)+\f{1}{3}\xt^{3/2}
-\int_0^\xt d\xt'
\ln\left[2\cosh(\sqrt{\xt'}/2)\right]\right\}.
\end{align}
Since $\cosh(\sqrt{\xt'})$ is analytic and positive at $\xt'=0$, the
analytic continuation of $\Fcal^\abc(\xt)$ in (\ref{G.Fcal.a.3}) from
positive $\xt$ to other $\xt$ is analytic at $\xt=0$.
Inspection of (\ref{G.Gcal.a.int}) shows that the integral there
is also analytic for all other $\xt>-\pi^2$.
Thus (\ref{G.Fcal.a.3}) is analytic in $\xt$ for all $\xt>-\pi^2$.
Computing the integral in (\ref{G.Gcal.a.int}) for $\xt=-\pi^2$ and
combining the result with (\ref{G.Fcal.a.3}) gives the result
(\ref{G.Fcal.a.cfilm.d3}).
Combining (\ref{G.Gcal.NN.DD.3}) and (\ref{G.Li2Li3}), we obtain for
$\xt\geq0$
\begin{align}
\label{G.Gcal.DD}
\Gcal^\DDbc(\xt)
&=
-\f{1}{16\pi}\left[\ze(3)+\int_0^{2\sqrt{\xt}}dzz\ln(1-e^{-z})\right]
\nn\\
&=
-\f{1}{16\pi}
\Bigg[\ze(3)-\f{4}{3}\xt^{3/2}+\xt(\ln\xt-1)
\nn\\
&\hspace{40pt}
+2\int_0^\xt d\xt'\ln\f{2\sinh(\sqrt{\xt'})}{\sqrt{\xt'}}\Bigg].
\end{align}
Since $\sinh(\sqrt{\xt'})/\sqrt{\xt'}$ is analytic and positive at
$\xt'=0$, the analytic continuation of (\ref{G.f.s.DD.regular.new}) is
analytic at $\xt=0$.
The integral in (\ref{G.Gcal.DD}) is also analytic for all other
$\xt>-\pi^2$.
Thus (\ref{G.f.s.DD.regular.new}) is analytic in $\xt$ for all
$\xt>-\pi^2$.
Using (\ref{G.Gcal.DD}) to expand (\ref{G.f.s.DD.regular.new}) around
$\xt=-\pi^2$ for $\xt\geq-\pi^2$ gives (\ref{G.It3.NN.smallsplusi2}),
with a finite value of (\ref{G.f.s.DD.regular.new}) at $\xt=-\pi^2$.

\end{document}